\documentclass[12pt, a4paper]{article}
\usepackage{hyperref}
\usepackage{graphicx}
\usepackage{mathrsfs}
\usepackage{amssymb}
\usepackage{color}
\usepackage{amsmath}
\usepackage{amsthm}
\usepackage{booktabs}
\usepackage{lscape}
\usepackage{dcolumn}
\usepackage{longtable}
\usepackage{dcolumn}
\usepackage{arydshln}
\usepackage{bm}
\usepackage{caption}
\usepackage{subfig}
\usepackage{setspace}
\usepackage{cases}
\usepackage{comment}
\usepackage{multirow}
\usepackage[square, sort,comma,numbers]{natbib}
\bibpunct[, ]{(}{)}{;}{a}{}{,}
\usepackage{url}
\usepackage{textcomp}
\usepackage{bigfoot}

\newcommand{\vx}{\mathbf{x}}

\newcommand{\eps}{\epsilon}

\newcommand{\vgamma}{\bm{\gamma}}

\newcommand{\vzeta}{\bm{\zeta}}
\newcommand{\vomega}{\bm{\omega}}

\DeclareMathOperator*{\ssum}{\Sigma}

\begin{document}

\title{
\Large{Pandemic Influenza and Gender Imbalance:\\ Mortality Selection before Births}
\thanks{\protect\linespread{1.0}\protect\selectfont
Kota Ogasawara:~Department of Industrial Engineering, School of Engineering, Tokyo Institute of Technology, 2-12-1, Ookayama, Meguro-ku, Tokyo 152-8552, Japan (E-mail: ogasawara.k.ab@m.titech.ac.jp).
I wish to thank Jo\"el Floris, Yuzuru Kumon, Barbara Petrongolo, Eric Schneider, and participants of the CSG seminar, Tokyo Tech seminars, ESSH conference, EHS conference, and SSHA conference for their helpful comments.
I also thank Minami Yumitori for research assistance.
A previous version of this paper was circulated under the title ``Pandemic influenza and the gender imbalance: Evidence from early twentieth century Japan''.
Funding was provided by the Japan Society for the Promotion of Science (grant 21K01611).
There are no conflicts of interest to declare.
All errors are my own.
}
}

\author{Kota Ogasawara
}
\date{\today}
\maketitle

\begin{abstract}
This study uses data from the 1918--1920 influenza pandemic in Japan along with newly digitized and complete census records on births and infant deaths to analyze mortality selection \textit{in utero}.
I find that fetal exposure to the influenza pandemic during the first trimester of the pregnancy decreases the proportion of males at birth.
The results from mechanism analysis suggest that this decline in male births is associated with the deterioration of fetal and infant health.
This result supports a wide range of existing literature on the long-run adverse effects of pandemic influenza.
\bigskip

\bigskip


\end{abstract}
\newpage
\section{Introduction}

Since \citet{Almond:2006va}'s seminal work, several studies have offered evidence on the long-run adverse effects of \textit{in utero} exposure to pandemic influenza \citep{Almond:2011jf, Currie:2013ke, Prinz2018}.
However, a recent related work by \citet{Beach:2018tm} has highlighted that the birth cohorts exposed to the pandemic influenza were born to parents with relatively lower socioeconomic statuses (SESs) owing to the positive conscription effects during World War I (WWI).
They revealed that \citet{Almond:2006va}'s main estimates are more likely to be attenuated and are statistically insignificant when the background differences are considered.
Therefore, their findings contradict previous studies in the literature since it indicates that fetal influenza exposure does not induce negative selection into births. In other words, the exposure does not increase the number of unhealthy babies.

The current study tests the prenatal selection mechanism behind the fetal exposure to the pandemic influenza from two different empirical settings.
First, the census-based vital statistics of pre-war Japan were utilized. Notably, Japan could serve as a useful experimental setting because it provided very few servicemen, and thus, the parental characteristics were minimally influenced by the war.
The total number of mobilized forces was approximately $8,904,467$ in the British Empire, $4,355,000$ in the U.S., and $800,000$ in Japan.
However, most Japanese servicemen were not sent to the battlefields. Hence, the death toll in Japan was $300$, whereas the death toll in the British Empire and the U.S. were approximately $908,400$ and $116,500$, respectively \citep{wwi}.
In addition, the sample selection issue could be essentially insignificant as the census-based systematic birth records are available in pre-war Japan.
The availability of systematic statistics on the spatiotemporal variation in the intensity of the pandemic also helped to improve identification within the framework of panel data analysis.
Second, the underlying mechanism of the prenatal mortality selection is clarified by employing a biological paradigm.
The Trivers--Willard hypothesis \citep{Trivers:1973fd} provides a clear illustration on how the fetal shock induces the gender imbalance (i.e., declines in the proportion of males) at birth.
This framework enables us to test the prenatal selection using the sex ratio at birth instead of the fetal death rate, which has been known as a problematic measure for testing the prenatal mortality selection \citep{Valente:2015ci}.

By linking the prefecture-month level census-based sex ratio at birth with the spatiotemporal variation in the influenza exposures, this study found that fetal exposure to pandemic influenza between 1918--1920 decreased the proportion of males at birth, and this culling effect was concentrated on exposure during the first trimester of the pregnancy.
The results from a mechanism analysis using the comprehensive vital statistics on the infant mortality provide evidence that such a reduction in male births was associated with a scarring mechanism under which the distribution of fetal health endowment shifted to the left.

While the broad literature describing reproductive suppression has investigated the mechanism behind the selection before births \citep{Bruckner2018ssmph}, it remains understudied in the economic literature.
A few exceptional economic studies are \citet{Valente:2015ci} as well as \citet{Sanders:2015iq}, who found that the fetal exposure to civil conflict in Nepal and the Clean Air Act Amendments of 1970 in the United States were associated with a lower and higher secondary sex ratio, respectively.
While this current study supports the evidence from both studies, to the best of our knowledge, it is the first to provide evidence on the occurrence of scarring mechanism during the pandemic, implying that the initial health endowment of the exposed cohorts decreased due to fetal influenza exposure.
Therefore, the results support the mechanism via the fetal shock argued in a wider range of previous studies originating from \citet{Almond:2006va}, which have found the adverse long-run effects of the pandemic influenza.

The remainder of this paper is organized as follows.
Section~\ref{sec:sec2} introduces the empirical setting.
Sections~\ref{sec:sec3} and \ref{sec:sec4} describe the data used and estimation strategy, respectively, and Section~\ref{sec:sec5} presents the main results.
Section~\ref{sec:sec6} assesses the mechanism behind fetal shocks on the gender imbalance.
Section~\ref{sec:sec_rob} examines the robustness of the results.
Finally, Section~\ref{sec:sec8} discusses the study's conclusion and potential external validity.

\section{Empirical Setting} \label{sec:sec2}

\subsection{Theoretical Framework} \label{sec:sec21}

\subsubsection*{\textit{Proportion of Male Births}} \label{sec:sec211}

An influential study undertaken by \citet{Trivers:1973fd} in the field of biology proposes a hypothesis about the mechanism behind the determinants of the secondary sex ratio (sex ratio at birth).
As \citet{Catalano:2006vv} illustrated, the intuition of the Trivers--Willard hypothesis can be explained using shifts in the distribution of a random variable.

Let $z_{b} \sim \mathcal{N}(\theta_{b},\,\sigma^{2})$ and $z_{g} \sim \mathcal{N}(\theta_{g},\,\sigma^{2})$ be the initial health endowments of boys and girls \textit{in utero}, respectively.
Owing to natural selection, a certain threshold ($\lambda$) exists, below which fetuses are culled before birth.
Since male fetuses are more vulnerable \textit{in utero} than female fetuses \citep{Kraemer:2000to}, the mean initial health endowment of girls is greater than that of boys: $\theta_{g} > \theta_{b}$.
When I consider the cumulative distribution function for boys ($F_{b}(\cdot)$) and girls ($F_{g}(\cdot)$), this initial assumption implies that the probability of having culled fetuses is always greater for males than females because of the following condition:
\begin{eqnarray}\label{pdf} 
\begin{split}
F_{b}(\lambda) > F_{g}(\lambda).
\end{split}
\end{eqnarray}

If fetuses are exposed to health shocks \textit{in utero}, the distribution shifts to the left or the survival threshold moves to the right.
The former is called the ``scarring'' mechanism, whereas the latter is called the ``selection'' mechanism.
In both cases, condition (\ref{pdf}) indicates that the male share at birth must decrease, which corresponds to the proposition implied by the Trivers--Willard hypothesis.
This study investigates whether this proposition holds for the influenza pandemic in the early 20th century in industrializing Japan.

\subsubsection*{\textit{Mechanism}} \label{sec:sec212}

\begin{figure}[]
\centering
\captionsetup{justification=centering,margin=1.5cm}
\subfloat[Scarring mechanism]{\label{fig:scarring}\includegraphics[width=0.55\textwidth]{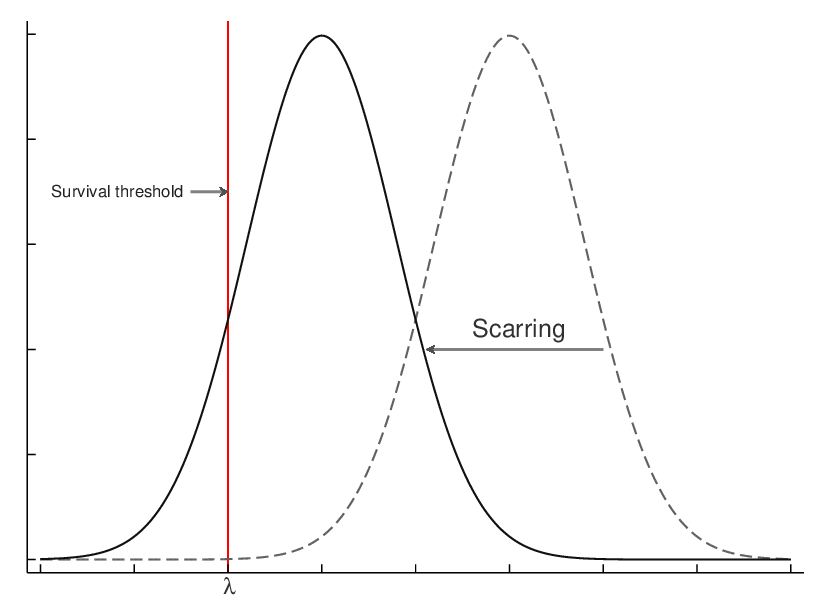}}\\
\subfloat[Selection mechanism]{\label{fig:selection}\includegraphics[width=0.55\textwidth]{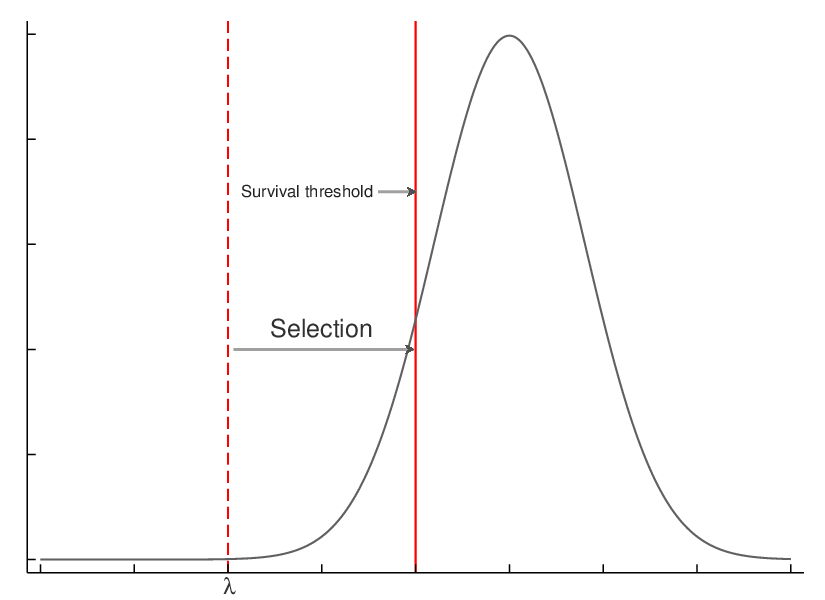}}
\caption{Scarring and selection mechanisms}
\label{fig:mechanism}
\scriptsize{\begin{minipage}{350pt}
\setstretch{0.85}
Notes: In Figure~\ref{fig:scarring}, the original probability density function is the dashed line, whereas the observed probability density function is the solid line.
In Figure~\ref{fig:selection}, the dashed line in red is the original survival threshold, whereas the solid line in red is the observed survival threshold.
Source: \citet{Catalano:2006vv}.
\end{minipage}}
\end{figure}

In contrast to culling before birth, the health status of an infant depends on the type of mechanism.
If the scarring mechanism works, the conditional mean of the truncated normal distribution shifts to the left because the original mean of the distribution moves to the left, as illustrated in Figure~\ref{fig:scarring}.
If the selection mechanism works instead, the conditional mean of the truncated normal distribution shifts to the right as the survival threshold moves to the right, cutting the lower tail of the original distribution, as illustrated in Figure~\ref{fig:selection}.
If both mechanisms work at the same time, the health status of an infant should therefore be unchanged.

Fortunately, the annual vital statistics reports provide complete figures on infant deaths in all prefectures (Section~\ref{sec:sec33}).
As the infant mortality rate can represent the health status at birth \citep{Almond:2006va}, I employ it as a proxy for the health status of infants and analyze the mechanism behind the observed sex ratio at birth during the influenza pandemic.
The fetal death rate cannot be used to test this mechanism due to the systematic issues in the statistics (Section~\ref{sec:sec31}).

\subsection{Maternal Stressor: Pandemic Influenza} \label{sec:sec22}

The Spanish influenza of 1918 infected 600 million people and killed 20--40 million patients worldwide \citep{Kilbourne:2006ce, Taubenberger:2006vq}.
After only five months of the first reported case of influenza in the United States, Spanish flu hit the Japanese archipelago between August 1918 and July 1920.
Figure~\ref{fig:ts_flu} shows the number of monthly deaths from pandemic influenza between 1918 and 1920 in Japan.
Similar to other Asian countries, there were two waves of the pandemic in Japan, with the first and second peaks observed in November 1918 and January 1920, respectively \citep{hayami2006}. 
Regarding the intensity of the pandemics, although the influenza mortality rate in Japan (4.5 per 1,000 people) between 1918 and 1919 was lower than that of other Asian countries, this rate was in a similar range to that of Western countries \citep{Rice:1993vp, hayami2010}.
Indeed, during the pandemic periods (August 1918--July 1919; September 1919--July 1920), more than one in every five people in Japan became infected with influenza \citep{csbhm1927}.

A key epidemical feature of the pandemic influenza is that it tends to affect young adults, which is a clear distinction from cyclical influenza \citep{Almond:2005uw, Erkoreka:2010fi}.
The official statistical reports in 1920 indicate that the most susceptible age range was 20--39.
Moreover, women aged 20--29 were more likely to be affected by the pandemic flu virus than men in the same age range (see Online Appendix~\ref{sec:seca_age} for details).
Since the average marrying age of Japanese women in the 1920s was 23 years old, the average age for childbearing is estimated to have been around 24--25 years old \citep{census1920v1, census1925v2}.
This means that during the pandemics, influenza tended to affect not only young adult women but also children in the womb via maternal infection.

Given these features of pandemics, a growing body of studies has employed pandemic influenza as a natural experiment to identify the long-term effects of fetal shocks on human capital formation \citep[e.g.,][]{Almond:2006va, Lin:2014bz}.
As these previous studies have found, pandemics show a certain random spatiotemporal distribution.
Figure~\ref{fig:map_flur} illustrates the spatial distribution of influenza death rates in the pandemic months of 1918--1920.
In the first wave, the epidemic cluster was generated in the southwestern region (Ky\=ush\=u and Shikoku) in November 1918 (Figure~\ref{fig:map_191811}); it then jumped to the northern and northeastern regions (Ch\=ubu and Toh\=oku) in the next month (Figure~\ref{fig:map_191812}) before moving to the central part of the main island (Kant\=o) (Figure~\ref{fig:map_19191}).
The second wave exhibits a more straightforward but not persistent transition.
The cluster was generated in the western region (Ch\=ugoku) in January 1920 (Figure~\ref{fig:map_19201}), and then transited to the northern region (Ch\=ubu) in the next month (Figure~\ref{fig:map_19202}).
It finally covered a broader region including the northeastern region (T\=ohoku) and Hokkaid\=o, a northeastern island.
The foregoing suggests that the patterns of the pandemics were not systematic or concentrated in a specific region.
This feature of the epidemics implies that there should be no systematic selection in the survivability of exposed pregnant women.
In other words, there must be a positive relationship during the pandemic wave if the influenza mortality were concentrated in unwealthy areas.
While Figure~\ref{fig:map_flur} does not support this sort of selection, I have further confirmed that the influenza mortality rates are not positively correlated with the lagged influenza mortality rates during each pandemic wave: November 1918--January 1919 for the first wave and January 1920--March 1920 for the second one.
Online Appendix~\ref{sec:seca6} presents a summary of these results.

A potential issue in the identification is the immediate response by the Japanese government.
However, due to the unavailability of a vaccine at the time, the government was unlikely to provide an efficient preventive policy during the pandemic.
Moreover, potential sorting might be an issue in the identification because this can cause measurement errors in the influenza death rate.
However, internal migration as an escaping strategy would not have worked because people could not have predicted the timing and place of the pandemics in the early 20th century \citep{hayami2006}.

\section{Data} \label{sec:sec3}

\subsection{Proportion of Male Births} \label{sec:sec31}

My main analysis uses a unique prefecture-month-level panel dataset on the male share at birth defined as the number of male births per 100 births.
I digitize the 1916--1922 editions of \textit{Nihonteikoku jink\=od\=otait\=okei} (Vital Statistics of Empire Japan, hereafter the VSEJ) to construct a dataset with the proportions of male births in 46 prefectures between January 1916 and December 1922.
Although I excluded Okinawa prefecture from the analyses because its vital statistics exhibit unnatural values in some cases \citep{Schneider:2018cx}, the data cover all births during the measured years because these vital statistics have been recorded based on the comprehensive national registration system (\textit{koseki}).
Online Appendices~\ref{sec:seca1} and \ref{sec:seca4} present the details of the vital statistics record, and the distribution of the male share at birth, respectively.
Panel A of Table~\ref{tab:sum} lists the summary statistics of the proportion of male births.
For the proportion of male births used in my analysis, several tests reject the null of unit root non-stationarity (Online Appendix~\ref{sec:secb_ts}).

The fetal death statistics are known to have critical challenges due to the reporting issues.
Although the VSEJ documents the number of fetal deaths by month, it does not record any information on the length of gestation period until fetal deaths.
Therefore, it is essentially difficult to match the timing of exposure to pandemic influenza with fetal death rates.
To avoid this issue, I use the male share at birth as the main dependent variable following the related previous studies \citep{Sanders:2015iq, Valente:2015ci}.
Despite this, using a complete census dataset on births in pre-war Japan as a case for a developing economy would contribute to the literature because data from household surveys on fetal losses used in related studies are more likely to suffer from unobserved selection issues due to small sample sizes \citep{Sanders:2015iq}.

\subsection{Influenza Mortality}\label{sec:sec32}

I use the influenza death rate, the number of deaths due to influenza per 10,000 people, as the key independent variable that captures the intensity of exposure to pandemic influenza.
The data on the monthly death tolls from influenza are obtained from the 1915--1922 editions of \textit{Nihonteikoku shiint\=okei} (Statistics of Causes of Death of the Empire of Japan, hereafter the SCDEJ), whereas the data on the population are taken from the official online database of the Ministry of Internal Affairs and Communications (Online Appendices~\ref{sec:seca2} and \ref{sec:seca3}).
Since fetal influenza exposure matters in my theoretical framework, I use the past nine-month average of influenza death rates in the regression analysis, given that the average pregnancy term was roughly nine months in prewar Japan \citep{tco1926}.
The impacts of fetal influenza exposure in each trimester are also considered in the flexible specification (Section~\ref{sec:sec4}).
Panel B of Table~\ref{tab:sum} shows the summary statistics of the average influenza death rates.
Although the minimum value is taken as  zero while considering the influenza death rates, these rates are not sparse.
The number of censored observations are approximately 50 (1.3\% of total observations) among all those rates.

\subsection{Infant Mortality Rates} \label{sec:sec33}

Analysis for testing the scarring and selection mechanisms requires data on the infant mortality by sex (Section~\ref{sec:sec21}).
The SCDEJ documented the complete censuses of the annual infant mortality for each prefecture. In contrast, the number of annual live births used as the denominator are available at the VSEJ (Online Appendices~\ref{sec:seca1} and~\ref{sec:seca2}).
I digitize the infant deaths and live births recorded in these documents to construct data on the infant mortality rates, that is, the number of infant deaths per $1,000$ live births, in 46 prefectures between 1916 and 1922.
Panel A of Table~\ref{tab:sum} shows the summary statistics of the infant mortality.

\subsection{Additional Control Variables} \label{sec:sec34}

To control for the observable factors, I include a set of available prefecture-year-level control variables.
The first set of controls are the indices of agricultural production.
Given the agrarian society at that time, a certain proportion of wealth can be captured by productivity in the agricultural sector.
I herein consider rice yield per hectare, soy yield per hectare, and milk production per capita as measures of potential wealth because these items were the main sources of carbohydrate and protein.
The data on these variables are digitized from \textit{Tod\=ofuken n\=ogy\=okisot\=okei} (Basic Statistics of Agriculture in Japanese Prefecture) edited by \citet{kayo1983}.
Another set of controls is access to medical care.
I include the share of medical doctors and midwives to control for access to medical care and related socioeconomic conditions and potential wealth level.
To obtain the data on medical access, I digitize the \textit{Nihonteikoku t\=okeinenkan} (Statistical Yearbook of the Japanese Empire, hereafter the SYEJ).
Online Appendices~\ref{sec:seca5} provides the details of the SYEJ.

Panel C of Table~\ref{tab:sum} presents the results of the simple balancing tests for these control variables because the pandemic severity might be correlated with the socioeconomic characteristics \citep{Correia:2020vx, Markel:2007uc}.
However for all the control variables, there are no systematic differences between the ad hoc treatment group that exceeds 75 percentile of the influenza death rate and the rest of the samples.
This result remains unchanged if I use the greater values of the thresholds.
This supports the evidence that the influenza severity measure used in the present study is plausibly exogenous as discussed in Section~\ref{sec:sec22}.

\section{Estimation Strategy} \label{sec:sec4}

A quasi-experimental approach using the spatiotemporal variation in influenza death rates is used to identify the impacts of fetal exposure to influenza on the sex ratio at birth.
My baseline specification is given as follows:
\begin{eqnarray}\label{bs} 
\footnotesize{
\begin{split}
y_{it} = \alpha + \beta \overline{\textit{FLUDR}}_{it} + \vx'_{ig_{t}} \vgamma + \nu_{i} + \phi_{t} + e_{it},
\end{split}
}
\end{eqnarray}
where $i$ indexes the prefecture, $t$ indexes the measured year-month, and $g_{t}$ indicates a group membership variable for the measured year (see Online Appendix~\ref{sec:secb_model} for more information on the model interpretation).
The variable $y$ is the proportion of male births, $\vx$ is a vector of the prefecture-year-level control variables, $\nu$ is the prefecture fixed effect, $\phi$ is the year-month-specific fixed effect, and $e$ is a random error term.
$\overline{\textit{FLUDR}}_{it}$ is the past nine-month average of influenza death rates.
My parameter of interest is $\beta$ and its estimate $\hat{\beta}$ captures the marginal effect of the influenza death rate on the proportion of male births.
Therefore, I expect $\hat{\beta}$ to be negative and statistically significant.

The first specification in equation~\ref{bs} assumes that the potential effects of fetal influenza exposure are constant regardless of the timing of exposure.
However, medical evidence suggests that fetuses are most susceptible to maternal stress in the first trimester when they experience rapid neuron differentiation and the proliferation of neuronal elements \citep{moore2013}.
This implies that the culling effects on male fetuses are much clearer in the first trimester than in the second and third trimesters.
Therefore, my preferred specification is as follows:
\begin{eqnarray}\label{fs} 
\footnotesize{
\begin{split}
y_{it} = \pi + \delta_{0} \overline{\textit{FLUDR}}_{it}^{\textit{First Trimester}} + \delta_{1} \overline{\textit{FLUDR}}_{it}^{Second Trimester} + \delta_{2} \overline{\textit{FLUDR}}_{it}^{\textit{Third Trimester}}\\
+ \vx'_{ig_{t}} \vzeta + \upsilon_{i} + \kappa_{t} + \eps_{it}.
\end{split}
}
\end{eqnarray}
The second to fourth terms on the right-hand side are the average influenza death rates during the first, second, and third trimesters, respectively.
For instance, the average influenza death rate during the first trimester is defined as: $\scriptsize{(\ssum_{j=7}^{9}\textit{Influenza deaths}_{it-j})/(\ssum_{j=7}^{9}\textit{Population}_{it-j})}$.
Therefore, I expect the estimates $\hat{\delta}_{1}$, $\hat{\delta}_{2}$, and $\hat{\delta}_{3}$ to be negative; among these, the estimate for the first trimester, $\hat{\delta}_{1}$, shows the clear adverse effects on the proportion of male births.

Since I use a within estimator for the fixed effect models in equations \ref{bs} and \ref{fs}, the identification depends on the sharp increases in influenza mortality during the pandemic years (Figure~\ref{fig:ts_flu}).
As discussed in Section~\ref{sec:sec22}, these influenza death rates are plausibly exogenous because no vaccination was available in the pre-war period and internal migration was unrealistic given the rapid spread of the virus.
Despite this preferable feature for the identification, I control for a large proportion of the unobservable factors and observable characteristics in the following ways.
First, I control for prefecture-specific time-invariant factors such as the baseline wealth level and geographical features using prefecture fixed effects.
In some regressions, I also consider the prefecture-specific liner time trend to allow the prefectures to have different trends in the proportion of males at births.
Since the sex ratio at birth is a biological measure rather than a socioeconomic outcome, the proportion of male births in both high- and low-exposed prefectures would have shown similar trends in the absence of the pandemic.
An event-study analysis shown in Online Appendix~\ref{sec:sec_es} provides evidence that the trends would have been plausibly similar.
In addition, Online Appendix~\ref{sec:secb_ct} confirmed that the prefectures with different intensities of influenza mortality during the pandemic indeed show similar trends in the proportion of male births.
Second, the macroeconomic shocks and cyclical effects of seasonal epidemics are captured using year-month fixed effects.
Using the year-month fixed effects also helps to  effectively control the potential influence of the macroeconomic wartime inflation.
After controlling for these fixed effects, the remaining potential confounding factors included in the error term that might be correlated with the influenza death rate include the time-varying wealth level and access to medical care.
To control for these factors, I further include the set of available prefecture-year-level control variables introduced in Section~\ref{sec:sec34}.

To address the potential spatial and prefecture-specific within correlations, I use the cluster-robust variance estimator (CRVE) and cluster standard errors at the 8-area level (Online Appendices~\ref{sec:secb_cluster} and \ref{sec:secb_serial} summarize the criterion of clusters and test results for the autocorrelation, respectively).
In other words, the method used in the present study allows for heteroskedasticity and arbitrary serial correlation within clusters and addresses the potential heteroskedasticity across clusters \citep{Arellano1987}.
Moreover, to address the small number of clusters in the CRVE, I adopt the wild cluster bootstrap-t method for statistical inference \citep{Cameron:2008ws}.
All the regressions are weighted by the average number of births over the sample period in each prefecture, given that the dependent variable is the proportion of male births (Section~\ref{sec:sec31}).

\section{Results} \label{sec:sec5}

Table~\ref{tab:r_main} presents the results.
Columns (1)--(4) present the results for the entire period (January 1916--December 1922).
Column (1) shows the results from the baseline specification in equation~\ref{bs}.
The estimate is negative but statistically insignificant.
This result is unchanged if I include the prefecture-specific time trend in column (2).
Column (3) shows the result from my preferred specification in equation~\ref{fs}.
The estimates listed in this column suggest that fetal influenza exposure in the first trimester has a statistically significantly negative effect on the proportion of male births, whereas that during the second and third trimesters does not have such an effect.
As explained, this finding is consistent with the fact that fetuses are more vulnerable in the first trimester than in the other trimesters.
In column (4), I find that this result is robust to including the prefecture-specific time trend as expected.

Columns (5) and (6) present the results for non-pandemic years (January 1916--December 1917 and January 1921--December 1922), whereas columns (7) and (8) present the results for pandemic years (January 1918--December 1920).
In columns (5) and (6), I find no statistically significant effects of fetal exposure to influenza on the proportion of male births during non-pandemic years.
By contrast, columns (7) and (8) show the clear significant adverse effects of fetal influenza exposure during pandemic years.
This implies that while seasonal influenza does not have any significant impacts on the secondary sex ratio, pandemic influenza does have such an effect.
The result of this placebo experiment supports the evidence that the identification in my within estimator using the sharp increase in influenza death rates during pandemic years seems to work well and should provide reliable estimates.

If I use the maximum average influenza death rate in the first trimester (Panel B of Table~\ref{tab:sum}) as the reference value to calculate the magnitude, the estimate in column (8) implies that fetal exposure to pandemic influenza decreased the proportion of male births by approximately 1.5 percentage points ($9.97 \times 0.1531$).
This magnitude is not very large but still non-negligible given that one standard deviation of the proportion of male births is 1.67 (Panel A of Table~\ref{tab:sum}).
The average influenza death rate during the pandemic periods is used herein because the identification should rely on the dramatic rise in the influenza mortality rates during those periods (Online Appendix~\ref{sec:secb_aff} presents a summary of the detailed discussion).
This means that the magnitude calculated herein is conservative because of the attenuation effects owing to the linear functional form assumption in equation~\ref{fs}.

\section{Mechanism} \label{sec:sec6}

\subsection{Estimation Strategy} \label{sec:sec61}

Previous results suggested that pandemic influenza can disturb the gender balance at birth.
Subsequently, I test whether the scarring or selection mechanism drove the gender imbalance at birth due to pandemic influenza by using the infant mortality rates.

Since infant mortality is measured at the annual level, I calculate a weighted influenza death rate using the monthly variations in the number of influenza deaths and live births to improve the assignments.
The weighted influenza death rate in prefecture $i$ in year $l$ is defined as follows:
\begin{eqnarray}\label{wdr}
\footnotesize{
\textit{Weighted FLUDR}_{il} = 
\frac{\sum_{m=\text{Jan}}^{\text{Dec}}\textit{Birth}_{ilm} \times \overline{\textit{FLUDR}}_{ilm}}
{\sum_{m=\text{Jan}}^{\text{Dec}}\textit{Birth}_{ilm}},
}
\end{eqnarray}
where $\textit{Birth}_{ilm}$ is the number of live births in month $m$ and $\overline{\textit{FLUDR}}_{ilm}$ is the past nine-month average of influenza mortality in month $m$.
This transformation takes both the severity of influenza exposure and the timing of birth into account: $\overline{\textit{FLUDR}}_{ilm}$ captures the treatment intensity, whereas the weight, $\textit{Birth}_{ilm}$, coordinates the differences in the timing of birth.
\citet{Ogasawara:2018hk} showed evidence that this transformation improves the treatment assignment to a certain extent if I compare it using a simple lagged influenza death rate.
The baseline specification is then given as follows:
\begin{eqnarray}\label{imrbs}
\footnotesize{
\begin{split}
h_{il} = \psi + \rho \text{\textit{Weighted FLUDR}}_{il} + \vx'_{il} \vomega + \theta_{i} + \iota_{l} + u_{il},
\end{split}
}
\end{eqnarray}
where $h$ is the infant mortality rate, $\vx$ is a vector of the same control variables introduced above, $\theta$ is the prefecture fixed effect, $\iota$ is the year-specific fixed effect, and $u$ is a random error term.
Online Appendix~\ref{sec:secb_ct} provides evidence that the prefectures which experienced different intensities of influenza mortality during the pandemic exhibit similar trends in infant mortality.
See Online Appendix~\ref{sec:secb_placebo} for the null results from the placebo tests to check the pre-treatment trends.
Despite this, I also consider a prefecture-specific time trend in some specifications, as in equation~\ref{bs}.
My parameter of interest is $\rho$ and its estimate $\hat{\rho}$ captures the marginal effect of the influenza death rate on the infant mortality rate.
As explained, I expect $\hat{\rho}$ to be negative and statistically significant if the selection mechanism works, whereas it should be statistically significantly positive if the scarring mechanism is relevant.

In the flexible specification, I consider the weighted influenza death rates for the first, second, and third trimester by replacing $\overline{\textit{FLUDR}}_{ilm}$ in equation~\ref{wdr} with the average influenza mortality rates for each trimester.
The flexible specification is given as follows:
\begin{eqnarray}\label{imrfs}
\footnotesize{
\begin{split}
h_{il} = \tau + \varrho_{0} \text{\textit{Weighted FLUDR}}_{il}^{\textit{First Trimester}} + \varrho_{1} \text{\textit{Weighted FLUDR}}_{il}^{\textit{Second Trimester}}\\
 + \varrho_{2} \text{\textit{Weighted FLUDR}}_{il}^{\textit{Third Trimester}} + \vx'_{il} \bm{\Gamma} + \xi_{i} + \varsigma_{l} + \varepsilon_{il}.
\end{split}
}
\end{eqnarray}
This specification allows us to investigate the most sensitive trimester for the impacts of fetal influenza exposure on infants' health.
One must be careful here, as in the regressions using infant mortality as a dependent variable, I do not necessarily expect the first trimester to be the most vulnerable for infants' health.
While fetuses are indeed relatively vulnerable during the first trimester, those affected by any shocks during this trimester are culled before birth.
In other words, surviving fetuses are positively selected into birth.
As described in Section~\ref{sec:sec21}, the conditional expectation of the truncated normal distribution is consistently greater than the original distribution.
Therefore, the observed (i.e., surviving) infants may be more sensitive to shocks during the second and/or third trimesters than those during the first trimester.
This natural selection mechanism suggests that the estimates $\hat{\varrho}_{1}$ and/or $\hat{\varrho}_{2}$ can be positive (negative) if the scarring (selection) mechanism works, whereas the estimate $\hat{\varrho}_{0}$ can be negative or statistically insignificant.

The inferences are conducted similarly for the specifications for the proportion of male births.
I use the CRVE and cluster the standard errors at the 8-area level to address the potential spatial and within correlations.
The wild cluster bootstrap-t method is employed for the statistical inference.
All the regressions are weighted by the average number of live births over the sample period in each prefecture, given that the dependent variable is the infant mortality rate (Section~\ref{sec:sec33}).

\subsection{Results} \label{sec:sec62}

Table~\ref{tab:r_main_imr} presents the results.
Panels A--C of this table present the results for the infant mortality rates for all infants, boys, and girls, respectively.
Columns (1) and (3) show the results from equations~\ref{imrbs} and \ref{imrfs}, respectively.
Columns (2) and (4) add the prefecture-specific time trend for both equations.

Column (1) of Panel A shows that the estimated effect of fetal influenza exposure on the infant mortality rate is positive and statistically significant.
This result is unchanged if I consider the prefecture-specific time trend in the infant mortality rate in column (2).
This implies that the scarring mechanism might have driven the gender imbalance at birth.
Column (3) of Panel A indicates that such an effect was concentrated on exposure during the third trimester as expected.
This result remains unchanged if I further control for the average influenza death rate during the ``fourth'' trimester, that is, between 1 and 3 months after birth (not reported).
This supports the evidence that the exposure during the third trimester is not a decent proxy for exposure in the neonatal period.
This result is still unchanged after controlling for the prefecture-specific time trend in column (4).
The estimate in column (4) indicates that a one standard deviation increase in the weighted influenza mortality rate increases the infant mortality rate by $15.1$ permil ($20.907 \times 0.72$).

Panels B and C of Table~\ref{tab:r_main_imr} show similar results for boys and girls.
An interesting gender difference can be highlighted: the estimates for girls are greater in magnitude than those for boys.
For example, if I compare column (2) of Panel B with that of Panel C, the estimate for girls is approximately $4.5$ permil greater than that for boys ($8.556-4.053$).
To test the gender difference, I pooled the infant mortality rates for boys and girls and interacted all the independent variables including fixed effects with the gender dummy (Online Appendix~\ref{sec:secb_gd} summarizes these results).
I then confirm that this difference in magnitude is statistically significant.
While this gender difference becomes statistically insignificant if I focus on the third-trimester effects reported in columns (3) and (4) of Panel B and C, this must be because the gender differences in the effects are generated by the cumulative effects of all trimesters.

This gender difference is considered to be consistent with the scarring mechanism because the total shift of the mean of the distribution induced by the scarring mechanism can be heterogeneous by gender.
As explained, the distribution of the fetal health endowment must shift to the left if the scarring mechanism works.
Suppose the distributions of both boys and girls shift to the left by same degree and that the survival threshold is fixed at $\lambda$ (in Figure~\ref{fig:mechanism}).
Then, the net shift of the distribution depends only on the degree of the selection effect due to the truncation (at $\lambda$).
Since the selection effect on the male fetus is always greater than that on the female fetus as condition~\ref{pdf} suggests, the net leftward shift of the distribution of girls can be greater than that of boys.
In fact, given that the conditional expectation of the truncated normal distribution of $z_{g}$ can be written as $E[z_{g}|z_{g}>\lambda]=\theta_{g}+\sigma f_{g}(\lambda)/(1-F_{g}(\lambda))$, the selection effect due to the truncation is expressed as $\sigma f_{g}(\lambda)/(1-F_{g}(\lambda))$.
The gender difference (boys minus girls) of this term can then be written as $\sigma (f_{b}(\lambda) - f_{g}(\lambda))/((1-F_{b}(\lambda)(1-F_{g}(\lambda))$, which is positive because $f_{b}(\lambda) > f_{g}(\lambda)$.
This means that the total shift of the mean caused by the scarring mechanism should be greater for girls than for boys.

The strong preferences for sons may be another potential interpretation of the result. Parents treated the exposed girls and boys differently during the post-natal period.
If this gender discrimination is true, the infant mortality rates of girls should always be greater than boys.
However, Panel A of Table~\ref{tab:sum} does not support this interpretation. The mean infant mortality rate of girls was $162$ permil, whereas that of boys was $179$ permil.

\section{Robustness}\label{sec:sec_rob}

\begin{figure}[]
\centering
\includegraphics[width=9cm]{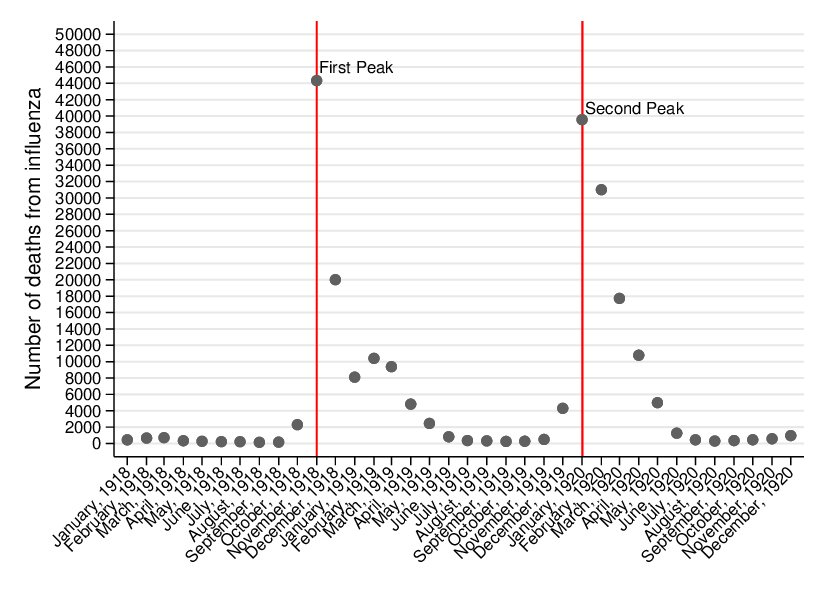}
\caption{The number of deaths from influenza between January 1918 and December 1920}
\label{fig:ts_flu}
\scriptsize{\begin{minipage}{350pt}
Notes: The red lines show the first and second peaks of the number of deaths due to influenza.
Source: Created by the author from \citet{scdej1918, scdej1919, scdej1920}.
\end{minipage}}
\end{figure}

I have conducted several sensitivity checks to confirm the robustness of the main results.
First, the results from a few placebo experiments within the framework of event study analysis provide evidence that the exposure variables do not capture any secular pre-trends in the outcome variables.
Online Appendices~\ref{sec:sec_es} and \ref{sec:secb_placebo} summarize these results.
Second, I have summarized a set of robustness checks on the potential omitted variable bias in Online Appendix~\ref{sec:sec71}.
These results confirm that the main results are robust against including the additional control variables on the weather shocks (Online Appendix~\ref{sec:sec711}), air pollution (Online Appendix~\ref{sec:sec712}), and parental school enrollment rates (Online Appendix~\ref{sec:sec713}).
Third, I assess the possibility that the influenza mortality might have been correlated with the unobservable spatio-temporal variations.
I have confirmed that my main results are materially similar if equation~\ref{fs} (\ref{imrfs}) uses the prefecture-by-year-by-quarter (prefecture-by-period) fixed effect rather than the prefecture and year fixed effects (Online Appendix~\ref{sec:sec714}).
Finally, I stratify the sample to confirm that the adverse effects are observed only in the pandemic years between 1918 and 1920.
Suppose my empirical setting was valid in capturing the overall effects of fetal influenza exposure on the proportion of male births. In that case, the estimated effect in the first wave must be clearer than in the second wave because the scale and term of epidemics were greater during the first wave than in the second wave (Figure~\ref{fig:ts_flu}). 
The results indeed support the evidence that my estimates respond to the intensity of epidemics well.
Online Appendix~\ref{sec:sec_str} summarizes these results.

\section{Conclusion}\label{sec:sec8}

The present study uses influenza pandemic as a natural experiment to investigate mortality selection \textit{in utero} by utilizing the comprehensive birth registration records from the beginning of the 20th century of Japan.
I find that fetal influenza exposure during the first trimester of the pregnancy period has a negative impact on the proportion of males at birth.
Mechanism analysis provides evidence that the reduction in male births is associated with a scarring mechanism under which the distribution of fetal health endowment shifted to the left.
Therefore, the present study reveals the channel to support a growing body of literature exploring the adverse long-run effects of fetal influenza exposure.

Using aggregate prefecture-level data, however, makes it difficult to precisely identify the actual assignments of the exposure at the individual level.
Although I use an appropriate set of weights for the analyses, the estimated effects found in this study are therefore considered to be the lower bounds rather than the true unobserved treatment effects.
Similarly, the aggregation also makes it difficult to separate the scarring effect via infection and selection effect via maternal mental stress.
In addition, this study neither theorizes the case of multi-fetus pregnancy nor of fragile females, which has been recently accepted argument in biological science \citep{Catalano2021emph, Kolk:2016ajhb, Wu2021hr}.
One should also be careful that the recent reproductive suppression literature is less likely to support the male-specific selection induced by exogenous stressors in the early gestation period \citep{Orzack2015pnas}.
Notwithstanding these limitations, my main results show the non-negligible impacts of the pandemic on the fetal health endowment that are robust to the confounding factors and potential endogenous responses.

The study findings also yield other insights into the phenomenon.
First, testing the mechanism in the mortality selection before birth is useful as a sanity robustness check in future studies focusing on the short- and/or long-run impacts of fetal exposure to pandemic influenza or other infectious diseases, such as COVID-19 (see \citet{Arthi2021eeh} and \citet{Beach:2021jel} for the discussions on the associations between the COVID-19 pandemic and historical pandemics).
Additional data on the health status of infants can offer sound evidence on the estimated adverse effects of the fetal shocks if the scarring mechanism is supported.
Second, the proportion of males in the birth cohorts exposed to the pandemic influenza in their post neonatal period is lower than those of the surrounding cohorts.
Thus, future research can further examine the potential long-term effects of fetal exposure to the pandemics (or even the other fetal shocks) on the gender imbalance in adulthood, which may cause disruption in the functions of marriage and labor markets \citep{Angrist:2010tz, Chiappori:2002bx, Abramitzky:2011bu, Bethmann:2012il, Brainerd:2017bw}.
In Online~\ref{sec:secc}, I have provided some evidence on the long-run impacts of the early-life influenza exposures on the childhood sex ratio using the official reports of the population censuses.

Finally, the potential external validity of this study should be discussed.
During the pandemic years, Japan experienced similar severity to the Western countries \citep{Rice:1993vp}.
The gender and age biases in infections explained in Section~\ref{sec:sec22} were also observed in these countries \citep{RICHARD:2009eb, Reid:2005vm}.
Although potentially influential institutions in Japan might be the infanticides \citep{Bruckner:2011ajhb}, the number of infanticides has been substantially decreased since the early 20th century \citep{Drixler2016}. This implies that the results obtained in this paper are unlikely to be disturbed by this institution.
Therefore, the influenza epidemic used as a natural experiment in this study should not be a peculiar event in prewar Japan.
In this light, this study might offer useful suggestions regarding future pandemics for today's developing countries having similar per capita GDP to Japan during the pandemic years, such as Cameroon, Kenya, Sudan, and Tanzania, based on the 2018 per capita GDP \citep{Bolt2020}.
However, one must carefully pay attention to the differences, at least in the medical technologies, initial population health, and disease environments, when applying the findings from this study to these countries.

\begingroup
	\setlength{\bibsep}{6pt}
	\setstretch{0.90}
	\bibliographystyle{plainnat}
	\bibliography{paper.bib}

\begin{thebibliography}{56}
\providecommand{\natexlab}[1]{#1}
\providecommand{\url}[1]{\texttt{#1}}
\expandafter\ifx\csname urlstyle\endcsname\relax
  \providecommand{\doi}[1]{doi: #1}\else
  \providecommand{\doi}{doi: \begingroup \urlstyle{rm}\Url}\fi

\bibitem[Abramitzky et~al.(2011)Abramitzky, Delavande, and
  Vasconcelos]{Abramitzky:2011bu}
Ran Abramitzky, Adeline Delavande, and Luis Vasconcelos.
\newblock {Marrying Up: The Role of Sex Ratio in Assortative Matching}.
\newblock \emph{American Economic Journal: Applied Economics}, 3\penalty0
  (3):\penalty0 124--157, July 2011.

\bibitem[Almond(2006)]{Almond:2006va}
Douglas Almond.
\newblock {Is the 1918 Influenza Pandemic Over? Long-Term Effects of In Utero
  Influenza Exposure in the Post-1940 U.S. Population}.
\newblock \emph{Journal of Political Economy}, 114\penalty0 (4):\penalty0
  1--41, 2006.

\bibitem[Almond and Currie(2011)]{Almond:2011jf}
Douglas Almond and Janet Currie.
\newblock {Killing Me Softly: The Fetal Origins Hypothesis}.
\newblock \emph{Journal of Economic Perspectives}, 25\penalty0 (3):\penalty0
  153--172, August 2011.

\bibitem[Almond and Mazumder(2005)]{Almond:2005uw}
Douglas Almond and Bhashkar Mazumder.
\newblock {Almond and Mazumder 2005 [AERPP]}.
\newblock \emph{American Economic Review}, 95\penalty0 (2):\penalty0 258--262,
  May 2005.

\bibitem[Angrist(2002)]{Angrist:2010tz}
Josh Angrist.
\newblock {How do sex ratios affect marriage and labor markets? Evidence from
  America's second generation}.
\newblock \emph{Quarterly Journal of Economics}, 117\penalty0 (3):\penalty0
  997--1038, August 2002.

\bibitem[Arellano(1987)]{Arellano1987}
Manuel Arellano.
\newblock {Computing standard errors for robust within-groups estimators}.
\newblock \emph{Oxford Bulletin of Economics and Statistics}, 49:\penalty0
  431--434, 1987.

\bibitem[Beach et~al.(2022{\natexlab{a}})Beach, Brown, Ferrie, Saavedra, and
  Thomas]{Beach:2018tm}
Brian Beach, Ryan Brown, Joseph Ferrie, Martin Saavedra, and Duncan Thomas.
\newblock {Re-evaluating the Long-Term Impact of In Utero Exposure to the 1918
  Influenza Pandemic}.
\newblock \emph{Journal of Political Economy}, pages 1--40, 2022{\natexlab{a}}.

\bibitem[Beach et~al.(2022{\natexlab{b}})Beach, Clay, and
  Saavedra]{Beach:2021jel}
Brian Beach, Karen Clay, and Martin Saavedra.
\newblock {The 1918 Influenza Pandemic and Its Lessons for COVID-19}.
\newblock \emph{Journal of Economic Literature}, 60\penalty0 (1),
  2022{\natexlab{b}}.

\bibitem[Bethmann and Kvasnicka(2012)]{Bethmann:2012il}
Dirk Bethmann and Michael Kvasnicka.
\newblock {World War II, Missing Men and Out of Wedlock Childbearing}.
\newblock \emph{The Economic Journal}, 123\penalty0 (567):\penalty0 162--194,
  August 2012.

\bibitem[Bolt and van Zanden(2020)]{Bolt2020}
Jutta Bolt and Jan~Luiten van Zanden.
\newblock {Maddison Project Database, version 2020}.
\newblock 2020.
\newblock
  \url{https://www.rug.nl/ggdc/historicaldevelopment/maddison/publications/wp15.pdf}.

\bibitem[Brainerd(2017)]{Brainerd:2017bw}
Elizabeth Brainerd.
\newblock {The Lasting Effect of Sex Ratio Imbalance on Marriage and Family:
  Evidence from World War II in Russia}.
\newblock \emph{Review of Economics and Statistics}, 99\penalty0 (2):\penalty0
  229--242, May 2017.

\bibitem[Bruckner et~al.(2011)Bruckner, Subbaraman, and
  Catalano]{Bruckner:2011ajhb}
T~A Bruckner, M~Subbaraman, and RA~Catalano.
\newblock {Transient cultural influences on infant mortality: Fire-Horse
  daughters in Japan}.
\newblock \emph{American Journal of Human Biology}, 23\penalty0 (5):\penalty0
  586--591, 2011.

\bibitem[Bruckner and Catalano(2018)]{Bruckner2018ssmph}
Tim~A Bruckner and Ralph Catalano.
\newblock {Selection in utero and population health\_ Theory and typology of
  research}.
\newblock \emph{SSM - Population Health}, 5:\penalty0 101 -- 113, 08 2018.

\bibitem[Cameron et~al.(2008)Cameron, Gelbach, and Miller]{Cameron:2008ws}
A~Colin Cameron, Jonah~B Gelbach, and Douglas~L Miller.
\newblock {Bootstrap-based improvements for inference with clustered errors}.
\newblock \emph{Review of Economics and Statistics}, 90\penalty0 (3):\penalty0
  414--427, July 2008.

\bibitem[Catalano and Bruckner(2006)]{Catalano:2006vv}
Ralph Catalano and Tim Bruckner.
\newblock {Secondary sex ratios and male lifespan: Damaged or culled cohorts}.
\newblock \emph{PNAS}, 103\penalty0 (5):\penalty0 1639--1643, 2006.

\bibitem[Catalano et~al.(2021)Catalano, Bruckner, Casey, Gemmill, Margerison,
  and Hartig]{Catalano2021emph}
Ralph Catalano, Tim Bruckner, Joan~A Casey, Alison Gemmill, Claire Margerison,
  and Terry Hartig.
\newblock {Twinning during the pandemic: Evidence of selection in utero}.
\newblock \emph{Evolution, Medicine, and Public Health}, 9\penalty0
  (1):\penalty0 374--382, 10 2021.

\bibitem[{Central Sanitary Bureau of the Home Ministry}(1927)]{csbhm1927}
{Central Sanitary Bureau of the Home Ministry}.
\newblock \emph{{Ry\=uk\=osei kanb\=o (Influenza). [in Japanese]}}.
\newblock Central Sanitary Bureau of the Home Ministry, Tokyo, 1927.

\bibitem[Chiappori et~al.(2002)Chiappori, Fortin, and
  Lacroix]{Chiappori:2002bx}
Pierre~Andr\'e Chiappori, Bernard Fortin, and Guy Lacroix.
\newblock {Marriage Market, Divorce Legislation, and Household Labor Supply}.
\newblock \emph{Journal of Political Economy}, 110\penalty0 (1):\penalty0
  37--72, February 2002.

\bibitem[Correia et~al.(2020)Correia, Luck, and Verner]{Correia:2020vx}
Sergio Correia, Stephan Luck, and Emil Verner.
\newblock {Pandemics depress the economy, public health interventions Do Not:
  Evidence from the 1918 Flu}.
\newblock \emph{SSRP Working Paper}, pages 1--56, June 2020.

\bibitem[Currie and Vogl(2013)]{Currie:2013ke}
Janet Currie and Tom Vogl.
\newblock {Early-Life Health and Adult Circumstance in Developing Countries}.
\newblock \emph{Annual Review of Economics}, 5\penalty0 (1):\penalty0 1--36,
  August 2013.

\bibitem[Drixler(2016)]{Drixler2016}
Fabian~F Drixler.
\newblock {Hidden in Plain Sight: Stillbirths and Infanticides in Imperial
  Japan}.
\newblock \emph{Journal of Economic History}, 76\penalty0 (3):\penalty0
  651--696, September 2016.

\bibitem[Erkoreka(2010)]{Erkoreka:2010fi}
Anton Erkoreka.
\newblock {The Spanish influenza pandemic in occidental Europe (1918-1920) and
  victim age}.
\newblock \emph{Influenza and Other Respiratory Viruses}, 4\penalty0
  (2):\penalty0 81--89, March 2010.

\bibitem[Hayami(2006)]{hayami2006}
Akira Hayami.
\newblock \emph{{Nihon wo osotta supein infuruenza (Spanish influenza struck
  down to Japan). [in Japanese]}}.
\newblock Fujiwara shoten, Tokyo, 2006.

\bibitem[Hayami(2010)]{hayami2010}
Akira Hayami.
\newblock {An estimation of Spanish influenza mortality in Imperial Japan:
  1918--20}.
\newblock In Satomi Kurosu, Tommy Bengtsson, and Cameron Campbell, editors,
  \emph{Demographic Responses to Economic and Environmental Crises}, pages
  282--300. Reitaku University Press, Kashiwa, 2010.

\bibitem[Kayo(1983)]{kayo1983}
Nobufumi Kayo.
\newblock \emph{{Tod\=ofuken n\=ogy\=okisot\=okei (Basic Statistics of
  Agriculture in Japanese Prefecture). [in Japanese]}}.
\newblock N\=orint\=okeiky\=okai, Tokyo, 1983.

\bibitem[Kilbourne(2006)]{Kilbourne:2006ce}
Edwin~D Kilbourne.
\newblock {Influenza Pandemics of the 20th Century}.
\newblock \emph{Emerging Infectious Diseases}, 12\penalty0 (1):\penalty0 9--14,
  January 2006.

\bibitem[Kolk and Schnettler(2016)]{Kolk:2016ajhb}
Martin Kolk and Sebastian Schnettler.
\newblock {Socioeconomic status and sex ratios at birth in Sweden: no evidence
  for a Trivers–Willard effect for a wide range of status indicators}.
\newblock \emph{American Journal of Human Biology}, 28\penalty0 (1):\penalty0
  67--73, 2016.

\bibitem[Kraemer(2000)]{Kraemer:2000to}
Sebastian Kraemer.
\newblock {The fragile male}.
\newblock \emph{British Medical Journal}, 321\penalty0 (7276):\penalty0
  1609--1612, December 2000.

\bibitem[Lin and Liu(2014)]{Lin:2014bz}
Ming-Jen Lin and Elaine~M Liu.
\newblock {Does in utero exposure to Illness matter? The 1918 influenza
  epidemic in Taiwan as a natural experiment}.
\newblock \emph{Journal of Health Economics}, 37:\penalty0 152--163, September
  2014.

\bibitem[Markel et~al.(2007)Markel, Lipman, Navarro, Sloan, Michalsen, Stern,
  and Cetron]{Markel:2007uc}
Howard Markel, Harvey~B Lipman, J~Alexander Navarro, Alexandra Sloan, Joseph~R
  Michalsen, Alexandra~Minna Stern, and Martin~S Cetron.
\newblock {Nonpharmaceutical Interventions Implemented by US Cities During the
  1918-1919 Influenza Pandemic}.
\newblock \emph{Journal of American Medical Association}, 298\penalty0
  (6):\penalty0 644--654, August 2007.

\bibitem[Moore et~al.(2013)Moore, T.V.N.~(Vid), and Torchia]{moore2013}
Keith~L Moore, Persaud T.V.N.~(Vid), and Mark~G Torchia.
\newblock \emph{{The Developing Human: Clinically Oriented Embryology, 11th
  edition}}.
\newblock Elsevier, Saunders, Philadelphia, 2013.

\bibitem[Ogasawara(2018)]{Ogasawara:2018hk}
Kota Ogasawara.
\newblock {The long-run effects of pandemic influenza on the development of
  children from elite backgrounds: Evidence from industrializing Japan}.
\newblock \emph{Economics and Human Biology}, 31:\penalty0 125--137, September
  2018.

\bibitem[Orzack et~al.(2015)Orzack, Stubblefield, Akmaev, Colls, Munné,
  Scholl, Steinsaltz, and Zuckerman]{Orzack2015pnas}
Steven~Hecht Orzack, J~William Stubblefield, Viatcheslav~R Akmaev, Pere Colls,
  Santiago Munné, Thomas Scholl, David Steinsaltz, and James~E Zuckerman.
\newblock {The human sex ratio from conception to birth}.
\newblock \emph{Proceedings of the National Academy of Sciences}, 112\penalty0
  (16):\penalty0 E2102 -- E2111, 04 2015.

\bibitem[Prinz et~al.(2018)Prinz, Cutler, and Frakt]{Prinz2018}
Daniel~M Prinz, David~D Cutler, and Austin Frakt.
\newblock {Health and economic activity over the lifecycle: Literature review}.
\newblock \emph{NBER Working Paper Series}, 24865:\penalty0 1--128, 2018.

\bibitem[Reid(2005)]{Reid:2005vm}
Alice Reid.
\newblock {The Effects of the 1918{\textendash}1919 Influenza Pandemic on
  Infant and Child Health in Derbyshire}.
\newblock \emph{Medical History}, 49:\penalty0 29--54, 2005.

\bibitem[Rice and Palmer(1993)]{Rice:1993vp}
Geoffrey~W Rice and Edwina Palmer.
\newblock {Pandemic influenza in Japan, 1918-19: Mortality patterns and
  official responses}.
\newblock \emph{Journal of Japanese Studies}, 19\penalty0 (2):\penalty0
  389--420, 1993.

\bibitem[Richard et~al.(2009)Richard, Sugaya, Simonsen, Miller, and
  Viboud]{RICHARD:2009eb}
S~A Richard, N~Sugaya, L~Simonsen, M~A Miller, and C~Viboud.
\newblock {A comparative study of the 1918{\textendash}1920 influenza pandemic
  in Japan, USA and UK: mortality impact and implications for pandemic
  planning}.
\newblock \emph{Epidemiology and Infection}, 137\penalty0 (08):\penalty0
  1062--18, February 2009.

\bibitem[Royde-Smith(2021)]{wwi}
John~G Royde-Smith.
\newblock \emph{{The World War I}}.
\newblock Encyclop\ae dia Britannica, London, 2021.

\bibitem[Sanders and Stoecker(2015)]{Sanders:2015iq}
Nicholas~J Sanders and Charles Stoecker.
\newblock {Where have all the young men gone? Using sex ratios to measure fetal
  death rates}.
\newblock \emph{Journal of Health Economics}, 41:\penalty0 30--45, May 2015.

\bibitem[Schneider and Ogasawara(2018)]{Schneider:2018cx}
Eric~B Schneider and Kota Ogasawara.
\newblock {Disease and child growth in industrialising Japan: Critical windows
  and the growth pattern, 1917-39}.
\newblock \emph{Explorations in Economic History}, 69:\penalty0 64--80, July
  2018.

\bibitem[{Statistics Bureau of the Cabinet}(1914--1927)]{syej33_46}
{Statistics Bureau of the Cabinet}.
\newblock \emph{{Nihonteikoku t\=okeinenkan (The statistical yearbook of the
  Japanese Empire, volumes 33--46). [in Japanese]}}.
\newblock Statistics Bureau of the Cabinet, Tokyo, 1914--1927.

\bibitem[{Statistics Bureau of the Cabinet}(1921{\natexlab{a}})]{scdej1918}
{Statistics Bureau of the Cabinet}.
\newblock \emph{{Nihonteikoku shiint\=okei (The statistics of causes of death
  of the empire Japan, 1918 edition). [in Japanese]}}.
\newblock Statistics Bureau of the Cabinet, Tokyo, 1921{\natexlab{a}}.

\bibitem[{Statistics Bureau of the Cabinet}(1921{\natexlab{b}})]{vsej1918}
{Statistics Bureau of the Cabinet}.
\newblock \emph{{Nihonteikoku jink\=od\=otait\=okei (The vital statistics of
  empire Japan, 1918 edition). [in Japanese]}}.
\newblock Statistics Bureau of the Cabinet, Tokyo, 1921{\natexlab{b}}.

\bibitem[{Statistics Bureau of the Cabinet}(1922{\natexlab{a}})]{scdej1919}
{Statistics Bureau of the Cabinet}.
\newblock \emph{{Nihonteikoku shiint\=okei (The statistics of causes of death
  of the empire Japan, 1919 edition). [in Japanese]}}.
\newblock Statistics Bureau of the Cabinet, Tokyo, 1922{\natexlab{a}}.

\bibitem[{Statistics Bureau of the Cabinet}(1922{\natexlab{b}})]{vsej1919}
{Statistics Bureau of the Cabinet}.
\newblock \emph{{Nihonteikoku jink\=od\=otait\=okei (The vital statistics of
  empire Japan, 1919 edition). [in Japanese]}}.
\newblock Statistics Bureau of the Cabinet, Tokyo, 1922{\natexlab{b}}.

\bibitem[{Statistics Bureau of the Cabinet}(1923)]{scdej1920}
{Statistics Bureau of the Cabinet}.
\newblock \emph{{Nihonteikoku shiint\=okei (The statistics of causes of death
  of the empire Japan, 1920 edition). [in Japanese]}}.
\newblock Statistics Bureau of the Cabinet, Tokyo, 1923.

\bibitem[{Statistics Bureau of the Cabinet}(1924)]{vsej1920}
{Statistics Bureau of the Cabinet}.
\newblock \emph{{Nihonteikoku jink\=od\=otait\=okei (The vital statistics of
  empire Japan, 1920 edition). [in Japanese]}}.
\newblock Statistics Bureau of the Cabinet, Tokyo, 1924.

\bibitem[{Statistics Bureau of the Cabinet}(1926)]{census1925v2}
{Statistics Bureau of the Cabinet}.
\newblock \emph{{Taish\=ojy\=uyonen Kokuseich\=osah\=okoku, dainikan (1925
  Population Census of Japan, volume 2). [in Japanese]}}.
\newblock Statistics Bureau of the Cabinet, Tokyo, 1926.

\bibitem[{Statistics Bureau of the Cabinet}(1928)]{census1920v1}
{Statistics Bureau of the Cabinet}.
\newblock \emph{{Taish\=oky\=unen Kokuseich\=osah\=okoku, daiikkan (1920
  Population Census of Japan, volume 1). [in Japanese]}}.
\newblock Statistics Bureau of the Cabinet, Tokyo, 1928.

\bibitem[Taubenberger(2006)]{Taubenberger:2006vq}
Jeffery~K Taubenberger.
\newblock {1918 Influenza: the Mother of All Pandemics}.
\newblock \emph{Emerging Infectious Diseases}, 12\penalty0 (1):\penalty0
  15--22, January 2006.

\bibitem[{Tokyo City Office}(1926)]{tco1926}
{Tokyo City Office}.
\newblock \emph{{Ny\=ujishib\=och\=osa (Survey on the infant mortality rate).
  [in Japanese]}}.
\newblock Tokyo City Office, Tokyo, 1926.

\bibitem[Trivers and Willard(1973)]{Trivers:1973fd}
R~L Trivers and D~E Willard.
\newblock {Natural selection of parental ability to vary the sex ratio of
  offspring}.
\newblock \emph{Science}, 179\penalty0 (4068):\penalty0 90--92, January 1973.

\bibitem[Valente(2015)]{Valente:2015ci}
Christine Valente.
\newblock {Civil conflict, gender-specific fetal loss, and selection: A new
  test of the Trivers{\textendash}Willard hypothesis}.
\newblock \emph{Journal of Health Economics}, 39:\penalty0 31--50, January
  2015.

\bibitem[Vellore and Parman(2021)]{Arthi2021eeh}
Arthi Vellore and John Parman.
\newblock {Disease, downturns, and wellbeing: Economic history and the long-run
  impacts of COVID-19}.
\newblock \emph{Explorations in Economic History}, 79:\penalty0 101381, January
  2021.

\bibitem[Wooldridge(2002)]{wooldridge}
Jeffrey~M Wooldridge.
\newblock \emph{{Econometric Analysis of Cross Section and Panel Data}}.
\newblock MIT Press, Cambridge, MA, 2002.

\bibitem[Wu(2021)]{Wu2021hr}
Hanbo Wu.
\newblock {Maternal stress and sex ratio at birth in Sweden over two and a half
  centuries: a retest of the Trivers–Willard hypothesis}.
\newblock \emph{Human Reproduction}, 36\penalty0 (10):\penalty0 2782--2792,
  June 2021.

\end{thebibliography}


\begin{thebibliography}{}
{\small

\item
Akihiko Kawana, Go Naka, Yuji Fujikura, Yasuyuki Kato, Yasutaka Mizuno, Tatsuya Kondo, and Koichiro Kudo.
Spanish Influenza in Japanese Armed Forces, 1918–1920.
\textit{Emerging Infectious Diseases}, 13(4):590–593, April 2007.

\item
Andrew Goodman-Bacon.
Difference-inDifferences with Variation in Treatment Timing.
\textit{Journal of Econometrics}, 225(2):254–277, 2021.

\item
Brantly Callaway and Pedro H.C. Sant'Anna.
Difference-in-differences with multiple time periods.
\textit{Journal of Econometrics}, 225(2): 200–230, 2021.

\item
Brantly Callaway, Andrew Goodman-Bacon, and Pedro H.C. San' Anna.
Difference-in- differences with a continuous treatment.
Working Paper, pages 1–74, 2021.

\item
Brian Beach and W Walker Hanlon.
Coal Smoke and Mortality in an Early Industrial Economy. 
\textit{The Economic Journal}, 128(615):2652–2675, November 2017. 

\item
Brian Beach, Ryan Brown, Joseph Ferrie, Martin Saavedra, and Duncan Thomas.
Re-evaluating the Long-Term Impact of In Utero Exposure to the 1918 Influenza Pandemic.
\textit{Journal of Political Economy}, pages 1–40, 2022. 

\item
Cl'ement de Chaisemartin and Xavier D'Haultfœuille.
Two-way fixed effects estimators with heterogeneous treatment effects.
\textit{American Economic Review}, 110(9):2964–96, September 2020.

\item
Cheng Hsiao.
\textit{Analysis of panel data, 3rd edition}.
Cambridge University Press, 2014.

\item
Department of Agriculture and Commerce.
\textit{N\=osh\=omush\=ot\=okeisho} (The statistical yearbook of agriculture and commerce, volumes 30-35). [in Japanese].

\item
Eric B Schneider and Kota Ogasawara.
Disease and child growth in industrialising Japan: Critical windows and the growth pattern, 1917-39.
\textit{Explorations in Economic History}, 69:64–80, July 2018.

\item
Fabian F Drixler.
Hidden in Plain Sight: Stillbirths and Infanticides in Imperial Japan.
\textit{Journal of Economic History}, 76(3):651–696, September 2016.

\item
In Choi.
Unit root tests for panel data.
\textit{Journal of International Money and Finance}, 20 (2): 249--272, April 2001.

\item
Ito Shigeru.
An analysis of prefectural differentials on fertility in early modern Japan [in Japanese].
Obihirochikusandaigaku gakujyutukenky\=uh\=okoku I, 15(2):145–155, 1987.

\item
Karen Clay, Joshua Lewis, and Edson Severnini.
Pollution, Infectious Disease, and Mortality: Evidence from the 1918 Spanish Influenza Pandemic.
\textit{Journal of Economic History}, 78(4):1179–1209, December 2018. 

\item
Kota Ogasawara.
Persistence of pandemic influenza on the development of children: Evidence from industrializing Japan.
\textit{Social Science \& Medicine}, 181: 43--53, May 2017.

\item
Kota Ogasawara.
The long-run effects of pandemic influenza on the development of children from elite backgrounds: Evidence from industrializing Japan.
\textit{Economics and Human Biology}, 31: 125--137, September 2018.

\item
Liyang Sun and Sarah Abraham.
Estimating dynamic treatment effects in event studies with heterogeneous treatment effects.
\textit{Journal of Econometrics}, 225(2), 2021.

\item
Mabel Andal\'on, Jo\~ao Pedro Azevedo, Carlos Rodr\'iguez-Castel\'an, Viviane Sanfelice, and Daniel Valderrama-Gonz\'alez.
Weather Shocks and Health at Birth in Colombia.
World Development, 82(C):69--82, June 2016.

\item
Ministry of Education.
\textit{Dainihon teikoku monbush\=onenp\=o} (The Annual Report of the Japanese Imperial Ministry of Education, 23--54 editions). [in Japanese].

\item
Nobufumi Kayo.
Tod\=ofuken n\=ogy\=okisot\=okei (Basic Statistics of Agriculture in Japanese Prefecture). [in Japanese].
N\=orint\=okeiky\=okai, Tokyo, 1983.


\item
Olivier Desch\^enes, Michael Greenstone, and Jonathan Guryan.
Climate Change and Birth Weight.
\textit{American Economic Review}, 99(2):211–217, April 2009.

\item
Oswaldo Molina and Victor Saldarriaga.
The perils of climate change: In utero exposure to temperature variability and birth outcomes in the Andean region.
\textit{Economics and Human Biology}, 24:111–124, February 2017.

\item
Secretariat of Agriculture and Commerce.
\textit{K\=ojy\=ot\=okeihy\=o} (Census of manufactures, 1919--1925 editions). [in Japanese].

\item
Sonoko Hijikata.
Kindainihon no gakk\=o to chiikishakai (Schools and community in prewar Japan). [in Japanese].
University of Tokyo Press, Tokyo, 1994.

\item
Statistics Bureau of the Cabinet.
\textit{Nihonteikoku t\=okeinenkan} (The statistical yearbook of the Japanese Empire, volumes 33--46). [in Japanese].
Statistics Bureau of the Cabinet, Tokyo, 1914--1927.

\item
Statistics Bureau of the Cabinet.
\textit{Nihonteikoku jink\=od\=otait\=okei} (The vital statistics of empire Japan, 1912 edition). [in Japanese].
Statistics Bureau of the Cabinet, Tokyo, 1916c.

\item
Statistics Bureau of the Cabinet.
\textit{Nihonteikoku jink\=od\=otait\=okei} (The vital statistics of empire Japan, 1913 edition). [in Japanese].
Statistics Bureau of the Cabinet, Tokyo, 1917.

\item
Statistics Bureau of the Cabinet.
\textit{Nihonteikoku jink\=od\=otait\=okei} (The vital statistics of empire Japan, 1914 edition). [in Japanese].
Statistics Bureau of the Cabinet, Tokyo, 1918c.

\item
Statistics Bureau of the Cabinet.
\textit{Nihonteikoku jink\=od\=otait\=okei} (The vital statistics of empire Japan, 1915 edition). [in Japanese].
Statistics Bureau of the Cabinet, Tokyo, 1918d.

\item
Statistics Bureau of the Cabinet.
\textit{Nihonteikoku jink\=od\=otait\=okei} (The vital statistics of empire Japan, 1916 edition). [in Japanese].
Statistics Bureau of the Cabinet, Tokyo, 1919b.

\item
Statistics Bureau of the Cabinet.
\textit{Nihonteikoku jink\=od\=otait\=okei} (The vital statistics of empire Japan, 1917 edition). [in Japanese].
Statistics Bureau of the Cabinet, Tokyo, 1920b.

\item
Statistics Bureau of the Cabinet.
\textit{Nihonteikoku jink\=od\=otait\=okei} (The vital statistics of empire Japan, 1918 edition). [in Japanese].
Statistics Bureau of the Cabinet, Tokyo, 1921b.

\item
Statistics Bureau of the Cabinet.
\textit{Nihonteikoku jink\=od\=otait\=okei }(The vital statistics of empire Japan, 1919 edition). [in Japanese].
Statistics Bureau of the Cabinet, Tokyo, 1922b.

\item
Statistics Bureau of the Cabinet.
\textit{Nihonteikoku jink\=od\=otait\=okei} (The vital statistics of empire Japan, 1920 edition). [in Japanese].
Statistics Bureau of the Cabinet, Tokyo, 1924b.

\item
Statistics Bureau of the Cabinet.
\textit{Nihonteikoku jink\=od\=otait\=okei} (The vital statistics of empire Japan, 1921 edition). [in Japanese].
Statistics Bureau of the Cabinet, Tokyo, 1924c.

\item
Statistics Bureau of the Cabinet.
\textit{Nihonteikoku jink\=od\=otait\=okei} (The vital statistics of empire Japan, 1922 edition). [in Japanese].
Statistics Bureau of the Cabinet, Tokyo, 1924d.

\item
Statistics Bureau of the Cabinet.
\textit{Nihonteikoku jink\=od\=otait\=okei} (The vital statistics of empire Japan, 1923 edition). [in Japanese].
Statistics Bureau of the Cabinet, Tokyo, 1925d.

\item
Statistics Bureau of the Cabinet.
\textit{Nihonteikoku jink\=od\=otait\=okei} (The vital statistics of empire Japan, 1924 edition). [in Japanese].
Statistics Bureau of the Cabinet, Tokyo, 1925e.

\item
Statistics Bureau of the Cabinet.
\textit{Nihonteikoku jink\=od\=otait\=okei} (The vital statistics of empire Japan, 1925 edition). [in Japanese].
Statistics Bureau of the Cabinet, Tokyo, 1926c.

\item
Statistics Bureau of the Cabinet.
\textit{Nihonteikoku jink\=od\=otait\=okei} (The vital statistics of empire Japan, 1926 edition). [in Japanese].
Statistics Bureau of the Cabinet, Tokyo, 1927b.

\item
Statistics Bureau of the Cabinet.
\textit{Nihonteikoku jink\=od\=otait\=okei} (The vital statistics of empire Japan, 1927 edition). [in Japanese].
Statistics Bureau of the Cabinet, Tokyo, 1928c.

\item
Statistics Bureau of the Cabinet.
\textit{Nihonteikoku shiint\=okei} (The statistics of causes of death of the empire Japan, 1912 edition). [in Japanese].
Statistics Bureau of the Cabinet, Tokyo, 1916a.

\item
Statistics Bureau of the Cabinet.
\textit{Nihonteikoku shiint\=okei} (The statistics of causes of death of the empire Japan, 1913 edition). [in Japanese].
Statistics Bureau of the Cabinet, Tokyo, 1916b

\item
Statistics Bureau of the Cabinet.
\textit{Nihonteikoku shiint\=okei} (The statistics of causes of death of the empire Japan, 1914 edition). [in Japanese].
Statistics Bureau of the Cabinet, Tokyo, 1918a.

\item
Statistics Bureau of the Cabinet.
\textit{Nihonteikoku shiint\=okei} (The statistics of causes of death of the empire Japan, 1915 edition). [in Japanese].
Statistics Bureau of the Cabinet, Tokyo, 1918b.

\item
Statistics Bureau of the Cabinet.
\textit{Nihonteikoku jink\=od\=otait\=okei} (The vital statistics of empire Japan, 1914 edition). [in Japanese].
Statistics Bureau of the Cabinet, Tokyo, 1918c.

\item
Statistics Bureau of the Cabinet.
\textit{Nihonteikoku jink\=od\=otait\=okei} (The vital statistics of empire Japan, 1915 edition). [in Japanese].
Statistics Bureau of the Cabinet, Tokyo, 1918d.

\item
Statistics Bureau of the Cabinet.
\textit{Nihonteikoku shiint\=okei} (The statistics of causes of death of the empire Japan, 1916 edition). [in Japanese].
Statistics Bureau of the Cabinet, Tokyo, 1919a.

\item
Statistics Bureau of the Cabinet.
\textit{Nihonteikoku jink\=od\=otait\=okei} (The vital statistics of empire Japan, 1916 edition). [in Japanese].
Statistics Bureau of the Cabinet, Tokyo, 1919b.

\item
Statistics Bureau of the Cabinet.
\textit{Nihonteikoku shiint\=okei} (The statistics of causes of death of the empire Japan, 1917 edition). [in Japanese].
Statistics Bureau of the Cabinet, Tokyo, 1920a.

\item
Statistics Bureau of the Cabinet.
\textit{Nihonteikoku shiint\=okei} (The statistics of causes of death of the empire Japan, 1918 edition). [in Japanese].
Statistics Bureau of the Cabinet, Tokyo, 1921a.

\item
Statistics Bureau of the Cabinet.
\textit{Nihonteikoku shiint\=okei} (The statistics of causes of death of the empire Japan, 1919 edition). [in Japanese].
Statistics Bureau of the Cabinet, Tokyo, 1922a.

\item
Statistics Bureau of the Cabinet.
\textit{Nihonteikoku shiint\=okei} (The statistics of causes of death of the empire Japan, 1920 edition). [in Japanese].
Statistics Bureau of the Cabinet, Tokyo, 1923.

\item
Statistics Bureau of the Cabinet.
\textit{Nihonteikoku shiint\=okei} (The statistics of causes of death of the empire Japan, 1921 edition). [in Japanese].
Statistics Bureau of the Cabinet, Tokyo, 1924a.

\item
Statistics Bureau of the Cabinet.
\textit{Nihonteikoku shiin\=okei} (The statistics of causes of death of the empire Japan, 1922 edition). [in Japanese].
Statistics Bureau of the Cabinet, Tokyo, 1925a.

\item
Statistics Bureau of the Cabinet.
\textit{Nihonteikoku shiint\=okei} (The statistics of causes of death of the empire Japan, 1923 edition). [in Japanese].
Statistics Bureau of the Cabinet, Tokyo, 1925b.

\item
Statistics Bureau of the Cabinet.
\textit{Nihonteikoku shiint\=okei} (The statistics of causes of death of the empire Japan, 1924 edition). [in Japanese].
Statistics Bureau of the Cabinet, Tokyo, 1925c.

\item
Statistics Bureau of the Cabinet.
\textit{Nihonteikoku shiint\=okei} (The statistics of causes of death of the empire Japan, 1925 edition). [in Japanese].
Statistics Bureau of the Cabinet, Tokyo, 1926b.

\item
Statistics Bureau of the Cabinet.
\textit{Nihonteikoku shiint\=okei} (The statistics of causes of death of the empire Japan, 1926 edition). [in Japanese].
Statistics Bureau of the Cabinet, Tokyo, 1927a.

\item
Statistics Bureau of the Cabinet.
\textit{Nihonteikoku shiint\=okei} (The statistics of causes of death of the empire Japan, 1927 edition). [in Japanese].
Statistics Bureau of the Cabinet, Tokyo, 1928b.

\item
Statistics Bureau of the Cabinet.
\textit{Taish\=ojy\=uyonen Kokuseich\=osah\=okoku, fukenhen} (1925 Population Census of Japan, prefecture part). [in Japanese].
Statistics Bureau of the Cabinet, Tokyo, 1929.

\item
Statistics Bureau of the Cabinet.
\textit{Sh\=owagonen Kokuseich\=osah\=okoku, fukenhen} (1930 Population Census of Japan, prefecture part). [in Japanese].
Statistics Bureau of the Cabinet, Tokyo, 1933.

\item
Statistical Survey Department, Statistics Bureau, Ministry of Internal Affairs and Communications.
Population (\url{http://www.stat.go.jp/data/chouki/zuhyou/02-05.xls}, accessed on July 13, 2017). [in Japanese].

\item
Walker W Hanlon.
Coal smoke, city growth, and the. costs of the industrial revolution.
Working Paper, pages 1–95, February 2019. 
}
\end{thebibliography}
\endgroup
\begin{figure}[h]
\centering
\captionsetup{justification=centering,margin=0.5cm}
\subfloat[November 1918]{\label{fig:map_191811}\includegraphics[width=0.3\textwidth]{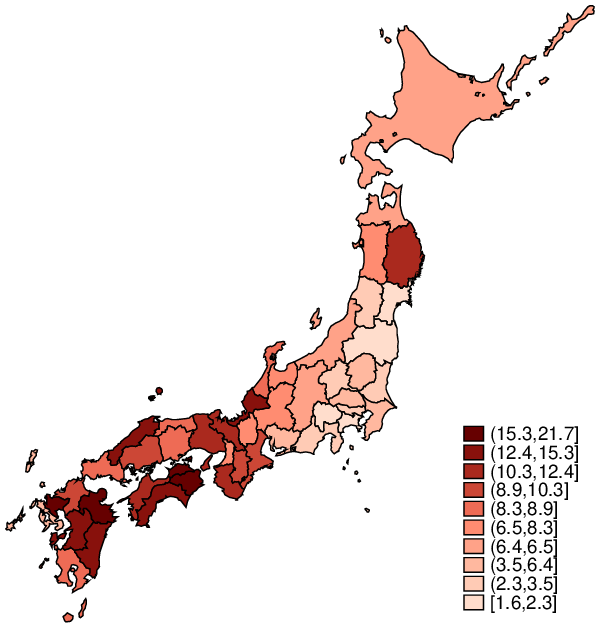}}
\subfloat[December 1918]{\label{fig:map_191812}\includegraphics[width=0.3\textwidth]{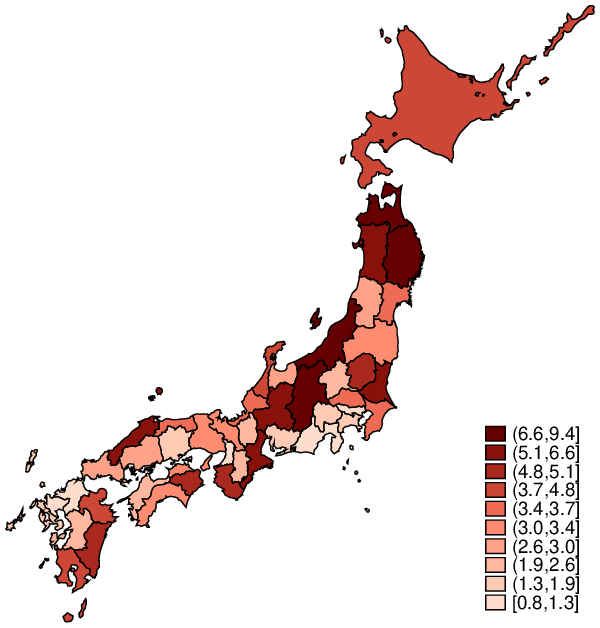}}
\subfloat[January 1919]{\label{fig:map_19191}\includegraphics[width=0.3\textwidth]{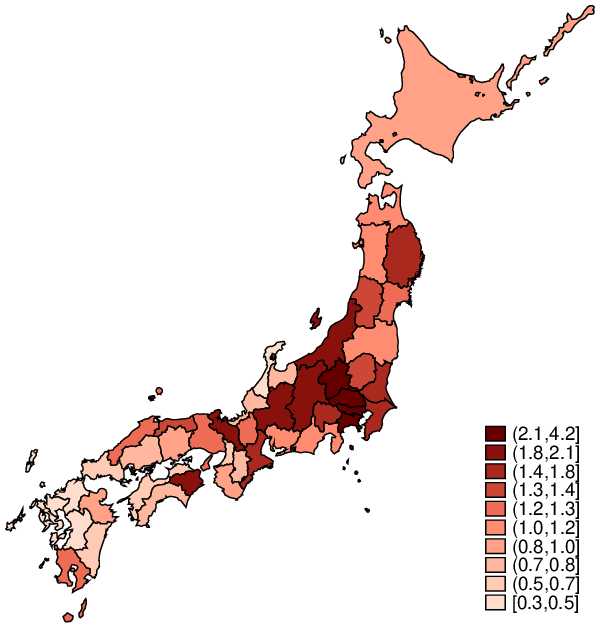}}\\
\subfloat[January 1920]{\label{fig:map_19201}\includegraphics[width=0.3\textwidth]{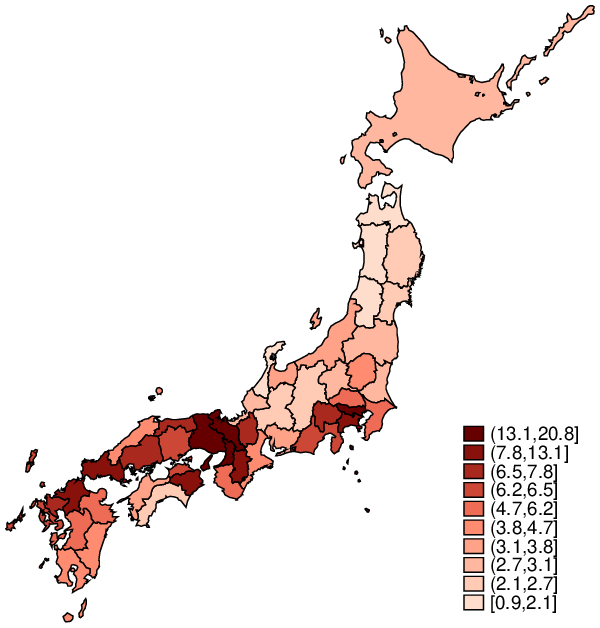}}
\subfloat[February 1920]{\label{fig:map_19202}\includegraphics[width=0.3\textwidth]{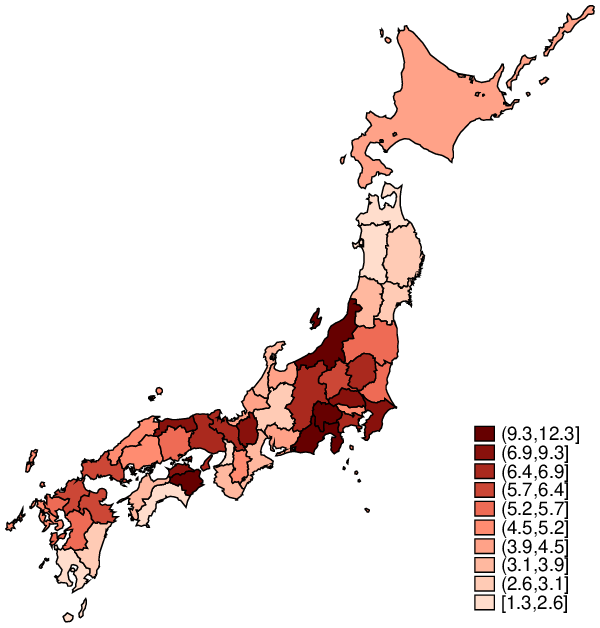}}
\subfloat[March 1920]{\label{fig:map_19203}\includegraphics[width=0.3\textwidth]{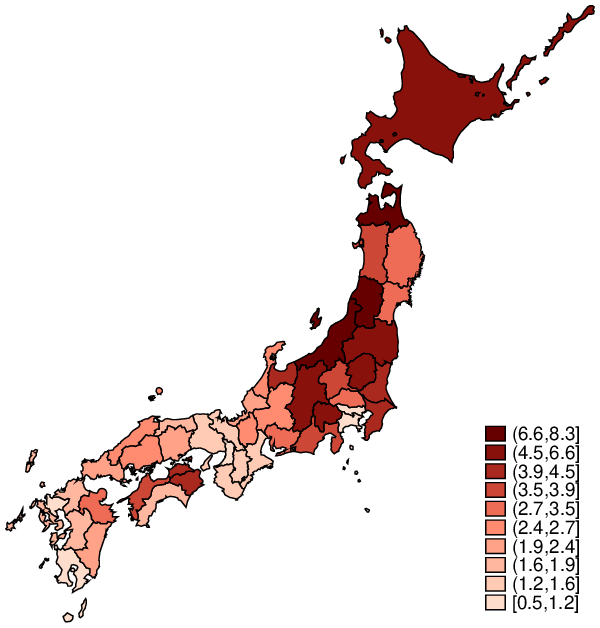}}
\caption{Spatiotemporal distributions of influenza death rates}
\label{fig:map_flur}
\scriptsize{\begin{minipage}{450pt}
\setstretch{0.85}
Notes: Figures~\ref{fig:map_191811}--\ref{fig:map_19191} illustrate the influenza death rates in the first pandemic period between November 1918 and January 1919, as shown in Figure~\ref{fig:ts_flu}.
Figures~\ref{fig:map_19201}--\ref{fig:map_19203} illustrate the influenza death rates in the second pandemic period between January 1920 and March 1920.
Okinawa prefecture is not included in the sample.
Sources: Created by the author from \citet{scdej1918, scdej1919, scdej1920}; \citet{vsej1918, vsej1919, vsej1920}; Statistical Survey Department, Statistics Bureau, Ministry of Internal Affairs and Communications (database).
\end{minipage}}
\end{figure}
\begin{landscape}
\begin{table}[h]
\def\arraystretch{1.0}
\begin{center}
\caption{Summary statistics}
\label{tab:sum}
\footnotesize
\scalebox{1.0}[1]{
\begin{tabular}{lcD{.}{.}{2}D{.}{.}{2}D{.}{.}{2}D{.}{.}{2}D{.}{.}{2}D{.}{.}{2}}
\toprule
&Frequency			&\multicolumn{1}{c}{Mean}&\multicolumn{1}{c}{Std. Dev.}&\multicolumn{1}{c}{Min}&\multicolumn{1}{c}{Max}&\multicolumn{1}{c}{Observations}\\\hline
Panel A: Dependent variables&&&&&&\\
\hspace{10pt}Male births (per 100 births)						&Monthly	&51.18		&1.67	&41.37		&61.39		&3,864\\
\hspace{10pt}Infant mortality rate (per 1,000 live births)			&Annual		&171.11	&29.88	&93.05		&252.21	&322\\
\hspace{20pt}Infant mortality rate (boys)						&Annual		&179.44	&30.33	&100.07	&263.44	&322\\
\hspace{20pt}Infant mortality rate (girls)							&Annual		&162.43	&29.71	&85.76		&246.99	&322\\
&&&&&&\\
Panel B: Influenza severity&&&&&&\\
\hspace{10pt}Influenza death rate (past nine-month average, per 10,000 people)	&Monthly	&0.53	&0.78	&0.00	&3.89	&3,864\\
\hspace{20pt}Influenza death rate (first trimester average)						&Monthly	&0.53	&1.23	&0.00	&9.97	&3,864\\
\hspace{20pt}Influenza death rate (second trimester average)					&Monthly	&0.53	&1.23	&0.00	&9.97	&3,864\\
\hspace{20pt}Influenza death rate (third trimester average)						&Monthly	&0.53	&1.23	&0.00	&9.97	&3,864\\
\hspace{10pt}Weighted influenza death rate (per 10,000 people)					&Annual		&0.50	&0.64	&0.01	&2.62	&322\\
\hspace{20pt}Weighted influenza death rate (first trimester average)				&Annual		&0.48	&0.59	&0.01	&2.57	&322\\
\hspace{20pt}Weighted influenza death rate (second trimester average)			&Annual		&0.48	&0.64	&0.01	&2.73	&322\\
\hspace{20pt}Weighted influenza death rate (third trimester average)				&Annual		&0.56	&0.72	&0.00	&3.21	&322\\
&&&&&&\\\hline
&\multicolumn{1}{c}{}&\multicolumn{2}{c}{Full sample}&\multicolumn{3}{c}{Balancing tests}\\
\cmidrule(rr){3-4}\cmidrule(rrr){5-7}
Panel C: Control variables
&\multicolumn{1}{c}{Frequency}
&\multicolumn{1}{c}{Mean}&\multicolumn{1}{c}{Std. Dev.}
&\multicolumn{1}{c}{$>$75pct}&\multicolumn{1}{c}{$\leq$ 75pct}&\multicolumn{1}{c}{Diff. [p-value]}\\\hline
\hspace{10pt}Rice yield per hectare (hectoliter)									&Annual		&34.26	&5.93	&34.90	&34.05	&\multicolumn{1}{c}{-0.84 [0.2711]}\\
\hspace{10pt}Soy yield per hectare (hectoliter)									&Annual		&16.34	&3.14	&16.59	&16.25	&\multicolumn{1}{c}{-0.34 [0.4056]}\\
\hspace{10pt}Milk production per capita (liter)									&Annual		&0.98	&0.82	&0.92	&1.00	&\multicolumn{1}{c}{0.08 [0.4451]}\\
\hspace{10pt}Coverage of doctors (per 100 people)								&Annual		&0.07	&0.02	&0.08	&0.07	&\multicolumn{1}{c}{-0.00 [0.5912]}\\
\hspace{10pt}Coverage of midwives (per 100 people)							&Annual		&0.06	&0.02	&0.06	&0.06	&\multicolumn{1}{c}{-0.00 [0.8403]}\\
\hspace{10pt}Observations														&			&\multicolumn{2}{c}{~~322}&\multicolumn{1}{c}{~~~~80}	&\multicolumn{1}{c}{~~~~242}	&\\\bottomrule
\end{tabular}
}
{\scriptsize
\begin{minipage}{620pt}
\setstretch{0.84}Notes:
Panel A reports the summary statistics for the prefecture-month- or prefecture-year-level dependent variables.
Panel B reports the summary statistics for the prefecture-month- or prefecture-year-level influenza death rates.
Panel C reports the summary statistics for the prefecture-year-level control variables.
In the balancing test, $>$75pct and $\leq$75pct indicate the influenza death rates above and less than or equal to the 75 percentile, respectively.
Prefecture-month-level data include the observations for 46 prefectures between January 1916 and December 1922.
Prefecture-year-level data include the observations for 46 prefectures between 1916 and 1922.
Sources: Dependent variables are from the VSEJ (1916--1922 editions).
Influenza death rates are from the SCDEJ (1915--1922 editions); VSEJ (1915--1922 editions); Statistical Survey Department, Statistics Bureau, Ministry of Internal Affairs and Communications (database).
The control variables are from \citet{kayo1983} and \citet{syej33_46}.
\end{minipage}
}
\end{center}
\end{table}
\end{landscape}
\begin{landscape}
\begin{table}[h]
\def\arraystretch{1.0}
\begin{center}
\captionsetup{justification=centering}
\caption{Effects of fetal influenza exposure on the proportion of male births (\%)}
\label{tab:r_main}
\footnotesize
\scalebox{0.89}[1]{
\begin{tabular}{lD{.}{.}{-2}D{.}{.}{-2}D{.}{.}{-2}D{.}{.}{-2}D{.}{.}{-2}D{.}{.}{-2}D{.}{.}{-2}D{.}{.}{-2}}
\toprule
&\multicolumn{4}{c}{Entire period}&\multicolumn{2}{c}{Non-pandemic years [placebo]}&\multicolumn{2}{c}{Pandemic years}\\
\cmidrule(rrrr){2-5}\cmidrule(rr){6-7}\cmidrule(rr){8-9}
Exposed trimesters
&\multicolumn{1}{c}{(1)}&\multicolumn{1}{c}{(2)}&\multicolumn{1}{c}{(3)}&\multicolumn{1}{c}{(4)}
&\multicolumn{1}{c}{(5)}&\multicolumn{1}{c}{(6)}&\multicolumn{1}{c}{(7)}&\multicolumn{1}{c}{(8)}\\\hline
All trimesters						&-0.1621*	&-0.1582*	&				&				&			&			&				&			\\
									&[0.1000]	&[0.0900]	&				&				&			&			&				&			\\
First trimester						&			&			&-0.1102$***$	&-0.1087$***$	&0.2087	&0.2277	&-0.1242$***$	&-0.1531$**$\\
									&			&			&[0.0050]		&[0.0050]		&[0.1500]	&[0.2700]	&[0.0050]		&[0.0150]	\\
Second trimester					&			&			&-0.0452		&-0.0438		&0.2858	&0.3538	&-0.0505		&-0.0792	\\
									&			&			&[0.2500]		&[0.2450]		&[0.1950]	&[0.2300]	&[0.3150]		&[0.2100]	\\
Third trimester						&			&			&-0.0047		&-0.0037		&-0.6393	&-0.5815	&-0.0007		&-0.0193	\\
									&			&			&[0.8450]		&[0.8800]		&[0.1350]	&[0.2100]	&[0.9500]		&[0.6050]	\\\hline
Time trend						
&\multicolumn{1}{c}{No}		&\multicolumn{1}{c}{Yes}
&\multicolumn{1}{c}{No}		&\multicolumn{1}{c}{Yes}
&\multicolumn{1}{c}{No}		&\multicolumn{1}{c}{Yes}
&\multicolumn{1}{c}{No}		&\multicolumn{1}{c}{Yes}\\
Period
&\multicolumn{1}{c}{Jan. 1916--}		&\multicolumn{1}{c}{Jan. 1916--}
&\multicolumn{1}{c}{Jan. 1916--}		&\multicolumn{1}{c}{Jan. 1916--}
&\multicolumn{1}{c}{Jan. 1916--}		&\multicolumn{1}{c}{Jan. 1916--}
&\multicolumn{1}{c}{Jan. 1918--}		&\multicolumn{1}{c}{Jan. 1918--}\\
&\multicolumn{1}{c}{Dec. 1922}		&\multicolumn{1}{c}{Dec. 1922}
&\multicolumn{1}{c}{Dec. 1922}		&\multicolumn{1}{c}{Dec. 1922}
&\multicolumn{1}{c}{Dec. 1917}		&\multicolumn{1}{c}{Dec. 1917}
&\multicolumn{1}{c}{Dec. 1920}		&\multicolumn{1}{c}{Dec. 1920}\\
&\multicolumn{1}{c}{}		&\multicolumn{1}{c}{}
&\multicolumn{1}{c}{}		&\multicolumn{1}{c}{}
&\multicolumn{1}{c}{Jan. 1921--}&\multicolumn{1}{c}{Jan. 1921--}
&&\\
&\multicolumn{1}{c}{}		&\multicolumn{1}{c}{}
&\multicolumn{1}{c}{}		&\multicolumn{1}{c}{}
&\multicolumn{1}{c}{Dec. 1922}&\multicolumn{1}{c}{Dec. 1922}
&&\\
Observations						
&\multicolumn{1}{c}{3864}		&\multicolumn{1}{c}{3864}
&\multicolumn{1}{c}{3864}		&\multicolumn{1}{c}{3864}
&\multicolumn{1}{c}{2208}		&\multicolumn{1}{c}{2208}
&\multicolumn{1}{c}{1656}		&\multicolumn{1}{c}{1656}\\
Number of prefectures
&\multicolumn{1}{c}{46}		&\multicolumn{1}{c}{46}
&\multicolumn{1}{c}{46}		&\multicolumn{1}{c}{46}
&\multicolumn{1}{c}{46}		&\multicolumn{1}{c}{46}
&\multicolumn{1}{c}{46}		&\multicolumn{1}{c}{46}\\
Number of months
&\multicolumn{1}{c}{84}		&\multicolumn{1}{c}{84}
&\multicolumn{1}{c}{84}		&\multicolumn{1}{c}{84}
&\multicolumn{1}{c}{48}		&\multicolumn{1}{c}{48}
&\multicolumn{1}{c}{36}		&\multicolumn{1}{c}{36}\\
Number of clusters
&\multicolumn{1}{c}{8}		&\multicolumn{1}{c}{8}
&\multicolumn{1}{c}{8}		&\multicolumn{1}{c}{8}
&\multicolumn{1}{c}{8}		&\multicolumn{1}{c}{8}
&\multicolumn{1}{c}{8}		&\multicolumn{1}{c}{8}\\\bottomrule
\end{tabular}
}
{\scriptsize
\begin{minipage}{640pt}
\setstretch{0.85}
***, **, and * represent statistical significance at the 1\%, 5\%, and 10\% levels based on the $p$-values from the wild cluster bootstrap resampling method in brackets, respectively.
The data are clustered at the 8-area level in the bootstrap procedure.
The number of replications is fixed to 400 for all the specifications.\\
Notes:
This table shows the results from equations~\ref{bs} and~\ref{fs}.
All the regressions include the prefecture fixed effect, year-month-specific fixed effect, and controls for rice yield, soy yield, milk production, coverage of doctors, and coverage of midwives.
All the regressions are weighted by the average number of births in each prefecture.
\end{minipage}
}
\end{center}
\end{table}
\end{landscape}
\begin{table}[]
\def\arraystretch{1.0}
\begin{center}
\captionsetup{justification=centering}
\caption{Effects of fetal influenza exposure on the infant mortality rate (\textperthousand)}
\label{tab:r_main_imr}
\footnotesize
\scalebox{1.0}[1]{
\begin{tabular}{lD{.}{.}{-2}D{.}{.}{-2}D{.}{.}{-2}D{.}{.}{-2}}
\toprule
&\multicolumn{4}{c}{Dependent variable: Infant mortality rate (\textperthousand)}\\
\cmidrule(rrrr){2-5}
Exposed trimesters
&\multicolumn{1}{c}{(1)}&\multicolumn{1}{c}{(2)}&\multicolumn{1}{c}{(3)}&\multicolumn{1}{c}{(4)}\\\hline
Panel A: All infants	&&&&\\
\hspace{10pt}All trimesters					&5.938$***$	&6.298$***$	&				&				\\
								&[0.000]		&[0.000]		&				&				\\
\hspace{10pt}First trimesters					&				&				&-13.842		&-13.637		\\
								&				&				&[0.147]		&[0.133]		\\
\hspace{10pt}Second trimesters				&				&				&-6.237		&-6.780		\\
								&				&				&[0.427]		&[0.427]		\\
\hspace{10pt}Third trimesters					&				&				&20.480$**$	&20.907$**$	\\
								&				&				&[0.013]		&[0.013]		\\
								&&&&\\
Panel B: Boys	&&&&\\
\hspace{10pt}All trimesters					&3.667$**$	&4.053$*$		&				&				\\
								&[0.013]		&[0.067]		&				&				\\
\hspace{10pt}First trimesters					&				&				&-11.006		&-11.023		\\
								&				&				&[0.240]		&[0.187]		\\
\hspace{10pt}Second trimesters				&				&				&-9.098		&-9.748		\\
								&				&				&[0.267]		&[0.267]		\\
\hspace{10pt}Third trimesters					&				&				&18.748$**$	&19.480$**$	\\
								&				&				&[0.013]		&[0.013]		\\
								&&&&\\
Panel C: Girls	&&&&\\
\hspace{10pt}All trimesters					&8.235$***$	&8.556$***$	&				&				\\
								&[0.000]		&[0.000]		&				&				\\
\hspace{10pt}First trimesters					&				&				&-16.609		&-16.189		\\
								&				&				&[0.120]		&[0.120]		\\
\hspace{10pt}Second trimesters				&				&				&-3.399		&-3.827		\\
								&				&				&[0.533]		&[0.533]		\\
\hspace{10pt}Third trimesters					&				&				&22.204$**$	&22.311$**$	\\
								&				&				&[0.013]		&[0.013]		\\\hline
Time trend						
&\multicolumn{1}{c}{No}		&\multicolumn{1}{c}{Yes}
&\multicolumn{1}{c}{No}		&\multicolumn{1}{c}{Yes}\\
Observations						
&\multicolumn{1}{c}{322}		&\multicolumn{1}{c}{322}
&\multicolumn{1}{c}{322}		&\multicolumn{1}{c}{322}\\
Number of prefectures
&\multicolumn{1}{c}{46}		&\multicolumn{1}{c}{46}
&\multicolumn{1}{c}{46}		&\multicolumn{1}{c}{46}\\
Number of years
&\multicolumn{1}{c}{7}		&\multicolumn{1}{c}{7}
&\multicolumn{1}{c}{7}		&\multicolumn{1}{c}{7}\\
Number of clusters
&\multicolumn{1}{c}{8}		&\multicolumn{1}{c}{8}
&\multicolumn{1}{c}{8}		&\multicolumn{1}{c}{8}\\\bottomrule
\end{tabular}
}
{\scriptsize
\begin{minipage}{395pt}
\setstretch{0.85}
***, **, and * represent statistical significance at the 1\%, 5\%, and 10\% levels based on the $p$-values from the wild cluster bootstrap resampling method in brackets, respectively.
The data are clustered at the 8-area level in the bootstrap procedure.
The number of replications is fixed to 150 for all the specifications.\\
Notes: 
This table shows the results from equations~\ref{imrbs} and~\ref{imrfs}.
All the regressions include the prefecture fixed effect, year fixed effect, and controls for rice yield, soy yield, milk production, coverage of doctors, and coverage of midwives.
All the regressions are weighted by the average number of live births (of boys in Panel B; of girls in Panel C) in each prefecture.
\end{minipage}
}
\end{center}
\end{table}
\clearpage
\thispagestyle{empty}

\begin{center}
\qquad

\qquad

\qquad

\qquad

\qquad

\qquad

{\LARGE \textbf{Appendices
}}
\end{center}

\clearpage
\appendix
\def\thesection{Appendix~\Alph{section}}
\def\thesubsection{\Alph{section}.\arabic{subsection}}

\setcounter{page}{1}
\section{Data Appendix}\label{sec:seca}
\setcounter{figure}{0} \renewcommand{\thefigure}{A.\arabic{figure}}
\setcounter{table}{0} \renewcommand{\thetable}{A.\arabic{table}}

\subsection{Gender and Age Cohort Risks}\label{sec:seca_age}

\begin{figure}[!h]
\centering
\includegraphics[width=9cm]{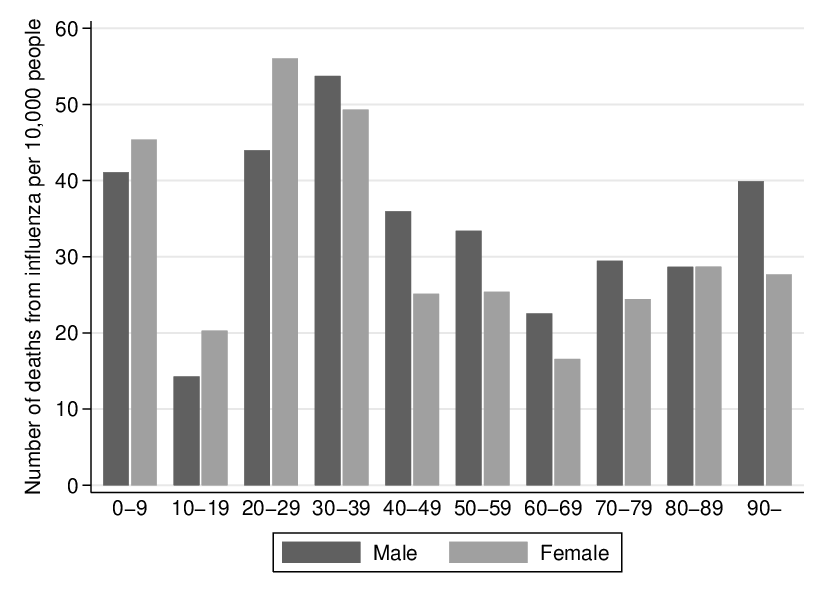}
\caption{Distribution of influenza mortality by gender and age cohort}
\label{fig:risk}
\scriptsize{\begin{minipage}{400pt}
Notes:
The age cohort death rate is defined as the number of deaths from influenza per 10,000 people in 1920.
Source: Ogasawara (2017).

\end{minipage}}
\end{figure}

Figure~\ref{fig:risk} shows the influenza mortality rates by gender and age-cohort calculated using the official reports of the 1920 Population Census and Vital Statistics.
As shown, adults aged 20--39 experienced the highest mortality risk.
Regarding gender differences, females aged 20-29 had suffered the highest risk.
As described in the conclusion section, this feature of the biased infection risks has also been observed in Western countries (Reid 2005).

\subsection{Vital Statistics Records}\label{sec:seca1}

\begin{figure}[h]
\centering
\includegraphics[width=8cm]{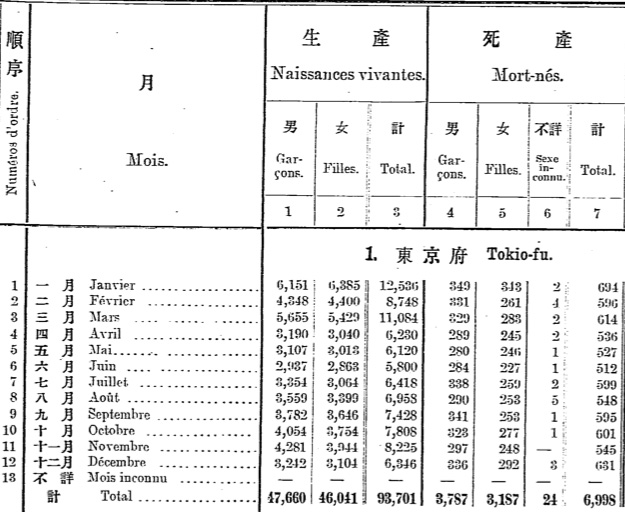}
\caption{Example of the vital statistics record from the VSEJ}
\label{fig:vsej}
\scriptsize{\begin{minipage}{350pt}
Notes: This image shows an example of the VSEJ of 1915.
Source: Statistics Bureau of the Cabinet (1918d, pp.44).
\end{minipage}}
\end{figure}

The data on the number of births by gender are from the 1916--1922 editions of \textit{Nihonteikoku jink\=od\=otait\=okei} (Vital Statistics of Empire Japan, hereafter the VSEJ) (Statistics Bureau of the Cabinet 1919b, 1920b, 1921b, 1922b, 1924b,c,d).
Figure~\ref{fig:vsej} shows an example image of the VSEJ that I mainly use to construct the dataset.
These vital statistics have been recorded based on the comprehensive national registration system (\textit{koseki}).
Thus, the data cover all births during the measured years.
The data on the number of annual live births used as the denominator of the infant mortality rate are also obtained from the VSEJ (Statistics Bureau of the Cabinet 1919b, 1920b, 1921b, 1922b, 1924b,c,d).
To prepare the weighted influenza death rates in 1913--1925 used in Online~\ref{sec:secc}, I additionally digitize the 1912--1914 and 1923--1927 editions of the VSEJ (Statistics Bureau of the Cabinet 1916c, 1917, 1918c, 1924d, 1925d,e, 1926c, 1927b, 1928c).
Although the quality of Japanese fetal death records that begun in 1900 was not high in the initial stage, the records became reliable around the 1920s (Kawana et al. 2007; Ito 1987).
See Drixler (2016) for a more in-depth discussion on birth records in prewar Japan.

\subsection{Statistics of Causes of Death}\label{sec:seca2}

The data on the monthly death tolls from influenza are obtained from the 1915--1922 editions of \textit{Nihonteikoku shiint\=okei} (Statistics of Causes of Death of the Empire of Japan, hereafter the SCDEJ) published by the Statistics Bureau of the Cabinet (Statistics Bureau of the Cabinet 1918b, 1919a, 1920a, 1921a, 1922a, 1923, 1924a, 1925a).
The SCDEJ takes a similar style to the VSEJ in Figure~\ref{fig:vsej}.
The data on the annual infant deaths are obtained from the 1916--1922 editions of the SCDEJ (Statistics Bureau of the Cabinet 1919a, 1920a, 1921a, 1922a, 1923, 1924a, 1925a).
In Online~\ref{sec:secc}, I additionally digitize the 1912--1914 and 1923--1927 editions of the SCDEJ (Statistics Bureau of the Cabinet 1916a,b, 1918a, 1925b,c, 1926b, 1927a, 1928b) to calculate the weighted influenza death rates in 1913--1925.

\subsection{Monthly Population}\label{sec:seca3}

The data on the population are obtained from the official online database of the Statistical Survey Department, Statistics Bureau, Ministry of Internal Affairs and Communications.
In the dataset, the population in month $j$ ($=1,2,...,12$) of year $t$ is calculated as $\tilde{P}^{j}_{t} = P_{t}+B^{j}_{t}-D^{j}_{t} + \{P_{t+1}-(P_{t} + \sum_{j=1}^{12}B^{j}_{t}-\sum_{j=1}^{12}D^{j}_{t})\}/12$, where $P_{t}$, $B^{j}_{t}$, and $D^{j}_{t}$ are the annual population, number of live births, and number of deaths, respectively.
The data on the number of live births and deaths are from the 1915--1922 editions of the VSEJ (Statistics Bureau of the Cabinet 1918d, 1919b, 1920b, 1921b, 1922b, 1924b,c,d).

\subsection{Histograms of the dependent variables}\label{sec:seca4}

\begin{figure}[h]
\centering
\captionsetup{justification=centering,margin=0.2cm}
\subfloat[Male births (\%)]{\label{fig:hist_mbs}\includegraphics[width=0.34\textwidth]{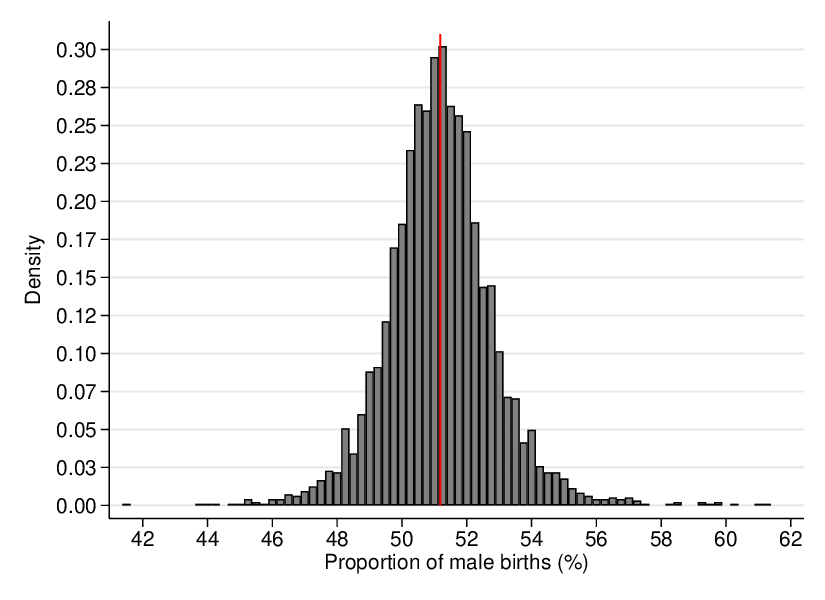}}
\subfloat[Infant mortality rate (\textperthousand)]{\label{fig:hist_imr}\includegraphics[width=0.34\textwidth]{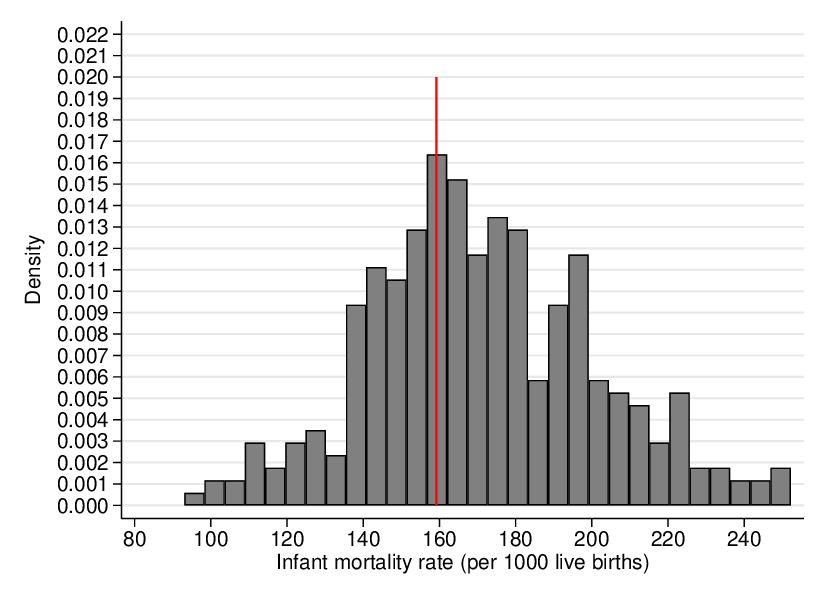}}
\subfloat[Boys (\%)]{\label{fig:hist_pmale}\includegraphics[width=0.34\textwidth]{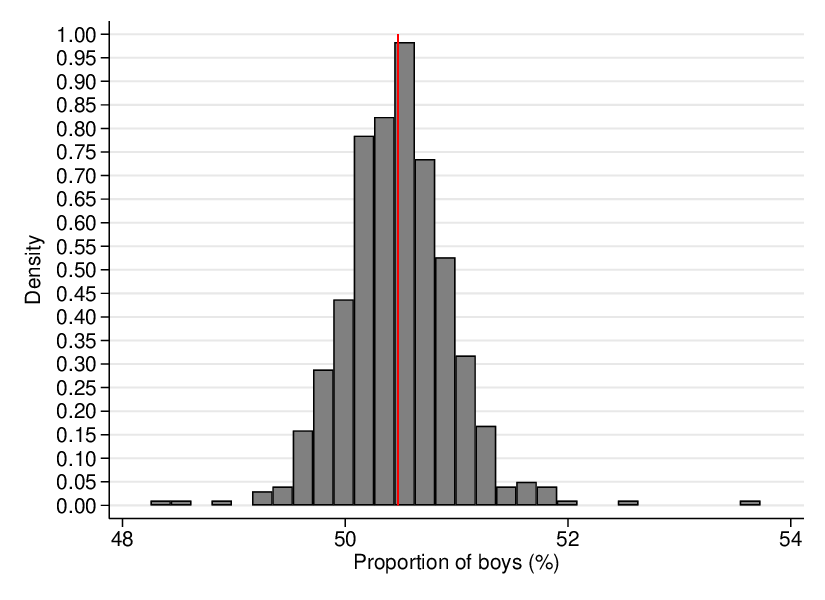}}
\caption{Distributions of the dependent variables}
\label{fig:hist}
\scriptsize{\begin{minipage}{450pt}
\setstretch{0.85}
Notes: Figure~\ref{fig:hist_mbs} shows the histogram of the proportion of male births (\%).
Figure~\ref{fig:hist_imr} shows the histogram of the infant mortality rates (\textperthousand).
Figure~\ref{fig:hist_pmale} shows the histogram of the proportion of boys aged 5--7 and 10--12 (\%).
The red lines indicate the mean values of each variable.
Sources: Created by the authors using Statistics Bureau of the Cabinet (1917, 1918c,d, 1919b, 1920b, 1921b, 1922b, 1924b,c,d, 1925d,e, 1926c).
\end{minipage}}
\end{figure}

Figure~\ref{fig:hist} shows the histograms of the dependent variables used: the proportion of male births (Figure~\ref{fig:hist_mbs}), infant mortality rate (Figure~\ref{fig:hist_imr}, and proportion of boys aged 5--7 and 10--12 (Figure~\ref{fig:hist_pmale}).

The male share at birth can be simplified as $\textit{LB}_{boy}/(\textit{LB}_{boy}+\textit{LB}_{girl})$, whereas the sex ratio is defined as $\textit{LB}_{boy} / \textit{LB}_{girl}$, where $\textit{LB}_{boy}$ and $\textit{LB}_{girl}$ are the number of male and female live births, respectively. Thus, all the empirical results obtained by using the male share at birth are virtually identical to those from the sex ratio at birth. As the Trivers--Willard hypothesis focuses on the vulnerability of male fetuses, it is natural to use the male share at birth because it is apparently more useful than the sex ratio at birth, which can facilitate a more intuitive interpretation of the results.

\subsection{Statistical Yearbook}\label{sec:seca5}

In order to obtain the data on the number of medical doctors and midwives, I digitize the volumes 36--43 of \textit{Nihonteikoku t\=okeinenkan} (Statistical Yearbook of the Japanese Empire, hereafter the SYEJ) (Statistics Bureau of the Cabinet 1914--1927).
The data on the population used as the denominator are taken from the official database of the Statistical Survey Department, Statistics Bureau, Ministry of Internal Affairs and Communications.

\subsection{Dynamics of the influenza mortality during the pandemic}\label{sec:seca6}

\begin{table}[h!]
\def\arraystretch{1.0}
\begin{center}
\captionsetup{justification=centering}
\caption{Dynamic relationships in the influenza mortality rates during the pandemics}
\label{tab:appa1}
\scriptsize
\scalebox{1.0}[1]{
\begin{tabular}{lcc}
\toprule
Dependent variable: Influenza mortality rate& (1) November 1918--January 1919&(2) January 1920--March 1920\\\hline
Lagged influenza mortality rate	&-0.0424		&0.1281	\\
							&[0.6800]		&[0.4000]	\\\hline
Number of prefectures		&46				&46	\\
Number of months			&3				&3	\\
Number of observations		&92				&92	\\
$R$-squared					&0.6604			&0.7460	\\\bottomrule

\end{tabular}
}
{\scriptsize
\begin{minipage}{430pt}
\setstretch{0.90}
The $p$-values from the wild cluster bootstrap resampling method are in brackets.
The data are clustered at the 8-area level in the bootstrap procedure.
The number of replications is fixed at 150 for all the specifications.\\
Notes: There are no control variables but both prefecture and year-month fixed effects are included.
$R$-squared is obtained from the least-squares dummy variable regression.
All the regressions are weighted by the average number of people in each prefecture.
\end{minipage}
}
\end{center}
\end{table}

In Figure~\ref{fig:map_flur} in Section~\ref{sec:sec22}, I have confirmed that there were no systematic spatiotemporal correlations among the influenza mortality rates during the first and second pandemic waves.
In this subsection, I further test whether the influenza mortality rates were correlated with the lagged rates in each wave.
If there were statistically significantly positive correlations between the influenza mortality rates and their lagged values, the randomness in the influenza mortality during the pandemics is considered weak, and thus there might have been systematic endogeneity.

Table~\ref{tab:appa1} presents the estimation results from the simple dynamic panel data analysis.
For each pandemic wave, I regress the influenza mortality rate on the lagged influenza mortality rate.
Both prefecture and year-month fixed effects are included in both specifications because I also use those fixed effects in my models above (see equations~\ref{bs} and~\ref{imrbs}).
Columns (1) and (2) in Table~\ref{tab:appa1} show the results for the first and second waves illustrated in Figure~\ref{fig:map_flur}, respectively.
The estimated coefficients of the lagged influenza mortality rates are statistically insignificant in both cases.
This provides evidence that my key exposure variable, the influenza mortality rate, has a certain random nature.

\subsection{Reports of the Population Census}\label{sec:seca7}

To obtain the data on the number of people by gender,  I digitize a number of Kokuse- ic\=osah\=okoku (Reports of the Population Census).
Since the prefecture editions of the Population Census were published for each prefecture, I use $92$ issues (46 prefectures $\times$ 2 census years) to construct the dataset.
For simplicity, I refer to those issues as Statistics Bureau of the Cabinet (1929) and Statistics Bureau of the Cabinet (1933).

\subsection{JMA database and Location Information}\label{sec:seca8}

The Japan Meteorological Agency (JMA) reports the number of days with a temperature above or below a certain threshold to record heat and cold waves
The data are available from the JMA database at \url{http://www.data.jma.go.jp/gmd/risk/obsdl/index.php} (accessed on March 30, 2018).
Data on latitude and longitude are taken from the database of the Geospatial Information Authority of Japan: \url{http://www.gsi.go.jp/KOKUJYOHO/kenchokan.html}, accessed on August 20, 2017.

\subsection{Coal consumption}\label{sec:seca9}

The data on the coal consumption between 1913 and 1918 are obtained from the \textit{N\=osh\=omush\=o t\=okei sho} (The statistical yearbook of agriculture and commerce, volumes 30--35) published by Department of Agriculture and Commerce.
The data between 1919 and 1925 are obtained from the \textit{K\=ojy\=o t\=okei hy\=o} (Census of manufactures , 1919--1925 editions) published by Secretariat of Agriculture and Commerce.
Each document records the total weights of coals that are consumed in all the factories that have more than or equal to five workers in a specific prefecture-year.
The unit of weights is measured in \textit{kin} (approximately $600$ grams), and thus I convert the unit in metric tons (i.e., $1,000$ kilograms).
There are only four missing values in Gunma, Kanagawa, Fukui, and Nara in 1922.
Although I linearly interoperate these cells to balance the panel dataset, this imputation does not influence the results.
To calculate per capita figure, I divide the coal consumption by the number of people in each prefecture-year cell that is taken from the database of the Statistical Survey Department, Statistics Bureau, Ministry of Internal Affairs and Communications (Online Appendix~\ref{sec:seca3}).

\subsection{School Enrollment Rate}\label{sec:seca10}

The statistics on the school enrollment rate are obtained from the Annual Report of the Japanese Imperial Ministry of Education (23-54 editions) published by the Ministry of Education.
According to Hijikata (1994,p.13), the school enrollment rate is defined as:
\begin{eqnarray*}\label{ser}
\footnotesize{
\text{School enrolment rate} = 100\times \frac{\text{Students aged 6--13 years}+\text{Graduates aged 6--13 years}}{\text{Children aged 6--13 years}}
 }.
\end{eqnarray*}
As the local governments comprehend all the number of children under the official registration and educational systems (\textit{koseki} and \textit{gakusei}), this rate provides a representative figure of the school enrollments of children at that time.
I use school enrollment rates from 20 years before the year of birth as a proxy of the average school enrollment rate of the parental generation.
This is based on the evidence that the mean age of mothers at the time of their first birth is in the 20s in prewar Japan (Ogasawara 2017).

\section{Empirical Analysis Appendix}\label{sec:secc}
\setcounter{table}{0} \renewcommand{\thetable}{B.\arabic{table}}
\setcounter{figure}{0} \renewcommand{\thefigure}{B.\arabic{figure}}

\subsection{Model Interpretation}\label{sec:secb_model}

A recent growing body of econometrics literature has investigated the interpretation of the estimator from regression difference-in-differences (DID) models within the framework of the fixed-effect model with two-way error components (TWFE) (Callaway and Sant'Anna 2021; de Chaisemartin and D'Haultfoeuille 2020; Goodman-Bacon 2021; Sun and Abraham 2021). While there are a few defined ways of interpreting the estimates from the TWFE with staggered treatment variable, the interpretation for the TWFE with continuous treatment variable has not been fixed yet (Callaway et al. 2021). Given this, I did not consider equation~\ref{bs} as a DID-styled estimation strategy. Instead, I interpreted the results from the within estimator in a conventional way.

\subsection{Criterion of Clusters}\label{sec:secb_cluster}

I use the most representative geographical classification in Japan for the clustering.
It divides 46 prefectures into 8 areas: Hokkaid\=o (northernmost), T\=ohoku (eastern), Kant\=o (east-central), Ch\=ubu (west-central), Kansai (south-central), Ch\=ugoku (westernmost), Shikoku (southwest of the main island), and Ky\=ush\=u (southwest island).
Therefore, by using the CRVE, the regressions allow for heteroskedasticity and correlations within clusters as well as the potential heteroskedasticity across clusters (Section~\ref{sec:sec4}).
In addition, there is an alternative classification dividing 16 different areas. I prefer to use the 8-area classification because the 16-area classification is evidently more narrow (i.e., including a small number of prefectures in a cluster) than the former classification and thus, postulates stronger assumptions for the inference.

\subsection{Testing Autocorrelations}\label{sec:secb_serial}

Table~\ref{tab:serial} presents the results of Wooldridge tests for serial correlation in the proportion of male births used in the main empirical analysis \citep{wooldridge}.
Column (1) shows the results for the specification excluding the flu exposure variables, whereas column (2) shows the results for baseline specifications, including the exposure variables (equation~\ref{fs}).
The null hypothesis of no serial correlation was not rejected at the conventional level in both specifications.

\begin{table}[h!]
\def\arraystretch{1.0}
\begin{center}
\captionsetup{justification=centering}
\caption{Results of the Wooldridge tests}
\label{tab:serial}
\footnotesize
\scalebox{1.0}[1]{
\begin{tabular}{lcccc}
\toprule
&\multicolumn{4}{c}{Specification}\\
\cmidrule(rrrr){2-5}
										&~~~~~~~~&(1)		&~~~~~~~~&(2)		\\\hline
$F$-statistic $p$-value					&~~~~~~~~&0.1476	&~~~~~~~~&0.1826	\\\hline
Exposure variables						&~~~~~~~~&No		&~~~~~~~~&Yes		\\
Control variables						&~~~~~~~~&Yes		&~~~~~~~~&Yes		\\
Fixed effects							&~~~~~~~~&Yes		&~~~~~~~~&Yes		\\\bottomrule

\end{tabular}
}
{\scriptsize
\begin{minipage}{240pt}
\setstretch{0.90}
Notes:
This table reports the results of Wooldridge tests for serial correlation based on the first-differenced panel regression during the pandemic.
$p$-value for the null hypothesis that there is no serial correlation is reported in each specification. 
See \citet{wooldridge} for the details of the tests.
\end{minipage}
}
\end{center}
\end{table}

\subsection{Testing Stationarity}\label{sec:secb_ts}

Table~\ref{tab:urt} presents the results of the panel unit root tests for the proportion of male births used in the main empirical analyses, confirming the stationarity of my panel dataset on the secondary sex ratio.

\begin{table}[h!]
\def\arraystretch{1.0}
\begin{center}
\captionsetup{justification=centering}
\caption{Results of the unit root tests for the proportion of male births}
\label{tab:urt}
\scriptsize
\scalebox{1.0}[1]{
\begin{tabular}{lcccc}
\toprule
&\multicolumn{2}{c}{January 1916--December 1922}&\multicolumn{2}{c}{January 1918--December 1920}\\
\cmidrule(rr){2-3}\cmidrule(rr){4-5}
Test statistics					&(1)	&(2)	&(3)	&(4)	\\\hline
$P$-statistic $p$-value			&0.0000	&0.0000	&0.0000	&0.0000	\\
$Z$-statistic $p$-value			&0.0000	&0.0000	&0.0000	&0.0000	\\
$L^{*}$-statistic $p$-value		&0.0000	&0.0000	&0.0000	&0.0000	\\
$P_{m}$-statistic $p$-value		&0.0000	&0.0000	&0.0000	&0.0000	\\\hline
Number of prefectures		&46		&46		&46		&46		\\
Number of periods			&84		&84		&36		&36		\\
Number of lagged differences	&1		&3		&1		&3		\\\bottomrule

\end{tabular}
}
{\scriptsize
\begin{minipage}{360pt}
\setstretch{0.90}
Notes: The results of the Fisher-type panel unit root tests based on augmented Dickey--Fuller (ADF) tests are reported in this table.
The null hypothesis is that all the panels contain unit roots, whereas the alternative hypothesis is that at least one panel is stationary. 
In all the specifications, the process under the null hypothesis is assumed to be a random walk with drift. 
The demeaned data are used to address the effect of cross-sectional dependence. 
Although the number of lagged differences in the ADF regression equation reported is set as either one or three, the results are not affected by the number of lagged differences.
See Choi (2001) for the details of the tests.
\end{minipage}
}
\end{center}
\end{table}

\subsection{Time-series plots of the proportion of male births and infant mortality}\label{sec:secb_ct}

\begin{figure}[h]
\centering
\captionsetup{justification=centering,margin=0.2cm}
\subfloat[Male births (\%)]{\label{fig:ts_mbs}\includegraphics[width=0.45\textwidth]{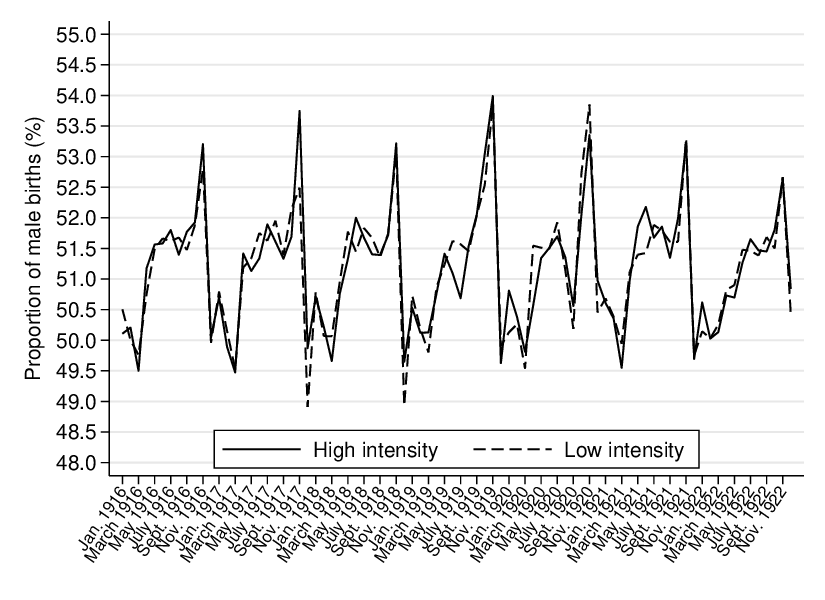}}
\subfloat[Infant mortality rate (\textperthousand)]{\label{fig:ts_imr}\includegraphics[width=0.45\textwidth]{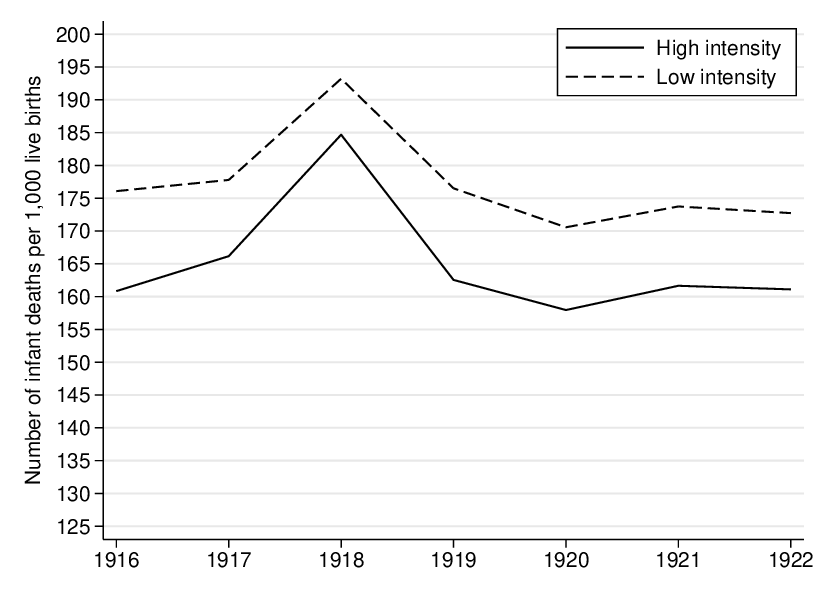}}
\caption{Time-series plots of the male births (\%) and infant mortality (\textperthousand)}
\label{fig:ts}
\scriptsize{\begin{minipage}{450pt}
\setstretch{0.85}
Notes: Figure~\ref{fig:ts_mbs} shows the monthly average time-series plots for the proportion of male births (\%).
In Figure~\ref{fig:ts_mbs}, ``High intensity'' (``Low intensity'') refers to the prefectures that experienced the monthly average influenza mortality above (less than) median: $4.72$ permyriad during the first and second waves.
The first and second waves are the periods between November 1918 and January 1919 and between January 1920 and March 1920, respectively (see Figure~\ref{fig:map_flur}).
Figure~\ref{fig:ts_imr} shows the annual average time-series plots of the infant mortality rates (\textperthousand).
In Figure~\ref{fig:ts_imr}, ``High intensity'' (``Low intensity'') refers to the prefectures that experienced an annual average influenza mortality above (below) the median, that is, $1.27$ permyriad in 1918 and 1920.
Sources: Created by the authors using data from Statistics Bureau of the Cabinet (1917, 1918c,d, 1919b, 1920b, 1921b, 1922b, 1924b,c,d, 1925d,e, 1926c).
\end{minipage}}
\end{figure}

Figures~\ref{fig:ts_mbs} and \ref{fig:ts_imr} present the time-series plots of the proportion of male births and infant mortality rates between 1916 and 1922.
In Figure~\ref{fig:ts_mbs}, ``High intensity'' (``Low intensity'') refers to the prefectures that experienced a monthly average influenza mortality above (below) the median, that is, $4.72$ permyriad during the first and second waves (November 1918--January 1919 and January 1920--March 1920, respectively).
As shown, the two series overlap in most year-month cells and the trends in both groups can be considered to be very similar.
The same is true for the infant mortality shown in Figure~\ref{fig:ts_imr}.
In Figure~\ref{fig:ts_imr}, ``High intensity'' (``Low intensity'') refers to the prefectures that experienced a monthly average influenza mortality above (below) the median, that is, $1.27$ permyriad in 1918 and 1920.
Clearly, both series exhibit very similar trends.
Note that the seasonality in the proportion of male births (shown in Figure~\ref{fig:ts_mbs}) is effectively captured by using the year-month fixed effects in my specifications (Section~\ref{sec:sec32}).

\subsection{Flexible Estimates}\label{sec:secb_aff}

Table~\ref{tab:secb_aff} presents the results from alternative specification that uses the intensity indicator variables based on the influenza death rates instead of the continuous average influenza death rates.
For example, [50pctl, 75pctl) listed in the ``First trimester'' panel is an indicator variable that takes a value of one if the average influenza death rate in the first trimester is equal to or more than 50 percentile and less than 75 percentile.
Column (1) includes the control variables, prefecture fixed effects, and year-month-specific fixed effects, whereas column (2) further contains the linear time trend.
Evidently, the estimated coefficients become larger in an absolute sense as the percentile becomes larger.
In column (2), the estimated coefficient on the [50pctl, 75pctl), [75pctl, 90pctl), and [90pctl, Max] indicator variables of the first trimester are $-0.1381$, $-0.3374$, and $-0.6461$, respectively.
This is reasonable because an individual's risk of infection increases at an increasing rate as the number of infected individuals goes up (Ogasawara 2018).
This means that my baseline specification of equation~\ref{fs} provides a conservative estimate in the framework of linear functional form because the identification relies on the dramatic rise in the influenza mortality rates during those periods.
Nonetheless, I prefer to use the feasible linear functional form, which can provide the intuitive estimates because the semi-parametric panel data technique is still underdeveloped and rather difficult to be implemented in the case of clustering regressions (Hsiao 2014, pp.461--463).

\subsection{Testing the Gender Difference}\label{sec:secb_gd}

In this subsection, I investigate the gender difference in the estimates reported in Table~\ref{tab:r_main_imr}.
To test the difference, I pool the infant mortality rates for boys and girls and then interact all the independent variables including the fixed effects with the gender dummy.
Table~\ref{tab:r_main_imr_gender} presents the results.
As shown, the number of observations is now $644$ ($322 + 322$).
Column (1) indicates that the estimated effect of fetal influenza exposure on the infant mortality rate of boys is approximately 4.6\textperthousand~lower than that of girls.
This result is largely unchanged if I include the prefecture-specific time trend in column (2).
If I disaggregate the exposure variable into trimesters, however, such gender difference becomes less clear in a statistical sense (not reported).
This must be because the gender differences in the effects are generated by the cumulative effects of all trimesters.

\subsection{Results for Event Study Analysis}\label{sec:sec_es}

This subsection considers a placebo experiment within the framework of event study analysis.
The specification is as follows:
\begin{eqnarray}\label{es} 
\footnotesize{
\begin{split}
y_{it} = \tilde{\pi} + \sum_{z \in Z} \tilde{\delta}_{z} \overline{\textit{FLUDR}}_{it}^{\textit{$z$-th trimester}} + \vx'_{ig_{t}} \tilde{\vzeta} + \tilde{\upsilon}_{i} + \tilde{\kappa}_{t} + \bar{\upsilon}_{i}t + \tilde{\eps}_{it}
,
\end{split}
}
\end{eqnarray}
where $Z = \{-6, -2,..., 5, 9\}$, and $\overline{\textit{FLUDR}}_{it}^{\textit{$z$-th trimester}}$ is the average influenza death rate during the $z$-th trimester.\footnote{Note that the event study specification uses continuous exposure variables from $-6$ to $9$ trimesters, implying that the reference trimester should not exist. See also Sun and Abraham (2021) for recent discussion on the event study analysis.}
If the trends representing the proportion of males at birth are plausibly similar across different prefectures, the proportion of males at birth do not respond to the influenza death rates in the statistical sense except for the first trimester.
For instance, if the influenza death rates have systematically negative impacts on the proportion of males at birth, there must be an endogenous relationship between both variables via, for example, the potential wealth and public health levels.
Therefore, one must expect that $\forall z \neq 1, \hat{\tilde{\delta}}_{z} \approx 0$, and $\tilde{\delta}_{1}$ is statistically significantly negative.
Table~\ref{tab:r_event} presents the results.
Column (1) indicates that the estimated coefficient for the first trimester variable is negative and statistically significant, whereas the other estimates fluctuate unsystematically around zero and are statistically insignificant.
This result is unchanged if I include the prefecture-specific time trend in column (2).
The result obtained herein is consistent with the fact that my baseline results are not sensitive to include the prefecture-specific trend terms (Table~\ref{tab:r_main}).
\footnote{The estimated coefficient for the $-1$-trimester is insignificant, but positive and relatively large. This might be a phenomenon whereby maternal influenza exposure just before conception might have improved maternal immunity during pregnancy. An alternative explanation is that the pandemic exposure in the first trimester led to some premature male births in the second trimester. Early birth may seem to imply that the exposure was $-1$ trimester. However, this mechanism may not work because these premature babies are less likely to survive and thus, should not be included in the live birth records. In any case, it was confirmed that this value was never statistically significant if the alternative cluster definition was used, excluding the prefecture-specific time trend, and trimming the length of the trimesters, from $-3$ to $6$ trimesters. Thus, this does not indicate any systematic trend in the gender balance at birth.}

Note that it is technically difficult to conduct the event study analysis on the infant mortality rate because both outcomes are measured in the annual level. The weighted influenza death rate of a certain trimester is more likely to be correlated with the rates in the other trimesters given the feature of the weighting transformation (equation~\ref{wdr}).

\subsection{Omitted Variable Bias}\label{sec:sec71}

\subsubsection{\textit{Weather Shocks}}\label{sec:sec711}

First, I examine the robustness of my main results to the inclusion of measures of heat and cold waves during the sample period.
A glowing body of literature found the evidence that weather shocks can affect birth outcomes, and temperature might correlate with the risk of infectious diseases during epidemics (Andal\'on et al. 2016; Desch\^enes et al. 2009; Molina and Saldarriaga 2017).
To control for heat and cold waves, I draw temperature data between 1915 and 1922 for a maximum of three weather stations in each prefecture based on the official database of the Japanese Meteorological Agency (JMA).\footnote{Online Appendix~\ref{sec:seca8} provides the details of this database.}
Using the official definitions provided by the JMA, I define a heat wave as the annual average number of days on which the maximum temperature exceeded 30$^\circ$C and a cold wave as the annual average number of days on which the minimum temperature was below 0$^\circ$C.
Then, following Desch\^enes et al. (2009), I calculate the inverse distance-weighted average of all the valid measurements from these.\footnote{
Each prefecture's centroid is set as the city office because a large part of the population lives in the principal city in each prefecture.
The weighted average of the weather shock variable for prefecture $i$ in month $m$ is given as follows:
\begin{eqnarray}\label{idw}
\text{IDWS}_{im} = \frac{\sum_{j=1}^{3}\frac{w_{imj}}{d_{ij}}}{\sum_{j=1}^{3}\frac{1}{d_{ij}}},
\end{eqnarray}
where $w$ denotes the weather shock variable and $d$ denotes the geospatial distance from the centroid to station $j$.
Data on latitude and longitude are taken from the database of the Geospatial Information Authority of Japan (Online Appendix~\ref{sec:seca8}).}
As regards the average influenza death rates in Section~\ref{sec:sec3}, I calculate the average number of days of heat and cold waves for all trimesters by using the weighted average of the weather shock variable.
These weather shock variables are finally added to equations~\ref{fs} and \ref{imrfs}.
The results are presented in Panel A of Table~\ref{tab:r_robust1}.
All specifications include the baseline control variables and the heterogeneous trend terms.
Clearly, the estimated coefficients of the influenza exposure variables are close to the corresponding estimates in Tables~\ref{tab:r_main} and \ref{tab:r_main_imr}.\footnote{Note that although the estimate for the first trimester is statistically significantly negative in column (3) of panel A, this is not robust in the same column of panels A and B, as well as Table~\ref{tab:r_main_imr}.}

\subsubsection{\textit{Air Pollution}}\label{sec:sec712}

Another potential source that leads to bias might be the air pollution.
In the case of the early-twentieth century United States, the influenza mortality rate had been positively correlated with the degree of air pollution (Clay et al. 2018).
If this sort of correlation does matter in my regressions, my estimates may overstate the impacts of the influenza exposures.
Following previous studies, therefore, I use the coal consumption rate in factories in each prefecture as a plausible measure of air pollution (Beach and Hanlon 2017; Hanlon 2019).\footnote{Online Appendix~\ref{sec:seca9} provides the finer details of the coal consumption data.}
Similar to the weather shock analysis, I include the per capita coal consumption into equations~\ref{fs} and \ref{imrfs}.
Panel B of Table~\ref{tab:r_robust1} presents the results, indicating that my baseline results shown in Tables~\ref{tab:r_main} and \ref{tab:r_main_imr} are largely unchanged.

\subsubsection{\textit{School Enrollment Rates}}\label{sec:sec713}

As I explained in the introduction section, the positive selection effects due to WWI, as highlighted by Beach et al. (2022), should not be a critical issue in the case of prewar Japan.
As a sanity check, therefore, I include the school enrollment rate for the parental generation as a proxy of parental characteristics to test whether parental SES has been associated with pandemic severity.
To conduct the analysis, I transcribe the available documents on the school enrollment rates between 1893 and 1922 by using the Annual Report of Japanese Imperial Ministry of Education.\footnote{As local governments comprehend the number of children under the official registration and educational systems (\textit{koseki} and \textit{gakusei}, the rate provides a representative figure of school enrollments of children at that time. See Online Appendix~\ref{sec:seca10} for the finer details of the school enrollment data.}
Given the available range of the data, I use the 20-year lagged rate from the birth year as the school enrollment rate for the parental generation.
Panel C of Table~\ref{tab:r_robust1} presents the results, supporting that the main results are unchanged after controlling for this variable.

\subsubsection{\textit{Flexible Fixed Effects}}\label{sec:sec714}

The baseline specifications of equations~\ref{fs} and \ref{imrfs} include both prefecture and time fixed effects.
However, there potentially be the prefecture-specific-time-varying unobservable factors that might be correlated with the influenza mortality.
I further consider alternative specifications using flexible fixed effects to deal with this potential issue.
Specifically, I include the prefecture-by-year-by-quarter fixed effect and the month fixed effect instead of the prefecture and year-by-month fixed effects in equation~\ref{fs}.
Similarly, I include the prefecture-by-period fixed effects instead of the prefecture and year fixed effects in equation~\ref{imrfs}.
Here, the period is defined as a category variable that takes one for 1916 and 1917 (i.e., pre-pandemic period), two for 1918--1920 (i.e., pandemic period), and three for 1921--1922 (i.e., post-pandemic period).
This is the most plausible definition in the practical sense because we are interested in controlling for the correlations between influenza mortality and unobservables. It is necessary to include a category representing the pandemic period, say 1918--1920.
To summarize, these flexible specifications allow the prefecture-specific unobservable factors to vary over time, effectively controlling for the confounding factors to a considerable degree.

Table~\ref{tab:r_robust_flex} presents the results.
Column (1) shows the result for the proportion of male births, whereas columns (2) and (3) show the results for the infant mortality rates.
Note that all estimates are obtained from the least squares dummy variable (LSDV) regressions because all these regressions have time-varying fixed effects that cannot be removed using the within transformation.
For a similar reason, while I employed the CRVE for calculating the standard errors reported in the parentheses, I could not apply the bootstrap-$t$ procedure to these regressions because all these regressions include many fixed effects relative to the number of observations. For instance, the regression in column (1) considers $1,288$ ($46 \times 7 \times 4$) bilateral fixed effects and 12-month fixed effects.

Overall, the results shown in Table~\ref{tab:r_robust_flex} are similar to the baseline results reported in Tables~\ref{tab:r_main} and~\ref{tab:r_main_imr}.
This supports the evidence that my main results are robust to the unobservable time-varying confounding factors.
Rather, the estimate for the first trimester in column (1) is greater than that in column (3) or (4) of Tables~\ref{tab:r_main}.
Similarly, the estimates for the third trimester in columns (2) and (3) are greater than those in columns (3) and (4) in panels B and C of Table~\ref{tab:r_main_imr}.
Essentially, this implies that the baseline estimates are conservative rather than overestimating the impacts of the flu pandemics.
Although the estimated coefficient for the first trimester in column (3) is weakly statistically significantly negative, it does not disturb the interpretation as explained in the main text.
Moreover, as previously explained, this estimate is not robust(Table~\ref{tab:r_robust1}).

\subsection{Pretreatment Trends}\label{sec:secb_placebo}

As previously explained, it is technically difficult to conduct the event study analysis for regression using the annual data (i.e., the infant mortality rates and proportion of males in the primary school ages).
Therefore, as an alternative test, I include a weighted average influenza death rate during ``-1'' trimester, that is, representing the rates between 10 and 12 months \textit{before} a birth, in the specifications using the annual-level datasets (equations~\ref{imrfs} and \ref{pmale2}).
Given that a fetus does not exist \textit{in utero} before conception, a future mother's exposure to pandemic influenza during this pre-conception period should have no significant impact on the proportion of males at birth.
Table~\ref{tab:r_robust_pretrend} presents the results from the specifications, including the weighted average influenza death rate during the $-1$ trimester, that is, the rate between 10 to 12 months before conception.
Columns (1)--(2) show the results for the infant mortality rate, whereas columns (3)--(4) show the results for the proportion of males.
Evidently, the estimated coefficients of these pre-treatment variables are statistically insignificant.
This provides evidence that the exposure variables do not capture any secular pre-trends in both outcome variables.

\subsection{Stratification}\label{sec:sec_str}

In Table~\ref{tab:r_main}, I stratify the sample to confirm that the adverse effects are observed only in the pandemic years between 1918 and 1920.
If my empirical setting were valid to capture the overall effects of the fetal influenza exposure to the proportion of male births, the estimated effect in the first wave must be clearer than that in the second wave because the scale and term of epidemics were greater during the first-wave than the second-wave (see Figure~\ref{fig:ts_flu}), and might be because there might have been unobservable preventive behavioral responses by people, including hygiene practices during the second-wave.
Table~\ref{tab:r_robust2} further shows the stratification of the sample into the first- and second-wave subsamples: October 1918--December 1919 and December 1920, respectively.
Columns (1)--(2) in Panel C of Table~\ref{tab:r_robust2} show the results for the first-wave subsample, whereas columns (3)--(4) show the results for the second-wave subsample.
The estimated coefficient of the average influenza death rate during the first trimester in column (1) is negative and statistically significant, which does not change if I include the heterogenous time trend in each prefecture in column (2).
The estimated coefficients in column (3) are also negative, and it turns out to be statistically significant if the heterogenous time trend is included in column (4).
As expected, the estimated effect is larger in the first-wave subsample ($-0.2101$ in column (2)) than that in the second-wave subsample ($-0.1058$ in column (4)).
Because the number of observations becomes substantially reduced in these regressions from my baseline regressions (shown in Table~\ref{tab:r_main}), it is slightly problematic to make further discussions on the statistical inference given the number of parameters to be estimated.
However, this result indicates that the estimates respond to the intensity of epidemics well, which thereby supports the evidence that my baseline specification of equation~\ref{fs} captures the overall impacts of the pandemics.

\begin{table}[]
\def\arraystretch{1.0}
\begin{center}
\captionsetup{justification=centering}
\caption{Effects of fetal influenza exposure on the proportion of male births (\%):\\ Flexible estimates using the intensity indicator variables}
\label{tab:secb_aff}
\footnotesize
\scalebox{1.0}[1]{
\begin{tabular}{lD{.}{.}{-2}D{.}{.}{-2}}
\toprule
&\multicolumn{2}{c}{Dependent variable: Proportion of male births (\%)}\\
\cmidrule(rr){2-3}
&\multicolumn{1}{c}{(1)}&\multicolumn{1}{c}{(2)}				\\\hline
First trimester&&\\
\hspace{10pt}[50pctl, 75pctl)	&-0.1474		&-0.1388		\\
								&[0.2300]		&[0.2300]		\\
\hspace{10pt}[75pctl, 90pctl)	&-0.3116		&-0.3044		\\
								&[0.1550]		&[0.1650]		\\
\hspace{10pt}[90pctl, Max]		&-0.5615$***$	&-0.5948$**$	\\
								&[0.0100]		&[0.0250]		\\
Second trimester&&\\
\hspace{10pt}[50pctl, 75pctl)	&-0.0380		&-0.0269		\\
								&[0.8650]		&[0.8450]		\\
\hspace{10pt}[75pctl, 90pctl)	&-0.2174		&-0.2096		\\
								&[0.3750]		&[0.3650]		\\
\hspace{10pt}[90pctl, Max]		&-0.2111		&-0.2509		\\
								&[0.3800]		&[0.4050]		\\
Third trimester&&\\
\hspace{10pt}[50pctl, 75pctl)	&-0.1345		&-0.1492		\\
								&[0.3200]		&[0.2400]		\\
\hspace{10pt}[75pctl, 90pctl)	&-0.3617		&-0.3664		\\
								&[0.2900]		&[0.2800]		\\
\hspace{10pt}[90pctl, Max]		&-0.1586		&-0.2061		\\
								&[0.6250]		&[0.5200]		\\\hline
Control variables					
&\multicolumn{1}{c}{Yes}		&\multicolumn{1}{c}{Yes}\\
Fixed effects		
&\multicolumn{1}{c}{Yes}		&\multicolumn{1}{c}{Yes}\\
Time trend					
&\multicolumn{1}{c}{No}		&\multicolumn{1}{c}{Yes}\\
Period						
&\multicolumn{1}{c}{Jan. 1918--Dec. 1920}		&\multicolumn{1}{c}{Jan. 1918--Dec. 1920}\\
Observations						
&\multicolumn{1}{c}{1656}		&\multicolumn{1}{c}{1656}\\
Number of prefectures
&\multicolumn{1}{c}{46}		&\multicolumn{1}{c}{46}\\
Number of months
&\multicolumn{1}{c}{36}		&\multicolumn{1}{c}{36}\\
Number of clusters
&\multicolumn{1}{c}{8}		&\multicolumn{1}{c}{8}\\\bottomrule
\end{tabular}
}
{\scriptsize
\begin{minipage}{340pt}
\setstretch{0.90}
***, **, and * represent statistical significance at the 1\%, 5\%, and 10\% levels based on the $p$-values from the wild cluster bootstrap resampling method in brackets, respectively.
The data are clustered at the 8-area level in the bootstrap procedure.
The number of replications is fixed to 400 for all the specifications.\\
Notes: This table shows the results of the specifications using the influenza intensity indicator variables as the exposure variables in equation~\ref{fs}.
For instance, [50pctl, 75pctl) is an indicator variable that takes the value of one if the average influenza death rate in each trimester is equal to or more than 50 percentile and less than 75 percentile.
Similarly, [90pctl, Max] indicates an indicator variable that takes the value of one if the average influenza death rate in each trimester is equal to or more than 90 percentile.
The control variables include rice yield, soy yield, milk production, coverage of doctors, and coverage of midwives.
The fixed effects include both the prefecture and the year fixed effects.
The time trend indicates the prefecture-specific linear time trend.
All the regressions are weighted by the average number of births in each prefecture.
\end{minipage}
}
\end{center}
\end{table}
\begin{table}[]
\def\arraystretch{1.0}
\begin{center}
\captionsetup{justification=centering}
\caption{Effects of fetal influenza exposure on the infant mortality rate (\textperthousand):\\ Testing the gender difference}
\label{tab:r_main_imr_gender}
\footnotesize
\scalebox{1.0}[1]{
\begin{tabular}{lD{.}{.}{-2}D{.}{.}{-2}}
\toprule
&\multicolumn{2}{c}{Dependent variable: Infant mortality rate (\textperthousand)}\\
\cmidrule(rr){2-3}
&\multicolumn{1}{c}{(1)}&\multicolumn{1}{c}{(2)}				\\\hline
All trimesters					&8.252$***$	&8.573$***$	\\
								&[0.007]		&[0.007]		\\
All trimesters $\times$ Boys		&-4.595$**$	&-4.530$**$\\
								&[0.033]		&[0.033]		\\\hline
Control variables					
&\multicolumn{1}{c}{Yes}		&\multicolumn{1}{c}{Yes}\\
Control variables $\times$ Boys					
&\multicolumn{1}{c}{Yes}		&\multicolumn{1}{c}{Yes}\\
Fixed effects		
&\multicolumn{1}{c}{Yes}		&\multicolumn{1}{c}{Yes}\\
Fixed effects $\times$ Boys					
&\multicolumn{1}{c}{Yes}		&\multicolumn{1}{c}{Yes}\\
Time trend					
&\multicolumn{1}{c}{No}		&\multicolumn{1}{c}{Yes}\\
Time trend $\times$ Boys					
&\multicolumn{1}{c}{No}		&\multicolumn{1}{c}{Yes}\\
Observations						
&\multicolumn{1}{c}{644}		&\multicolumn{1}{c}{644}\\
Number of prefectures
&\multicolumn{1}{c}{46}		&\multicolumn{1}{c}{46}\\
Number of years
&\multicolumn{1}{c}{14}		&\multicolumn{1}{c}{14}\\
Number of clusters
&\multicolumn{1}{c}{8}		&\multicolumn{1}{c}{8}\\\bottomrule
\end{tabular}
}
{\scriptsize
\begin{minipage}{335pt}
\setstretch{0.90}
***, **, and * represent statistical significance at the 1\%, 5\%, and 10\% levels based on the $p$-values from the wild cluster bootstrap resampling method in brackets, respectively.
The data are clustered at the 8-area level in the bootstrap procedure.
The number of replications is fixed to 150 for all the specifications.\\
Notes: This table shows the results of the specifications using the sample pooling the infant mortality rates for boys and girls.
As shown, all the independent variables are interacted with an indicator variable for boys.
The control variables include rice yield, soy yield, milk production, coverage of doctors, and coverage of midwives.
The fixed effects include both the prefecture and the year fixed effects.
The time trend indicates the prefecture-specific linear time trend.
All the regressions are weighted by the average number of live births in each prefecture.
\end{minipage}
}
\end{center}
\end{table}
\begin{table}[h]
\def\arraystretch{1.0}
\begin{center}
\captionsetup{justification=centering}
\caption{Results for Event Study Analysis}
\label{tab:r_event}
\footnotesize
\scalebox{1.0}[1]{
\begin{tabular}{lD{.}{.}{-2}D{.}{.}{-2}}
\toprule
&\multicolumn{2}{c}{Proportion of Male Births (\%)}\\
\cmidrule(rr){2-3}
Exposed trimesters &\multicolumn{1}{c}{(1)}&\multicolumn{1}{c}{(2)}\\\hline
$-6$th trimesters					&0.0011	&0.0093	\\
									&[1.0000]	&[0.7800]	\\
$-5$th trimesters					&-0.0480	&-0.0403	\\
									&[0.5200]	&[0.6850]	\\
$-4$th trimesters					&-0.0063	&0.0009	\\
									&[0.8350]	&[0.9550]	\\
$-3$th trimesters					&-0.0303	&-0.0236	\\
									&[0.4550]	&[0.5100]	\\
$-2$th trimesters					&-0.0091	&-0.0035	\\
									&[0.7300]	&[0.8500]	\\
$-1$th trimesters					&0.1338	&0.1383	\\
									&0.0750	&[0.0850]	\\\hdashline
\textbf{First trimesters}				&-0.0856*	&-0.0813*	\\
									&[0.0400]	&[0.0400]	\\
Second trimesters					&-0.0365	&-0.0327	\\
									&[0.2600]	&[0.2300]	\\
Third trimesters						&-0.0044	&-0.0011	\\
									&[0.8150]	&[0.9350]	\\\hdashline
$4$th trimesters					&-0.0133	&-0.0100	\\
									&[0.8200]	&[0.8300]	\\
$5$ trimesters						&-0.0407	&-0.0373	\\
									&[0.4900]	&[0.4950]	\\
$6$ trimesters						&-0.0874	&-0.0850	\\
									&[0.1250]	&[0.1250]	\\
$7$ trimesters						&0.0418	&0.0427	\\
									&[0.5000]	&[0.5050]	\\
$8$ trimesters						&0.0297	&0.0287	\\
									&[0.5950]	&[0.5700]	\\
$9$ trimesters						&-0.0439	&-0.0454	\\
									&[0.2750]	&[0.2600]	\\\hline
Time trend						
&\multicolumn{1}{c}{No}		&\multicolumn{1}{c}{Yes}\\
Period
&\multicolumn{1}{c}{Jan. 1916--}		&\multicolumn{1}{c}{Jan. 1916--}\\
&\multicolumn{1}{c}{Dec. 1922}		&\multicolumn{1}{c}{Dec. 1922}\\
Observations						
&\multicolumn{1}{c}{3864}		&\multicolumn{1}{c}{3864}\\
Number of prefectures
&\multicolumn{1}{c}{46}		&\multicolumn{1}{c}{46}\\
Number of months
&\multicolumn{1}{c}{84}		&\multicolumn{1}{c}{84}\\
Number of clusters
&\multicolumn{1}{c}{8}		&\multicolumn{1}{c}{8}\\\bottomrule
\end{tabular}
}
{\scriptsize
\begin{minipage}{250pt}
\setstretch{0.85}
* represents statistical significance at the 10\% level based on the $p$-values from the wild cluster bootstrap resampling method in brackets, respectively.
The data are clustered at the 8-area level in the bootstrap procedure.
The number of replications is fixed to 400 for all the specifications.\\
Notes: 
This table shows the results from equation~\ref{es}.
All the regressions include the prefecture fixed effect, year-month-specific fixed effect, and controls for rice yield, soy yield, milk production, coverage of doctors, and coverage of midwives.
The prefecture-specific linear time trend is also included in column (2).
All the regressions are weighted by the average number of births in each prefecture.
\end{minipage}
}
\end{center}
\end{table}
\begin{table}[h]
\def\arraystretch{0.95}
\begin{center}
\captionsetup{justification=centering}
\caption{Robustness to additional control variables: Effects of exposure to influenza on gender imbalance at births (\%) and infant mortality rates (\textperthousand)}
\label{tab:r_robust1}
\footnotesize
\scalebox{1.0}[1]{
\begin{tabular}{lD{.}{.}{-2}D{.}{.}{-2}D{.}{.}{-2}}
\toprule
&\multicolumn{1}{c}{(1) Proportion of}&\multicolumn{2}{c}{Infant mortality rate (\textperthousand)}\\
\cmidrule(rr){3-4}
&\multicolumn{1}{c}{male births (\%)}&\multicolumn{1}{c}{(2) Boys}	&\multicolumn{1}{c}{(3) Girls}\\\hline
Panel A: Including weather shocks						&&&\\
\hspace{5pt}FLUDR (First trimester)						&-0.1710$**$	&-10.4585		&-16.4366$**$	\\
														&[0.0200]		&[0.1200]		&[0.0400]			\\
\hspace{5pt}FLUDR (Second trimester)					&-0.0906		&-6.7775		&-1.4254			\\
														&[0.1800]		&[0.3733]		&[0.7467]			\\
\hspace{5pt}FLUDR (Third trimester)						&-0.0132		&16.8561$*$	&20.7474$*$		\\
														&[0.6950]		&[0.0800]		&[0.0533]			\\
Heat and cold waves						
&\multicolumn{1}{c}{Yes}		&\multicolumn{1}{c}{Yes}&\multicolumn{1}{c}{Yes}		\\
Period
&\multicolumn{1}{c}{Jan. 1918--Dec. 1920}
&\multicolumn{1}{c}{1916--22}
&\multicolumn{1}{c}{1916--22}\\
Observations						
&\multicolumn{1}{c}{1656}		
&\multicolumn{1}{c}{322}
&\multicolumn{1}{c}{322}\\
Number of prefectures (clusters)
&\multicolumn{1}{c}{46 (8)}		
&\multicolumn{1}{c}{46 (8)}
&\multicolumn{1}{c}{46 (8)}\\
Number of months or years
&\multicolumn{1}{c}{36}		
&\multicolumn{1}{c}{7}
&\multicolumn{1}{c}{7}\\
												&&&\\
Panel B: Coal consumption						&&&\\
\hspace{5pt}FLUDR (First trimester)				&-0.1530$**$	&-10.9522		&-16.1740		\\
												&[0.0150]		&[0.1867]		&[0.1200]		\\
\hspace{5pt}FLUDR (Second trimester)			&-0.0791		&-9.6917		&-3.8152		\\
												&[0.2000]		&[0.2800]		&[0.5333]		\\
\hspace{5pt}FLUDR (Third trimester)				&-0.0190		&19.4637$**$	&22.3079$**$	\\
												&[0.6150]		&[0.0133]		&[0.0133]		\\
Coal consumption						
&\multicolumn{1}{c}{Yes}		&\multicolumn{1}{c}{Yes}&\multicolumn{1}{c}{Yes}		\\
Period
&\multicolumn{1}{c}{Jan. 1918--Dec. 1920}
&\multicolumn{1}{c}{1916--22}
&\multicolumn{1}{c}{1916--22}\\
Observations						
&\multicolumn{1}{c}{1656}		
&\multicolumn{1}{c}{322}
&\multicolumn{1}{c}{322}\\
Number of prefectures (clusters)
&\multicolumn{1}{c}{46 (8)}		
&\multicolumn{1}{c}{46 (8)}
&\multicolumn{1}{c}{46 (8)}\\
Number of months or years
&\multicolumn{1}{c}{36}		
&\multicolumn{1}{c}{7}
&\multicolumn{1}{c}{7}\\
&&&\\
Panel C: School enrollment rate					&&&\\
\hspace{5pt}FLUDR (First trimester)				&-0.1523$***$	&-10.6847		&-15.7024		\\
												&[0.0050]		&[0.2000]		&[0.1333]		\\
\hspace{5pt}FLUDR (Second trimester)			&-0.0796		&-8.5298		&-2.0466		\\
												&[0.2100]		&[0.3200]		&[0.5200]		\\
\hspace{5pt}FLUDR (Third trimester)				&-0.0206		&18.5555$**$	&20.9650$**$	\\
												&[0.5550]		&[0.0133]		&[0.0133]		\\
School enrollment rate						
&\multicolumn{1}{c}{Yes}		&\multicolumn{1}{c}{Yes}		&\multicolumn{1}{c}{Yes}\\
Period
&\multicolumn{1}{c}{Jan. 1918--Dec. 1920}
&\multicolumn{1}{c}{1916--22}
&\multicolumn{1}{c}{1916--22}\\
Observations						
&\multicolumn{1}{c}{1656}		
&\multicolumn{1}{c}{322}
&\multicolumn{1}{c}{322}		\\
Number of prefectures (clusters)
&\multicolumn{1}{c}{46 (8)}		
&\multicolumn{1}{c}{46 (8)}
&\multicolumn{1}{c}{46 (8)}		\\
Number of months or years
&\multicolumn{1}{c}{36}		
&\multicolumn{1}{c}{7}
&\multicolumn{1}{c}{7}\\\bottomrule
\end{tabular}
}
{\scriptsize
\begin{minipage}{420pt}
\setstretch{0.85}
***, **, and * represent statistical significance at the 1\%, 5\%, and 10\% levels based on the $p$-values from the wild cluster bootstrap resampling method in brackets, respectively.
The data are clustered at the 8-area level in the bootstrap procedure.
The number of replications is fixed to 400 (150) for the regressions using the proportion of male births (infant mortality rates or proportion of male) as a dependent variable.\\
Notes:
The regressions in column (1) include the prefecture fixed effect, year-month-specific fixed effect, and prefecture-specific time trend, and are weighted by the average number of births in each prefecture.
The regressions in columns (2)--(3) include the prefecture fixed effect, year fixed effect, and prefecture-specific time trend, and are weighted by the average number of live births (of boys for column (2); of girls for column (3)) in each prefecture.
The regressions in column (4) include the prefecture-measured year specific fixed effect, age fixed effect, and prefecture-specific time trend, and are weighted by the average number of people in each prefecture.
All the regressions include baseline controls for rice yield, soy yield, milk production, coverage of doctors, and coverage of midwives.
\end{minipage}
}
\end{center}
\end{table}
\begin{table}[h]
\def\arraystretch{0.95}
\begin{center}
\captionsetup{justification=centering}
\caption{Alternative specification including flexible fixed effects: Effects of exposure to influenza on gender imbalance at births (\%) and infant mortality rates (\textperthousand)}
\label{tab:r_robust_flex}
\footnotesize
\scalebox{1.0}[1]{
\begin{tabular}{lD{.}{.}{-2}D{.}{.}{-2}D{.}{.}{-2}}
\toprule
&\multicolumn{1}{c}{(1) Proportion of}&\multicolumn{2}{c}{Infant mortality rate (\textperthousand)}\\
\cmidrule(rr){3-4}
&\multicolumn{1}{c}{male births (\%)}&\multicolumn{1}{c}{(2) Boys}	&\multicolumn{1}{c}{(3) Girls}\\\hline
\hspace{5pt}FLUDR (First trimester)						&-0.1972$***$	&-26.8666		&-35.2942$*$	\\
														&(0.0480)		&(17.7260)		&(15.0120)			\\
\hspace{5pt}FLUDR (Second trimester)					&-0.0380		&-18.4616		&-15.8027			\\
														&(0.0238)		&(13.2633)		&(11.5176)			\\
\hspace{5pt}FLUDR (Third trimester)						&0.0012		&30.3582$*$	&38.5032$**$		\\
														&(0.0383)		&(13.2714)		&(12.2813)			\\
Prefecture-by-year-by-quarter FE						
&\multicolumn{1}{c}{Yes}		&\multicolumn{1}{c}{--}&\multicolumn{1}{c}{--}		\\
Prefecture-by-period FE						
&\multicolumn{1}{c}{--}		&\multicolumn{1}{c}{Yes}&\multicolumn{1}{c}{Yes}		\\
Period
&\multicolumn{1}{c}{Jan. 1916--Dec. 1922}
&\multicolumn{1}{c}{1916--22}
&\multicolumn{1}{c}{1916--22}\\
Observations						
&\multicolumn{1}{c}{3864}		
&\multicolumn{1}{c}{322}
&\multicolumn{1}{c}{322}\\
Number of prefectures (clusters)
&\multicolumn{1}{c}{46 (8)}		
&\multicolumn{1}{c}{46 (8)}
&\multicolumn{1}{c}{46 (8)}\\
Number of months or years
&\multicolumn{1}{c}{84}		
&\multicolumn{1}{c}{7}
&\multicolumn{1}{c}{7}\\\bottomrule
\end{tabular}
}
{\scriptsize
\begin{minipage}{400pt}
\setstretch{0.85}
***, **, and * represent statistical significance at the 1\%, 5\%, and 10\% levels, respectively.
Standard errors reported in the parentheses are clustered at the 8-area level.\\
Notes:
All the regressions include baseline controls for rice yield, soy yield, milk production, coverage of doctors, and coverage of midwives.
The regression in column (1) includes the prefecture-by-year-by-quarter fixed effect and month fixed effect and is weighted by the average number of live births in each prefecture.
The regressions in columns (2)--(3) include the prefecture-by-period fixed effect and are weighted by the average number of live births (of boys for column (2); of girls for column (3)) in each prefecture.
The least squares dummy variable (LSDV) regressions are used.
Therefore, in column (1), the regression is weighted by the number of births in each prefecture-year-month cell; in column (2) ((3)), the regression is weighted by the number of males (female) live births in each prefecture-year cell.
\end{minipage}
}
\end{center}
\end{table}
\begin{table}[h]
\def\arraystretch{1.0}
\begin{center}
\captionsetup{justification=centering}
\caption{Robustness: Testing pretreatment trend assumptions for the annual-level datasets (the infant mortality rate (\textperthousand) and proportion of males (\%))}
\label{tab:r_robust_pretrend}
\scriptsize
\scalebox{0.95}[1]{
\begin{tabular}{lD{.}{.}{-2}D{.}{.}{-2}D{.}{.}{-2}D{.}{.}{-2}}
\toprule
&\multicolumn{2}{c}{Infant Mortality Rate (\textperthousand)}&\multicolumn{2}{c}{Proportion of males (\%)}\\
\cmidrule(rr){2-3}\cmidrule(rr){4-5}
&\multicolumn{1}{c}{(1)}&\multicolumn{1}{c}{(2)}&\multicolumn{1}{c}{(3)}&\multicolumn{1}{c}{(4)}\\\hline
FLUDR ($-1$ trimester: exposed before conception)	&-7.5279		&-4.9118		&-0.0041	&-0.0217	\\
													&[0.1067]		&[0.1600]		&[0.9267]	&[0.7933]	\\
FLUDR (First trimester)								&-11.8809		&-16.7499		&&\\
													&[0.1333]		&[0.1467]		&&\\
FLUDR (Second trimester)							&-6.1927		&-1.5072		&&\\
													&[0.3733]		&[0.7200]		&&\\
FLUDR (Third trimester)								&21.7186$**$	&23.7711$**$	&&\\
													&[0.0133]		&[0.0133]		&&\\
FLUDR (exposed \textit{in utero})					&				&				&0.0043		&0.0115		\\
													&				&				&[0.9133]		&[0.9133]		\\
FLUDR (exposed at age 0)							&				&				&-0.1500$**$	&-0.1606$**$	\\
													&				&				&[0.0200]		&[0.0200]		\\
FLUDR (exposed at age 1)							&				&				&-0.0103		&-0.0036		\\
													&				&				&[0.7133]		&[0.8067]		\\\hline
Prefecture fixed effect						
&\multicolumn{1}{c}{Yes}&\multicolumn{1}{c}{Yes}&\multicolumn{1}{c}{}&\multicolumn{1}{c}{}\\
Year fixed effect						
&\multicolumn{1}{c}{Yes}&\multicolumn{1}{c}{Yes}&\multicolumn{1}{c}{}&\multicolumn{1}{c}{}\\
Prefecture-year specific fixed effect						
&\multicolumn{1}{c}{}&\multicolumn{1}{c}{}&\multicolumn{1}{c}{Yes}&\multicolumn{1}{c}{Yes}\\
Age fixed effect						
&\multicolumn{1}{c}{}&\multicolumn{1}{c}{}&\multicolumn{1}{c}{Yes}&\multicolumn{1}{c}{Yes}\\
Heterogeneous trend across prefectures						
&\multicolumn{1}{c}{No}&\multicolumn{1}{c}{Yes}&\multicolumn{1}{c}{No}&\multicolumn{1}{c}{Yes}\\
Measured years
&\multicolumn{1}{c}{1916--1922}&\multicolumn{1}{c}{1916--1922}&\multicolumn{1}{c}{1925 \& 1930}&\multicolumn{1}{c}{1925 \& 1930}\\
Observations						
&\multicolumn{1}{c}{322}&\multicolumn{1}{c}{322}&\multicolumn{1}{c}{552}&\multicolumn{1}{c}{552}\\
Number of prefectures
&\multicolumn{1}{c}{46}&\multicolumn{1}{c}{46}&\multicolumn{1}{c}{46}&\multicolumn{1}{c}{46}\\
Number of years
&\multicolumn{1}{c}{7}&\multicolumn{1}{c}{7}&\multicolumn{1}{c}{7}&\multicolumn{1}{c}{7}\\
Number of clusters
&\multicolumn{1}{c}{8}&\multicolumn{1}{c}{8}	&\multicolumn{1}{c}{8}&\multicolumn{1}{c}{8}\\\bottomrule
\end{tabular}
}
{\scriptsize
\begin{minipage}{445pt}
\setstretch{0.85}
***, **, and * represent statistical significance at the 1\%, 5\%, and 10\% levels based on the $p$-values from the wild cluster bootstrap resampling method in brackets, respectively.
The data are clustered at the 8-area level in the bootstrap procedure.
The number of replications is fixed to 150 for all the specifications.\\
Notes: All the regressions include controls for rice yield, soy yield, milk production, coverage of doctors, and coverage of midwives.
FLUDR ($-1$ trimester: exposed before conception) refers to the 10- to 12-month weighted average of the influenza mortality rates before a birth.
FLUDR (First trimester), FLUDR (Second trimester), and FLUDR (Third trimester) are the weighted averages of the influenza mortality rates during the first, second, and third trimesters, respectively.
FLUDR (exposed \textit{in utero}) refers to the 9-month weighted average of the influenza mortality rates before a birth.
FLUDR (exposed at age 0) refers to the 12-month weighted average of the influenza mortality rates after a birth.
FLUDR (exposed at age 1) refers to the 13- to 24-month weighted average of the influenza mortality rates after a birth.
The regressions in columns (1)--(2) are weighted by the average number of live births in each prefecture.
The regressions in columns (3)--(4) are weighted by the average number of children in each prefecture-year cell.
\end{minipage}
}
\end{center}
\end{table}
\begin{table}[h]
\def\arraystretch{1.0}
\begin{center}
\captionsetup{justification=centering}
\caption{Subsample analysis: Effects of exposure to influenza on gender imbalance at births for each pandemic wave}
\label{tab:r_robust2}
\footnotesize
\scalebox{1.0}[1]{
\begin{tabular}{lD{.}{.}{-2}D{.}{.}{-2}D{.}{.}{-2}D{.}{.}{-2}}
\toprule
&\multicolumn{4}{c}{Proportion of male births (\%)}\\
\cmidrule(rrrr){2-5}
&\multicolumn{1}{c}{(1) 1st wave}&\multicolumn{1}{c}{(2) 1st wave}	&\multicolumn{1}{c}{(3) 2nd wave}&\multicolumn{1}{c}{(4) 2nd wave}\\\hline
FLUDR (First trimester)			&-0.2792$**$	&-0.2101$**$	&-0.1345	&-0.1058$*$	\\
								&[0.0100]		&[0.0200]		&[0.1200]	&[0.0950]	\\
FLUDR (Second trimester)		&-0.1379		&-0.0302		&-0.1130	&-0.0503	\\
								&[0.3000]		&[0.7350]		&[0.1300]	&[0.6050]	\\
FLUDR (Third trimester)			&-0.0516		&-0.0566		&-0.0676	&-0.0095	\\
								&[0.5950]		&[0.6200]		&[0.3100]	&[0.8750]	\\
Time trend						
&\multicolumn{1}{c}{No}		&\multicolumn{1}{c}{Yes}
&\multicolumn{1}{c}{No}		&\multicolumn{1}{c}{Yes}\\
Period
&\multicolumn{1}{c}{Oct. 1918--}
&\multicolumn{1}{c}{Oct. 1918--}
&\multicolumn{1}{c}{Dec. 1919--}
&\multicolumn{1}{c}{Dec. 1919--}\\
&\multicolumn{1}{c}{Nov. 1920}	
&\multicolumn{1}{c}{Nov. 1920}
&\multicolumn{1}{c}{Dec. 1920}
&\multicolumn{1}{c}{Dec. 1920}\\
Observations						
&\multicolumn{1}{c}{644}		
&\multicolumn{1}{c}{644}
&\multicolumn{1}{c}{598}
&\multicolumn{1}{c}{598}		\\
Number of prefectures
&\multicolumn{1}{c}{46}		
&\multicolumn{1}{c}{46}
&\multicolumn{1}{c}{46}
&\multicolumn{1}{c}{46}		\\
Number of months
&\multicolumn{1}{c}{14}		
&\multicolumn{1}{c}{14}
&\multicolumn{1}{c}{14}
&\multicolumn{1}{c}{14}		\\
Number of clusters
&\multicolumn{1}{c}{8}		
&\multicolumn{1}{c}{8}
&\multicolumn{1}{c}{8}
&\multicolumn{1}{c}{8}		\\\bottomrule
\end{tabular}
}
{\scriptsize
\begin{minipage}{420pt}
\setstretch{0.85}
***, **, and * represent statistical significance at the 1\%, 5\%, and 10\% levels based on the $p$-values from the wild cluster bootstrap resampling method in brackets, respectively.
The data are clustered at the 8-area level in the bootstrap procedure.
The number of replications is fixed to 400 for all the specifications.\\
Notes:
All the regressions include the prefecture fixed effect, year-month-specific fixed effect, and controls for rice yield, soy yield, milk production, coverage of doctors, and coverage of midwives.
The regression is weighted by the average number of births in each prefecture.
\end{minipage}
}
\end{center}
\end{table}

\clearpage
\section{Persistency: Evidence from Population Censuses} \label{sec:secc}
\setcounter{table}{0} \renewcommand{\thetable}{C.\arabic{table}}
\setcounter{figure}{0} \renewcommand{\thefigure}{C.\arabic{figure}}

\subsection{Data and Specification} \label{sec:secc1}

Thus far, I have found that fetal exposure to pandemic influenza decreased the proportion of male births.
In this section, I assess whether the gender imbalance at birth persisted into their teens.
The Population Censuses conducted in 1925 and 1930 documented the population by age and gender in each prefecture.
To investigate the potential lasting effects of fetal influenza exposure on the sex ratio, I digitize the data from some \textit{Kokuseich\=osah\=okoku} (Reports of the Population Census) and calculate the proportion of males aged 0--20 for each prefecture.
Online Appendix~\ref{sec:seca7} provides the details of these documents.

\begin{figure}[h]
\centering
\captionsetup{justification=centering,margin=0.5cm}
\subfloat[1925 Population Census]{\label{fig:pmale1925}\includegraphics[width=0.51\textwidth]{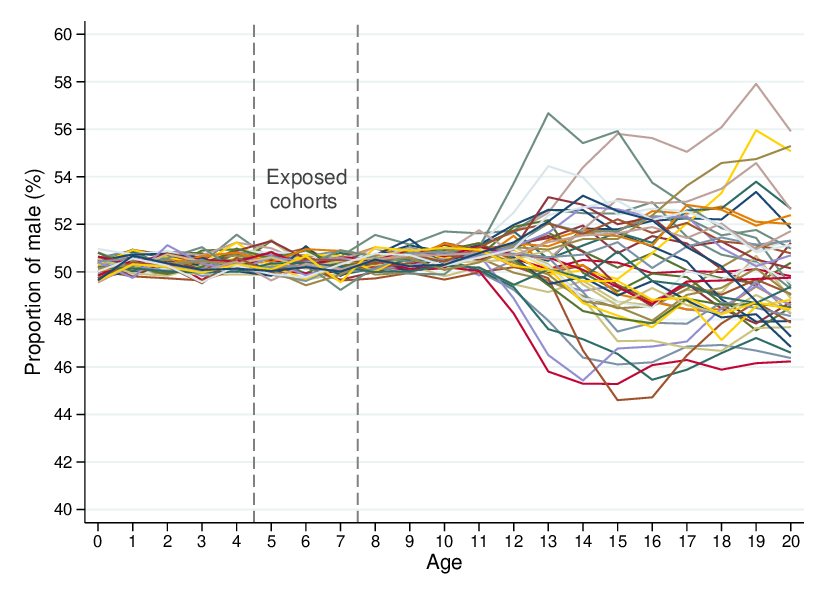}}
\subfloat[1930 Population Census]{\label{fig:pmale1930}\includegraphics[width=0.51\textwidth]{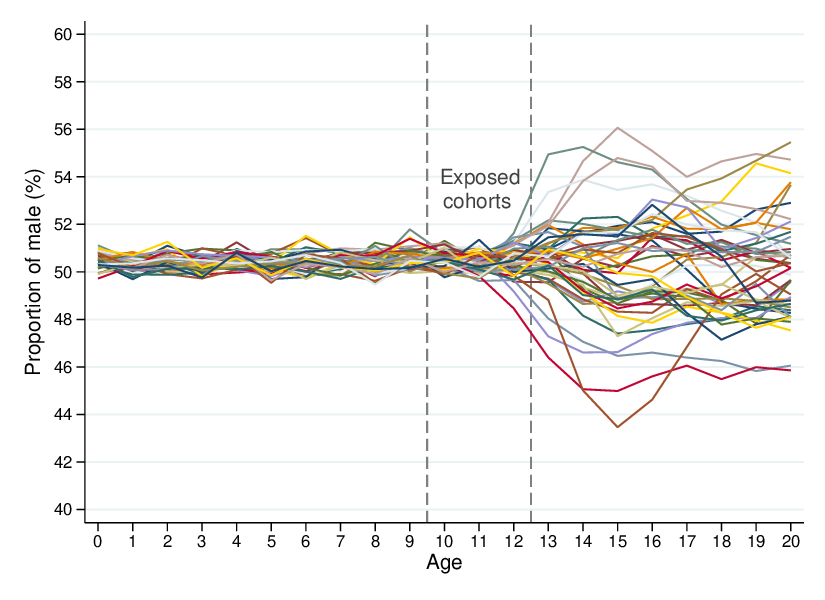}}
\caption{Proportion of males by prefecture and age (\%)}
\label{fig:pmale}
\scriptsize{\begin{minipage}{450pt}
\setstretch{0.85}
Notes: Figures~\ref{fig:pmale1925} and \ref{fig:pmale1930} illustrate the proportion of males in the 46 prefectures by age in 1925 and 1930, respectively.
Okinawa prefecture is not included in the sample.
Sources: Created by the authors from Statistics Bureau of the Cabinet (1929; 1933).
\end{minipage}}
\end{figure}

Figure~\ref{fig:pmale} shows the proportion of males in percentage points in each prefecture by age.
As shown, the variance in the sex ratio is stable until 12 years old because children graduate from primary school around then (Hijikata 1994).
After graduation, while some children go to higher schools, a large part of them begin to work, which creates a gender imbalance due to the flow of migrant workers.\footnote{The school enrollment rate for primary school was near 100\% and there were no significant differences in the rates across prefectures at that time in Japan. See Schneider and Ogasawara (2018) for finer details about primary school students in prewar Japan.}
This kind of internal migration after primary school age makes it difficult to analyze the potential long-run impacts of fetal influenza exposure on the sex ratio among teen workers because I use the prefecture-level aggregate dataset that does not have useful information on birthplace.
Therefore, I focus on the gender imbalance up to 12 years old.
I further trim the ages to improve identification.
Specifically, since children born between 1918 and 1920 were exposed to pandemic influenza, children aged 5--7 in the 1925 Population Census and 10--12 in the 1930 Population Census are defined as the exposed cohorts, as shown in Figures~\ref{fig:pmale1925} and \ref{fig:pmale1930}.
Considering this, I focus on the proportion of boys aged 5--7 and 10--12 years old in 1925 and 1930.
This means that my analytical sample includes children aged 5--7 years and 10--12 years born in 1913--1925.\footnote{Precisely, those children aged 5--7 years (10--12 years) in 1925 were born in 1918--1920 (1913--1916). Those children aged 5--7 years (10--12 years) in 1930 were born in 1918--1920 (1923--1925). Accordingly, to prepare the weighted influenza death rates in 1913--1925, I additionally digitize the 1912--1914 and 1923--1927 editions of the VSEJ and the SCDEJ (see Online Appendices~\ref{sec:seca1} and~\ref{sec:seca2}, respectively).}
Panel A of Table~\ref{tab:sum_sr} lists the summary statistics for the dependent variables used.

I begin my analysis by estimating the cohort effects of fetal exposure to pandemic influenza using the following specification:
\begin{eqnarray}\label{pmale1}
\footnotesize{
\begin{split}
s_{it-a} = \varpi + \chi \textit{Exposed}_{it-a} + \vx'_{it-a} \bm{\Delta} + \varphi_{it} + \eta_{a} + o_{it-a},
\end{split}
}
\end{eqnarray}
where $i$ indexes the prefecture, $t$ indexes the measured year, $a$ indexes the age, and thus $t-a$ indexes the cohort (birth year).
The variable $s$ is the proportion of males, $\textit{Exposed}$ is an indicator variable for the exposed cohorts (Figure~\ref{fig:pmale}), $\vx$ is a vector of the prefecture-birth year-level control variables, $\varphi$ is the prefecture-year-specific fixed effect, $\eta$ is the age fixed effect, and $o$ is a random error term.
I expect $\hat{\chi}$ to be negative and statistically significant, as it captures the cohort effects of fetal influenza exposure on the proportion of males.

My main specification is then designed to estimate the marginal effects of fetal influenza exposure on the proportion of males:
\begin{eqnarray}\label{pmale2}
\footnotesize{
\begin{split}
s_{it-a} = \Upsilon + \Xi_{0} \text{\textit{Weighted FLUDR}}^{\textit{in utero}}_{it-a} + \Xi_{1} \text{\textit{Weighted FLUDR}}^{\textit{Age 0}}_{it-a} + \Xi_{2} \text{\textit{Weighted FLUDR}}^{\textit{Age 1}}_{it-a}\\
 + \vx'_{it-a} \bm{\Theta} + \vartheta_{it} + \Omega_{a} + \Lambda_{it-a},
\end{split}
}
\end{eqnarray}
where $\textit{Weighted FLUDR}^{\textit{in utero}}$ is the weighted influenza death rate defined in equation \ref{wdr}.
$\textit{Weighted FLUDR}^{\textit{Age 0}}$ and $\textit{Weighted FLUDR}^{\textit{Age 1}}$ are the 12-month and 13- to 24-month weighted average of the rates after a birth, respectively.
I consider these rates because the sex ratio at primary school ages might have been affected by postnatal exposures to the pandemic influenza rather than by prenatal exposure, unlike the impacts on the secondary sex ratio and on the infant mortality rates.
In this specification, $\vx$ includes the same control variables used in equation~\ref{bs}: rice yield, soy yield, milk production, coverage of doctors, and coverage of midwives.
An important difference is that I use these variables to control for the variations in the \textit{birth} year (i.e., 1913--1925) rather than the \textit{measured} year (i.e., 1925 and 1930).
Therefore, these variables are used to control for the birth year heterogeneities in the potential wealth level and socioeconomic conditions that might be correlated with $\textit{Weighted FLUDR}$.
Panels B and C of Table~\ref{tab:sum_sr} show the summary statistics for the key and control variables, respectively.
On the contrary, the instantaneous effects, namely, any unobserved shocks in the prefecture-measured year cells such as local economic shocks, are captured by the prefecture-year-specific fixed effect $\vartheta$.
The age fixed effect, $\Omega$, captures the common trend in the proportion of males over time.
Thus, the identification assumption is that after controlling for these observed and unobserved factors, $\textit{Weighted FLUDR}$ is uncorrelated with the error term $\Lambda_{it-a}$.
Together with the randomness of the pandemics, my key variable is thus considered to be plausibly exogenous.
However, the specifications of both equations~\ref{pmale1} and \ref{pmale2} assume a common trend in the proportion of males across prefectures.
To relax this assumption, I therefore allow the trend of the dependent variable to vary across prefectures using the prefecture-specific trend, say $a\Psi_{i}$, in some of the specifications.

To address the potential spatial and prefecture-specific within correlations, I use the CRVE and cluster the standard errors at the 8-area level.
Since my data are a three-dimensional (i.e., prefecture-measured year-age) panel, this clustering can mitigate the potential correlations across cohorts.
To overcome the issue of the small number of clusters, I use the wild cluster bootstrap-t method for the statistical inference.
All the regressions are weighted by the average number of children in each prefecture-year cell.

\subsection{Results} \label{sec:secc2}

Table~\ref{tab:r_main_sr} presents the results.
Column (1) presents the estimates from equation~\ref{pmale1}, whereas column (5) shows the estimates from the specification including the prefecture-specific trend.
The estimate in column (1) shows that the exposed cohort, on average, exhibits a 0.16 percentage points lower proportion of males than the surrounding cohorts.
This becomes 0.17 percentage points if I relax the common trend assumption across prefectures in column (5), accounting for approximately 35\% of the standard deviation of the dependent variable (Panel A in Table~\ref{tab:sum_sr}).

The estimated coefficient of $\textit{Weighted FLUDR}^{\textit{in utero}}$ in column (2) is negative and statistically significant.
This result remains unchanged if I include the prefecture-specific trend in column (6), suggesting that \textit{in utero} exposure to the pandemic influenza can decrease the sex ratio of children at primary school ages.
However, the estimated coefficient of $\textit{Weighted FLUDR}^{\textit{in utero}}$ in column (3) is negative but statistically insignificant whereas the estimated coefficient of $\textit{Weighted FLUDR}^{\textit{Age 0}}$ is negative and statistically significant.
This result remains unchanged if I include the prefecture-specific trend in column (7).
Column (4) indicates that exposure to the pandemic influenza at age 1 did not have such a negative impact on the proportion of males in children of primary school ages.
Finally, column (8) confirms that this result is largely unchanged if I partially relax the common trend assumption.

The estimate in column (8) shows that a one standard deviation ($0.66$ in Panel B in Table~\ref{tab:sum_sr}) increase in the weighted influenza death rate (at age 0) decreases the proportion of males in these children by approximately $0.11$ percentage points ($0.66 \times 0.161$).
Further, the proportion of males decreases by approximately 0.43 percentage points ($2.7 \times 0.161$) in the case of exposure to the maximum influenza death rate.
This accounts for approximately 88\% of the standard deviation of the dependent variable and thus, is considered to be non-negligible in terms of its magnitude.

\subsection{Mechanism} \label{sec:secc3}

The foregoing results suggest that exposure to pandemic influenza during the first 12 months after birth has lasting effects on the sex ratio after birth, at least for children aged 5--12.
This implies that postnatal exposure to pandemic influenza might have increased the infant mortality rate, especially that of the exposed boys.
This is considered to be consistent with the results in the previous section.
In the last section, I found that \textit{in utero} exposure (note: not postnatal exposure) to the pandemic influenza had increased the infant mortality rate, especially that of the girls, under the scarring mechanism.
This means that the surviving female infants might have been healthier than their male counterparts.
In other words, male infants could be more vulnerable to postnatal exposures.
This type of interpretation seems reasonable because there were two waves of epidemics in the case of pre-war Japan (Figure~\ref{fig:ts_flu}): the infants exposed to the first wave \textit{in utero} were at age 0 when the second wave arrived.
To test this potential mechanism, I regress infant mortality on the measured influenza mortality using the 1920--1921 subsample that included the infants who were exposed to the first wave and who might have also been impacted by the second wave.
Table~\ref{tab:r_mechanism} shows the results.
The estimated coefficients are positive in both columns (1) and (2) but the estimate in column (2) are statistically insignificant.
This implies that the second wave might have increased the infant mortality rates of boys.

\begin{landscape}
\begin{table}[]
\def\arraystretch{1.0}
\begin{center}
\caption{Summary statistics: Estimating the persistent effects on the proportion of males aged 5--12}
\label{tab:sum_sr}
\footnotesize
\scalebox{1.0}[1]{
\begin{tabular}{lcD{.}{.}{2}D{.}{.}{2}D{.}{.}{2}D{.}{.}{2}D{.}{.}{2}D{.}{.}{2}}
\toprule
&Unit	&\multicolumn{1}{c}{Mean}&\multicolumn{1}{c}{Std. Dev.}&\multicolumn{1}{c}{Min}&\multicolumn{1}{c}{Max}&\multicolumn{1}{c}{Observations}\\\hline

Panel A: Dependent variable&&&&&&\\
\hspace{10pt}Proportion of males (per 100 people)							&Prefecture-year-age	&50.47	&0.49	&48.25	&53.73	&552\\
&&&&&&\\
Panel B: Influenza severity&&&&&&\\
\hspace{10pt}Exposed cohort (dummy variable)								&Prefecture-birth year	&0.50	&		&		&		&552\\
\hspace{10pt}Weighted FLUDR (exposed \textit{in utero})					&Prefecture-birth year	&0.54	&0.69	&0.00	&2.62	&552\\
\hspace{10pt}Weighted FLUDR (exposed at age 0) 							&Prefecture-birth year	&0.57	&0.66	&0.01	&2.70	&552\\
\hspace{10pt}Weighted FLUDR (exposed at age 1)							&Prefecture-birth year	&0.37	&0.53	&0.01	&2.71	&552\\
&&&&&&\\\hline
&\multicolumn{1}{c}{}&\multicolumn{2}{c}{Full sample}&\multicolumn{3}{c}{Balancing tests}\\
\cmidrule(rr){3-4}\cmidrule(rrr){5-7}
Panel C: Control variables
&\multicolumn{1}{c}{Frequency}
&\multicolumn{1}{c}{Mean}&\multicolumn{1}{c}{Std. Dev.}
&\multicolumn{1}{c}{$>$75pct}&\multicolumn{1}{c}{$\leq$ 75pct}&\multicolumn{1}{c}{Diff. [p-value]}\\\hline
\hspace{10pt}Rice yield per hectare (hectoliter)									&Annual		&33.65	&6.92	&34.52	&33.36	&\multicolumn{1}{c}{-1.15 [0.1426]}\\
\hspace{10pt}Soy yield per hectare (hectoliter)									&Annual		&16.03	&3.93	&16.65	&15.82	&\multicolumn{1}{c}{-0.83 [0.0637]}\\
\hspace{10pt}Milk production per capita (liter)									&Annual		&1.07	&1.10	&0.91	&1.13	&\multicolumn{1}{c}{0.21 [0.0858]}\\
\hspace{10pt}Coverage of doctors (per 100 people)								&Annual		&0.07	&0.03	&0.08	&0.07	&\multicolumn{1}{c}{-0.00 [0.6610]}\\
\hspace{10pt}Coverage of midwives (per 100 people)							&Annual		&0.06	&0.02	&0.06	&0.06	&\multicolumn{1}{c}{0.00 [0.5830]}\\
\hspace{10pt}Observations														&			&\multicolumn{2}{c}{~~414}&\multicolumn{1}{c}{~~~311}	&\multicolumn{1}{c}{~~~~103}	&\\\bottomrule
\end{tabular}
}
{\scriptsize
\begin{minipage}{545pt}
\setstretch{0.84}Notes:
Panel A reports the summary statistics for the proportion of males (\%) observed at the prefecture-year-age level.
The summary statistics for children aged 5--7 years and 10--12 years in both 1925 and 1930 are listed in this table.
Panel B reports the summary statistics for the prefecture-birth year-level indicator variable and weighted influenza death rate (per 10,000 people).
Regarding the indicator variable, children aged 5--7 in 1925 and 10--12 in 1930 are defined as the exposed cohort.
Weighted FLUDR (exposed \textit{in utero}) refers to the 9-month weighted average of the influenza mortality rates before a birth.
Weighted FLUDR (exposed at age 0) refers to the 12-month weighted average of the influenza mortality rates after a birth.
Weighted FLUDR (exposed at age 1) refers to the 13- to 24-month weighted average of the influenza mortality rates after a birth.
Panel C reports the summary statistics for the prefecture-birth year-level control variables.
The number of observations is 414, because there are 46 prefectures $\times$ 9 birth cohorts in my sample.
In the balancing test, $>$75pct and $\leq$75pct indicate the influenza death rates above and less than or equal to the 75 percentile, respectively.
Sources: The dependent variables are from the VSEJ (1912--1925 editions).
The influenza death rates are from the SCDEJ (1912--1927 editions); VSEJ (1912--1927 editions); Statistical Survey Department, Statistics Bureau, Ministry of Internal Affairs and Communications (database).
The control variables are from Kayo (1983) and Statistics Bureau of the Cabinet (1914--1927).
\end{minipage}
}
\end{center}
\end{table}
\end{landscape}
\begin{landscape}
\begin{table}[h]
\def\arraystretch{1.0}
\begin{center}
\captionsetup{justification=centering}
\caption{Effects of fetal influenza exposure on the proportion of males (\%)\\ between 5 and 12 years old}
\label{tab:r_main_sr}
\footnotesize
\scalebox{0.90}[1]{
\begin{tabular}{lD{.}{.}{-2}D{.}{.}{-2}D{.}{.}{-2}D{.}{.}{-2}D{.}{.}{-2}D{.}{.}{-2}D{.}{.}{-2}D{.}{.}{-2}}
\toprule
&\multicolumn{8}{c}{Dependent variable: Proportion of males (\%)}\\
\cmidrule(rrrrrrrrrr){2-9}
&\multicolumn{1}{c}{(1)}&\multicolumn{1}{c}{(2)}&\multicolumn{1}{c}{(3)}&\multicolumn{1}{c}{(4)}&\multicolumn{1}{c}{(5)}&\multicolumn{1}{c}{(6)}&\multicolumn{1}{c}{(7)}&\multicolumn{1}{c}{(8)}\\\hline
Exposed cohort								&-0.156$**$	&				&				&				&-0.170$**$	&				&				&			\\
											&[0.020]		&				&				&				&[0.020]		&				&				&			\\
FLUDR (exposed \textit{in utero})			&				&-0.070$**$	&0.003			&0.001			&				&-0.078$**$	&-0.005		&-0.005	\\
											&				&[0.020]		&[0.993]		&[1.000]		&				&[0.020]		&[0.673]		&[0.687]	\\
FLUDR (exposed at age 0)					&				&				&-0.156$**$	&-0.150$**$	&				&				&-0.162$**$	&-0.161$**$\\
											&				&				&[0.020]		&[0.020]		&				&				&[0.020]		&[0.020]	\\
FLUDR (exposed at age 1)					&				&				&				&-0.010		&				&				&				&-0.002	\\
											&				&				&				&[0.727]		&				&				&				&[0.767]	\\\hline
Prefecture-year-specific fixed effect			
&\multicolumn{1}{c}{Yes}&\multicolumn{1}{c}{Yes}&\multicolumn{1}{c}{Yes}&\multicolumn{1}{c}{Yes}
&\multicolumn{1}{c}{Yes}&\multicolumn{1}{c}{Yes}&\multicolumn{1}{c}{Yes}&\multicolumn{1}{c}{Yes}\\
Age fixed effect						
&\multicolumn{1}{c}{Yes}&\multicolumn{1}{c}{Yes}&\multicolumn{1}{c}{Yes}&\multicolumn{1}{c}{Yes}
&\multicolumn{1}{c}{Yes}&\multicolumn{1}{c}{Yes}&\multicolumn{1}{c}{Yes}&\multicolumn{1}{c}{Yes}\\
Heterogeneous trend across prefectures						
&\multicolumn{1}{c}{No}&\multicolumn{1}{c}{No}&\multicolumn{1}{c}{No}&\multicolumn{1}{c}{No}
&\multicolumn{1}{c}{Yes}&\multicolumn{1}{c}{Yes}&\multicolumn{1}{c}{Yes}&\multicolumn{1}{c}{Yes}\\
Observations				
&\multicolumn{1}{c}{552}&\multicolumn{1}{c}{552}&\multicolumn{1}{c}{552}&\multicolumn{1}{c}{552}
&\multicolumn{1}{c}{552}&\multicolumn{1}{c}{552}&\multicolumn{1}{c}{552}&\multicolumn{1}{c}{552}\\
Number of prefectures
&\multicolumn{1}{c}{46}&\multicolumn{1}{c}{46}&\multicolumn{1}{c}{46}&\multicolumn{1}{c}{46}
&\multicolumn{1}{c}{46}&\multicolumn{1}{c}{46}&\multicolumn{1}{c}{46}&\multicolumn{1}{c}{46}\\
Number of clusters
&\multicolumn{1}{c}{8}&\multicolumn{1}{c}{8}&\multicolumn{1}{c}{8}&\multicolumn{1}{c}{8}
&\multicolumn{1}{c}{8}&\multicolumn{1}{c}{8}&\multicolumn{1}{c}{8}&\multicolumn{1}{c}{8}\\
Measured years
&\multicolumn{1}{c}{1925 \& 1930}&\multicolumn{1}{c}{1925 \& 1930}&\multicolumn{1}{c}{1925 \& 1930}&\multicolumn{1}{c}{1925 \& 1930}
&\multicolumn{1}{c}{1925 \& 1930}&\multicolumn{1}{c}{1925 \& 1930}&\multicolumn{1}{c}{1925 \& 1930}&\multicolumn{1}{c}{1925 \& 1930}\\
Ages
&\multicolumn{1}{c}{5--7 \& 10--12}&\multicolumn{1}{c}{5--7 \& 10--12}&\multicolumn{1}{c}{5--7 \& 10--12}&\multicolumn{1}{c}{5--7 \& 10--12}
&\multicolumn{1}{c}{5--7 \& 10--12}&\multicolumn{1}{c}{5--7 \& 10--12}&\multicolumn{1}{c}{5--7 \& 10--12}&\multicolumn{1}{c}{5--7 \& 10--12}\\\bottomrule
\end{tabular}
}
{\scriptsize
\begin{minipage}{640pt}
\setstretch{0.85}
***, **, and * represent statistical significance at the 1\%, 5\%, and 10\% levels based on the $p$-values from the wild cluster bootstrap resampling method in brackets, respectively.
The data are clustered at the 8-area level in the bootstrap procedure.
The number of replications is fixed to 150 for all the specifications.\\
Notes:
This table shows the results from equations~\ref{pmale1} and~\ref{pmale2}.
Exposed cohorts are 5--7-year-olds in 1925 and 10--12-year-olds in 1930.
All the regressions include the prefecture-measured year-specific fixed effect, age fixed effect, and birth year controls for rice yield, soy yield, milk production, coverage of doctors, and coverage of midwives.
All the regressions are weighted by the average number of children in each prefecture-year cell.
\end{minipage}
}
\end{center}
\end{table}
\end{landscape}
\begin{table}[h]
\def\arraystretch{1.0}
\begin{center}
\captionsetup{justification=centering}
\caption{Testing mechanism behind the postnatal exposure effects}
\label{tab:r_mechanism}
\footnotesize
\scalebox{1.0}[1]{
\begin{tabular}{lD{.}{.}{-2}D{.}{.}{-2}}
\toprule
&\multicolumn{2}{c}{Infant mortality rate (\textperthousand) in the 2nd wave}\\
\cmidrule(rr){2-3}
Exposed trimesters
&\multicolumn{1}{c}{(1) Boys}	&\multicolumn{1}{c}{(2) Girls}\\\hline
Influenza mortality rate (in measured years)			&5.484$*$		&4.488		\\
													&[0.080]		&[0.120]	\\\hline
Measured years
&\multicolumn{1}{c}{1920--21}		&\multicolumn{1}{c}{1920--21}\\
Observations						
&\multicolumn{1}{c}{96}			&\multicolumn{1}{c}{96}\\
Number of prefectures
&\multicolumn{1}{c}{46}			&\multicolumn{1}{c}{46}\\
Number of years
&\multicolumn{1}{c}{2}			&\multicolumn{1}{c}{2}\\
Number of clusters
&\multicolumn{1}{c}{8}				&\multicolumn{1}{c}{8}\\\bottomrule
\end{tabular}
}
{\scriptsize
\begin{minipage}{400pt}
\setstretch{0.85}
***, **, and * represent statistical significance at the 1\%, 5\%, and 10\% levels based on the $p$-values from the wild cluster bootstrap resampling method in brackets, respectively.
The data are clustered at the 8-area level in the bootstrap procedure.
The number of replications is fixed to 50 for all the specifications.\\
Notes: 
This table shows the estimates from the regression of the infant mortality rate on the log-transformed influenza mortality rate (in measured years).
All the regressions include the prefecture fixed effect, year fixed effect, and controls for rice yield, soy yield, milk production, coverage of doctors, and coverage of midwives.
The regressions in columns (1)--(2) are weighted by the average number of live births (of boys in column (1); of girls in column (2)) in each prefecture.
\end{minipage}
}
\end{center}
\end{table}

\clearpage
\renewcommand{\refname}{{\large References, Documents, Statistical Reports, and Database}}
\begin{spacing}{0.91}

\end{spacing}

\end{document}